\documentclass[11pt,a4paper]{article}
\usepackage[margin=1in]{geometry}
\usepackage{amsthm}
\usepackage{amsmath}
\usepackage{newtxtext}
\usepackage{newtxmath}
\usepackage{setspace}
\usepackage{authblk}
\usepackage{subcaption}
\usepackage{natbib}
\setcitestyle{authoryear,round}
\usepackage[english]{babel}
\usepackage{graphicx}
\usepackage{float}
\usepackage[shortlabels]{enumitem}
\usepackage{tcolorbox}
\usepackage{tikz}
\usetikzlibrary{matrix}
\usepackage{hyperref}

\makeatletter

\newcommand{\Rmnum}[1]{\expandafter\@slowromancap\romannumeral #1@}
\makeatother

\theoremstyle{definition}

\title{Long-Range Dependence in Financial Markets: Empirical Evidence and Generative Modeling Challenges}
\author{Yifan He\thanks{Correspondence: \href{mailto:Yifan.He@ttu.edu}{Yifan.He@ttu.edu}}}
\author{Svetlozar T. Rachev}

\affil{\footnotesize Department of Mathematics and Statistics, Texas Tech University, Lubbock, TX 79409, USA}

\begin{document}

\date{\small June 9, 2026}
\maketitle

\begin{abstract}
\noindent This study provides an empirical investigation of long-range dependence (LRD) in financial markets and evaluates the ability of deep generative models to reproduce such temporal structures. Using daily data from three sectors--equity (S\&P 500, DAX, Nikkei 225), commodities (Wheat, Corn, Soybeans), and energy (UNG, USO, XLE)--we examine LRD through rescaled range (R/S) analysis, detrended fluctuation analysis (DFA), segmented multifractal analysis around the COVID-19 period, and an ARFIMA--FIGARCH model with Student's $t$-distributed innovations. The evidence suggests that while mean returns exhibit limited persistence, pronounced long memory is observed in conditional volatility across most assets, and equity-market scaling properties change non-negligibly after 2020. Building on these findings, we assess whether Quant Generative Adversarial Networks (Quant GANs) can learn and reproduce these stylized temporal dependencies against econometric and resampling benchmarks. Although the generated series reproduce heavy-tailed return distributions and aspects of volatility clustering, they do not consistently capture the magnitude and persistence structure of LRD observed in real data. These results highlight an important limitation of deep generative architectures in modeling slow-decaying dependence structures and underscore the need for explicit long-memory mechanisms when synthetic financial data are intended for risk management or long-horizon forecasting.\par\medskip\noindent
\end{abstract}

\section{Introduction}
The concept of long-range dependence (LRD), characterized by a slow, hyperbolic decay of autocorrelations, describes a persistent memory effect in stochastic processes whereby past realizations exert a lasting influence on future outcomes. Originally studied by Hurst in hydrology in the context of the Nile River's reservoir capacity~\citep{hurst1951long}, the notion was later formalized through rescaled range (R/S) analysis by~\cite{mandelbrot1968noah}. Since then, LRD has been widely investigated across disciplines including geophysics, telecommunications, and economics~\citep{doukhan2002theory, beran2017statistics}. In financial markets, the existence of LRD--particularly in volatility--poses challenges to the weak-form efficient market hypothesis and carries important implications for asset pricing, risk measurement, and portfolio management~\citep{lo1991long}.

Parallel to the development of statistical tools for measuring LRD, machine learning has experienced rapid progress in generative modeling. Among these advances, generative adversarial networks (GANs), introduced by~\cite{goodfellow2014generative}, have emerged as a powerful framework for learning complex data distributions through an adversarial training process between a generator and a discriminator. For sequential data, recent studies have adapted adversarial learning to preserve temporal structure through supervised embedding losses, causal optimal transport, signature-based Wasserstein metrics, self-attention mechanisms, and convolutional representation learning~\citep{yoon2019timegan, xu2020cotgan, ni2021sigwasserstein, jeha2021psagan, huang2023tcgan}. These developments show that GANs are not merely image-generation tools, but are increasingly used for time-series representation learning, data augmentation, classification, clustering, and scenario generation. This is especially relevant in financial applications, where synthetic data can support stress testing, scenario generation, privacy-preserving model development, and robustness checks under limited historical samples. This paradigm has been extended to financial time series by~\cite{wiese2020quant}, who proposed Quant GANs based on temporal convolutional networks (TCNs)~\citep{bai2018empirical}. Quant GANs are specifically designed to reproduce stylized facts of financial returns, such as heavy tails and volatility clustering, making them attractive tools for synthetic data generation in quantitative finance.

Despite their success in replicating distributional features and short-term dynamics, an important question remains largely unexplored: can deep generative models faithfully learn and reproduce LRD? Unlike heavy tails or volatility bursts, LRD is a structural temporal property manifested through slowly decaying correlations over extended horizons. Capturing such dependence requires the model to internalize multi-scale temporal information, which may not be explicitly enforced by standard adversarial training objectives. This gap motivates a systematic empirical evaluation of deep generative models from a long-memory perspective.

This paper addresses the above question through a two-stage empirical framework. First, we establish the existence and characteristics of LRD in a diversified set of financial assets spanning equity, commodity, and energy markets. This stage combines classical scaling diagnostics, segmented analysis around the COVID-19 period, and parametric long-memory volatility modeling. Second, we evaluate the capability of Quant GANs to reproduce these properties using simulated path analysis and benchmark comparisons against econometric and resampling procedures. The remainder of the paper is organized as follows. Section~\ref{sec:data} introduces the data and preprocessing procedures. Section~\ref{sec:non_normal_return} examines the non-normal nature of returns.
Section~\ref{sec:measure_LRD} presents the methodologies for measuring LRD, including R/S, DFA, segmented scaling analysis, and ARFIMA--FIGARCH models. Section~\ref{sec:learn_LRD} evaluates the empirical performance of Quant GANs, compares simulated series with real data and benchmark generators, and discusses the associated computational workload. Section~\ref{sec:conclusion} concludes with economic interpretations and implications for risk management and policy.

\section{The Data}\label{sec:data}
This section introduces the datasets used in this study. In Section~\ref{sec:data_source}, we describe the selection of assets and their data sources. In Section~\ref{sec:data_preprocessing}, we outline the preprocessing procedures applied to the raw price series prior to the empirical analysis.

\subsection{Data Sources and Selection}\label{sec:data_source}
All data employed in this study are obtained from Yahoo Finance\footnote{Official website: \url{https://finance.yahoo.com/}}, which provides publicly accessible and widely used financial time series. We consider a total of nine assets, drawn from three major asset classes: equity markets, commodity markets, and energy markets. For each asset class, three representative instruments are selected in order to ensure broad market coverage and cross-market comparability.

For the equity markets, we select three major stock indices: 
\begin{itemize}[noitemsep, topsep=0pt]
\item \textbf{S\&P 500} (ticker: \textasciicircum GSPC), representing the U.S. equity market,
\item \textbf{DAX} (ticker: \textasciicircum GDAXI), representing the European equity market, and
\item \textbf{Nikkei 225} (ticker: \textasciicircum N225), representing the Asia-Pacific equity market.
\end{itemize}
These indices are among the most liquid and influential benchmarks in their respective regions and together provide a comprehensive view of equity market dynamics across the world's major economic areas.

For the commodity market, we focus on three key agricultural commodities:
\begin{itemize}[noitemsep, topsep=0pt]
\item \textbf{Wheat} (ticker: WEAT),
\item \textbf{Corn} (ticker: CORN), and
\item \textbf{Soybeans} (ticker: SOYB).
\end{itemize}
These commodities play a central role in global agricultural production and trade, and their price dynamics are commonly used to characterize stylized facts in commodity markets.

For the energy market, we select the following instruments:
\begin{itemize}[noitemsep, topsep=0pt]
\item \textbf{United States Natural Gas Fund} (ticker: UNG), which tracks natural gas prices,
\item \textbf{United States Oil Fund} (ticker: USO), which tracks crude oil prices, and
\item \textbf{Energy Select Sector SPDR Fund} (ticker: XLE), which represents a diversified portfolio of major U.S. energy sector companies.
\end{itemize}
Together, these assets capture different segments of the energy market, including fossil fuels and equity-based exposure to the energy sector, and are widely regarded as representative benchmarks for energy-related financial dynamics.

\subsection{Data Preprocessing}\label{sec:data_preprocessing}
Due to differences in trading calendars across asset classes and geographical regions, trading days are not perfectly aligned across markets. In particular, equity markets in the United States, Europe, and the Asia-Pacific region observe different public holidays, which implies that a given calendar date may correspond to a trading day in one market but not in others. Similar issues arise across commodity and energy markets.

To ensure comparability and to facilitate cross-asset analysis, we synchronize all price series by retaining only those dates on which all nine assets are simultaneously traded. Observations corresponding to dates with missing values for any asset are removed. This procedure yields a common set of trading days shared across all assets and eliminates distortions arising from non-synchronous trading.

Following this alignment, the resulting dataset consists of 3,319 daily observations, spanning the period from September 20, 2011\footnote{September 20, 2011 is the earliest date on which all nine assets considered in this study have available trading data in Yahoo Finance.} to December 30, 2025.

Finally, to allow for meaningful comparison across assets with different price levels and units, each price series is normalized such that its initial value equals one. This normalization can be interpreted as assuming a simultaneous initial investment of \$1 in each asset at the beginning of the sample period. Subsequent analyses are therefore conducted on normalized price dynamics that are directly comparable across markets and asset classes.

\begin{figure}[htbp]
\centering
\includegraphics[width=0.75\textwidth]{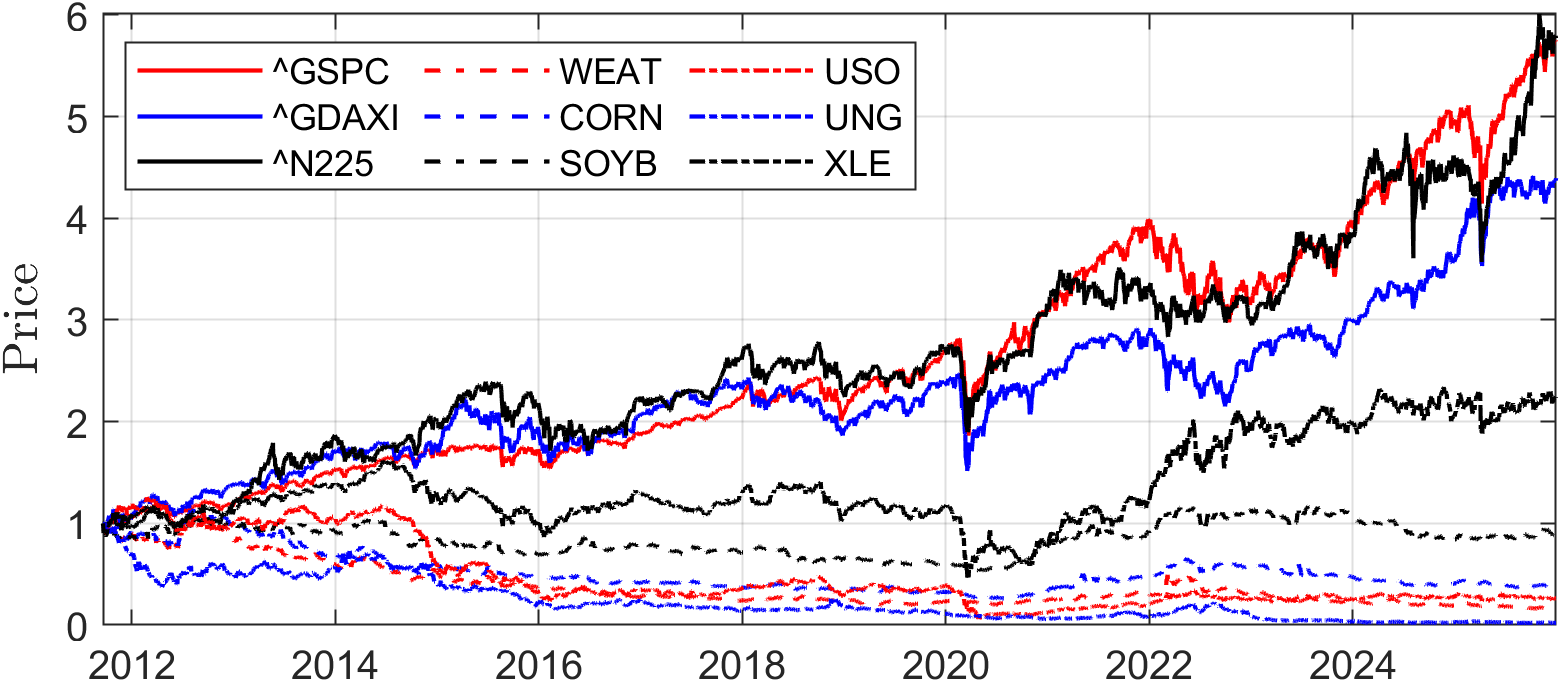}
\caption{Daily price trajectories of the nine assets, assuming an initial investment of \$1 on September 20, 2011}
\label{fig:price_evolution}
\end{figure}

Figure~\ref{fig:price_evolution} illustrates the normalized daily price trajectories of the nine selected assets from September 20, 2011 to December 30, 2025. Solid lines correspond to the three equity market indices, dashed lines represent the three commodity assets, and dash--dot lines denote the three energy market instruments.
The figure reveals substantial cross-asset heterogeneity. The three equity indices follow clear long-run upward trajectories: relative to their initial levels, both the S\&P 500 (\textasciicircum GSPC) and the Nikkei 225 (\textasciicircum N225) increase by nearly 500\%, while the DAX (\textasciicircum GDAXI) records an increase exceeding 300\%. Commodity and energy instruments display weaker and more uneven long-horizon performance. With the exception of the Energy Select Sector SPDR Fund (XLE), which more than doubles in value, the remaining five non-equity assets decline relative to their initial levels. The most pronounced case is the United States Natural Gas Fund (UNG), whose cumulative loss exceeds 95\% by the end of the sample.

Beyond these level differences, the solid-line equity trajectories display visible changes in slope and fluctuation amplitude around the onset of the COVID-19 period. Because such structural changes may affect scaling behavior and self-similarity, the full-sample LRD estimates are later complemented by a segmented scaling analysis for the three equity indices.

\section{The Non-Normal Nature of Returns}\label{sec:non_normal_return}
In Section~\ref{sec:data_preprocessing}, we briefly examined the daily price series of the nine selected assets. A subsequent discussion of their returns naturally follows.

There are two commonly used types of returns. The first is the \textit{arithmetic return}, also referred to as the \textit{simple return}:
\begin{equation*}
R_t := \frac{P_t - P_{t-1}}{P_{t-1}} = \frac{P_t}{P_{t-1}} - 1,
\end{equation*}
where $P_t$ denotes the asset's price at time $t$ and $R_t$ denotes the corresponding arithmetic return. The second type is the \textit{logarithmic return} (or \textit{log-return}), defined by
\begin{equation*}
r_t:= \ln\left(\frac{P_t}{P_{t-1}}\right) = \ln P_t - \ln P_{t-1},
\end{equation*}
where $r_t$ denotes the log-return at time $t$.

Log-returns are often preferred in quantitative finance for several reasons. First, they are time-additive, meaning that the log-return over multiple periods equals the sum of the log-returns of the individual periods, which simplifies the modeling and analysis~\citep{tsay2014introduction}. Second, log-returns approximate continuously compounded returns, making them more consistent with many financial models~\citep{chambers2015alternative,butler2016multinational}. Lastly, they tend to better handle large price fluctuations and are more suitable for statistical analysis under the assumption of normally distributed returns. In this study, unless otherwise specified, \textit{all returns refer to log-returns.}

\begin{figure}[htbp]
\centering
\begin{subfigure}[b]{0.32\textwidth}
\includegraphics[width=\textwidth]{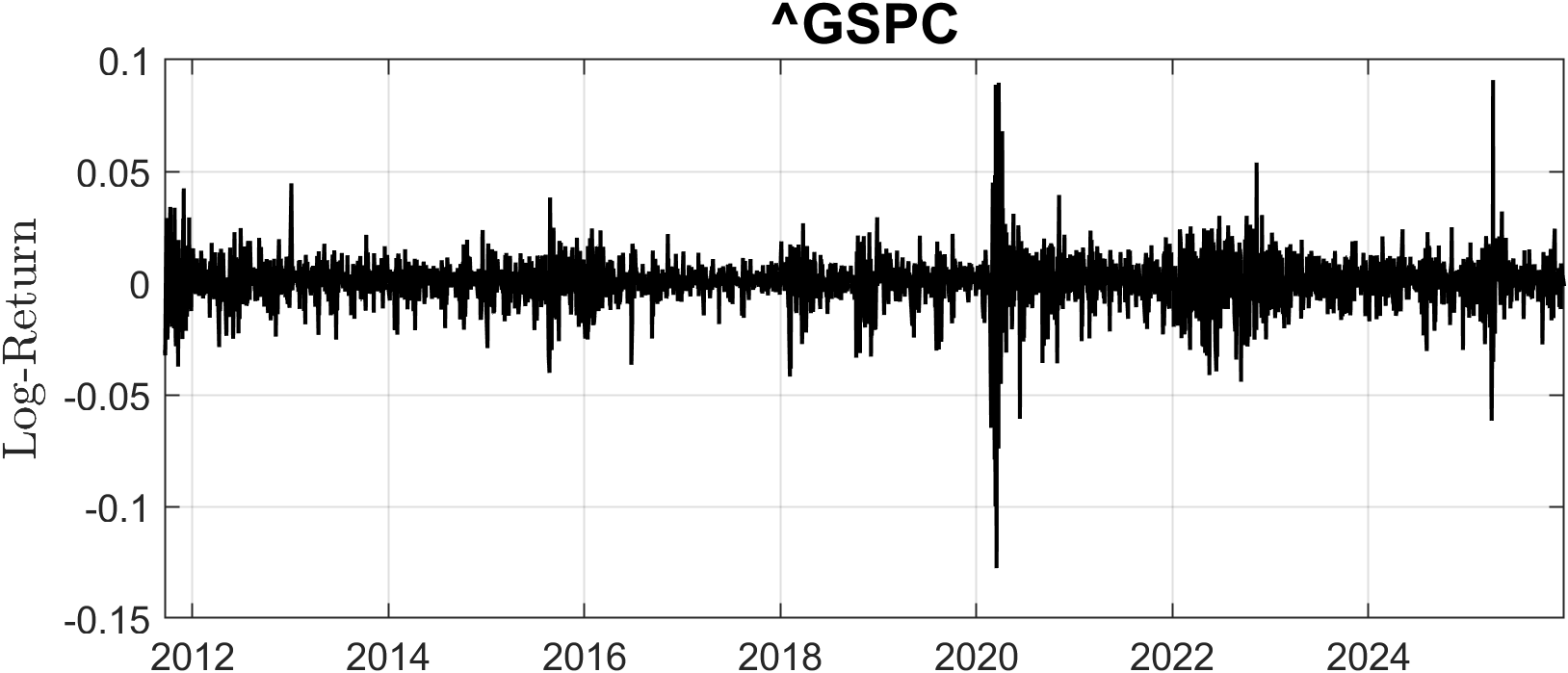}
\caption{\textasciicircum GSPC}
\end{subfigure}
\hfill
\begin{subfigure}[b]{0.32\textwidth}
\includegraphics[width=\textwidth]{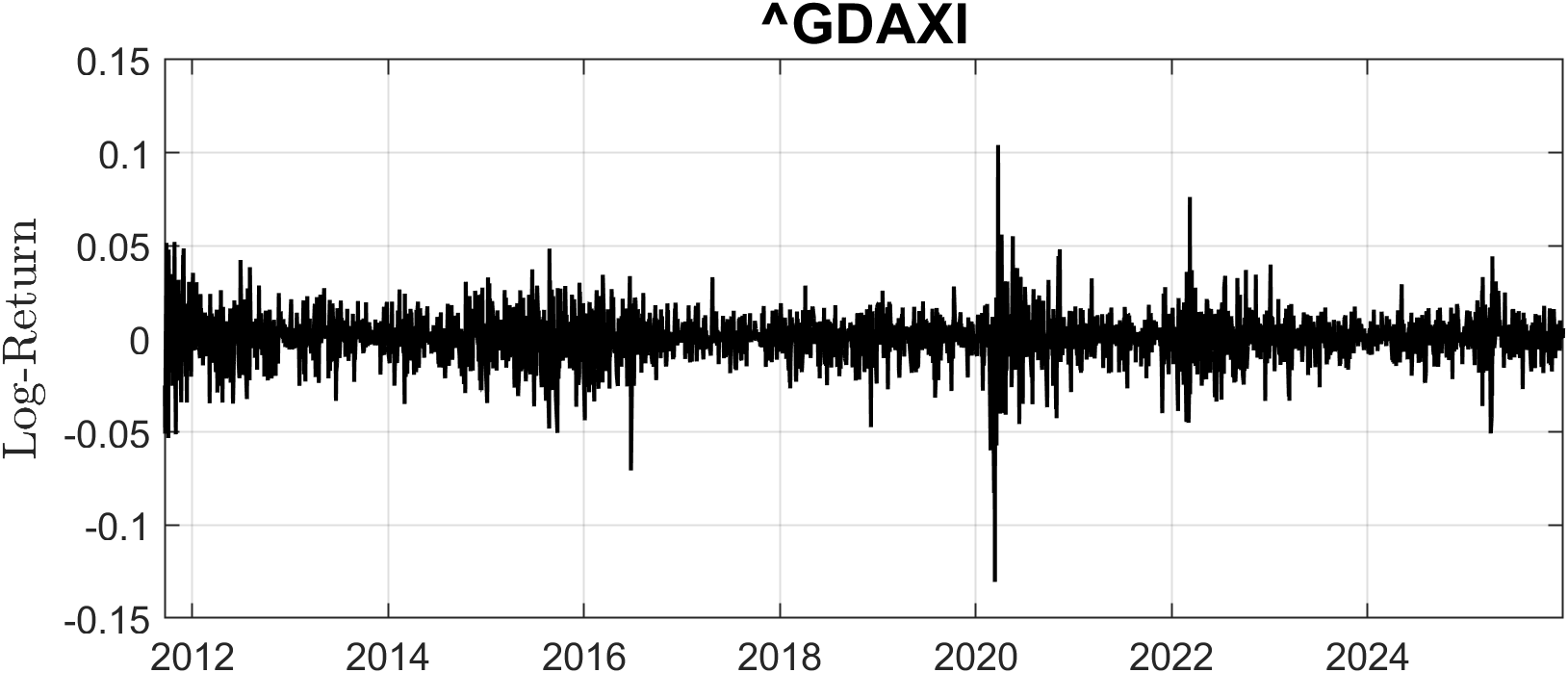}
\caption{\textasciicircum GDAXI}
\end{subfigure}
\hfill
\begin{subfigure}[b]{0.32\textwidth}
\includegraphics[width=\textwidth]{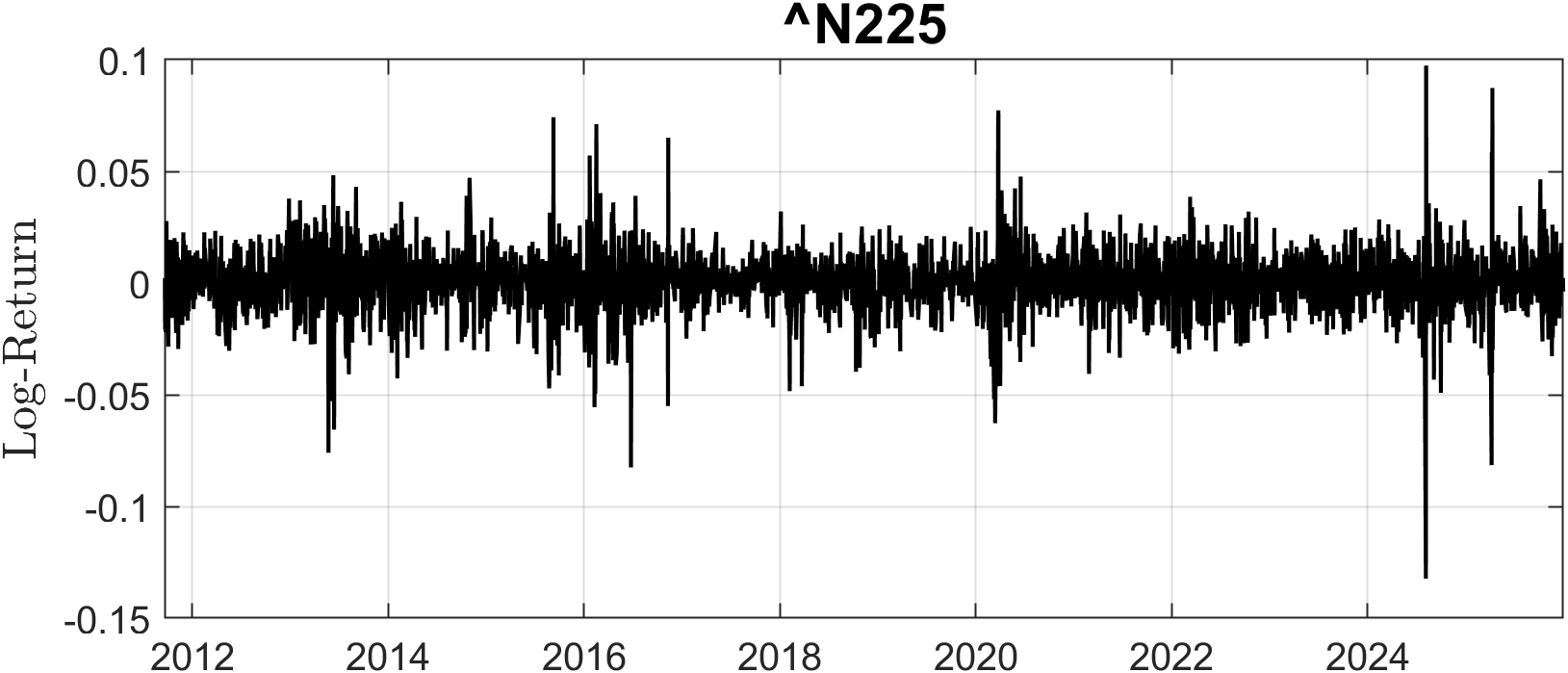}
\caption{\textasciicircum N225}
\end{subfigure}
\hfill
\begin{subfigure}[b]{0.32\textwidth}
\includegraphics[width=\textwidth]{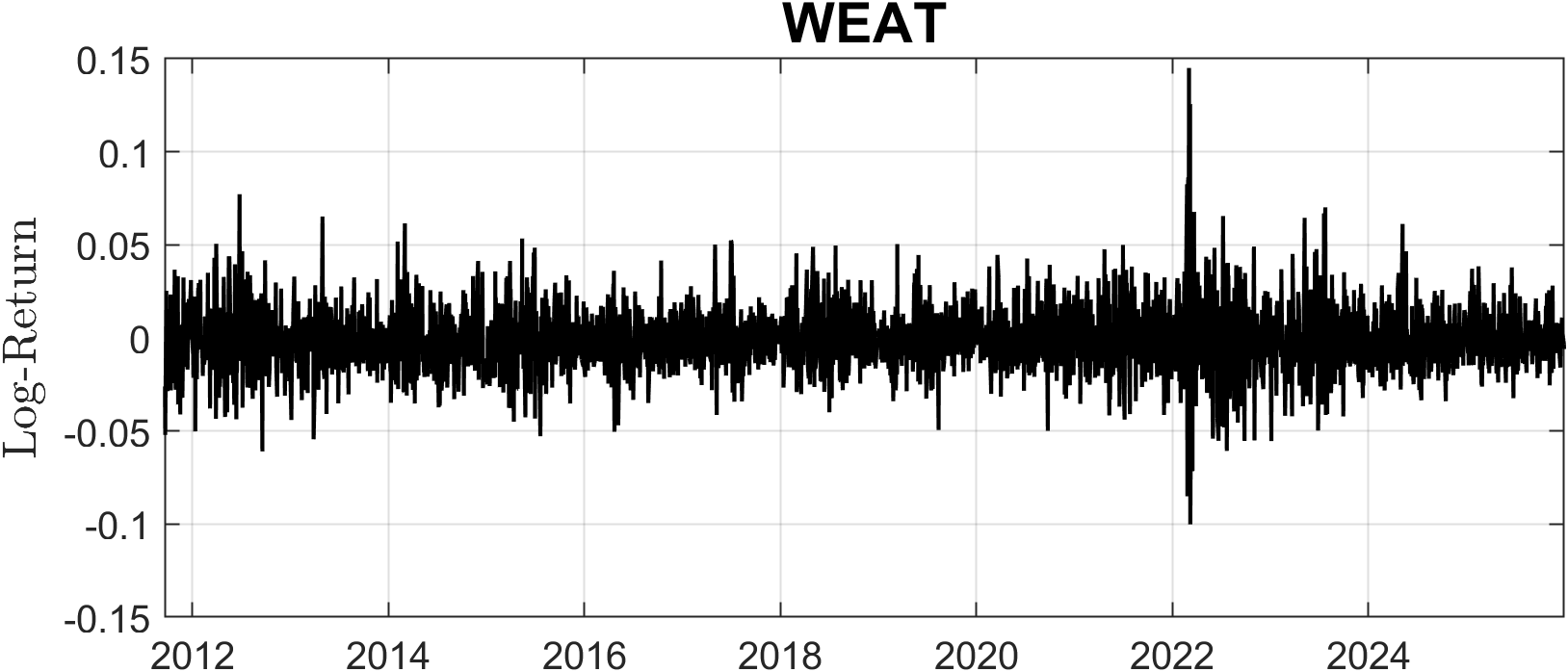}
\caption{WEAT}
\end{subfigure}
\hfill
\begin{subfigure}[b]{0.32\textwidth}
\includegraphics[width=\textwidth]{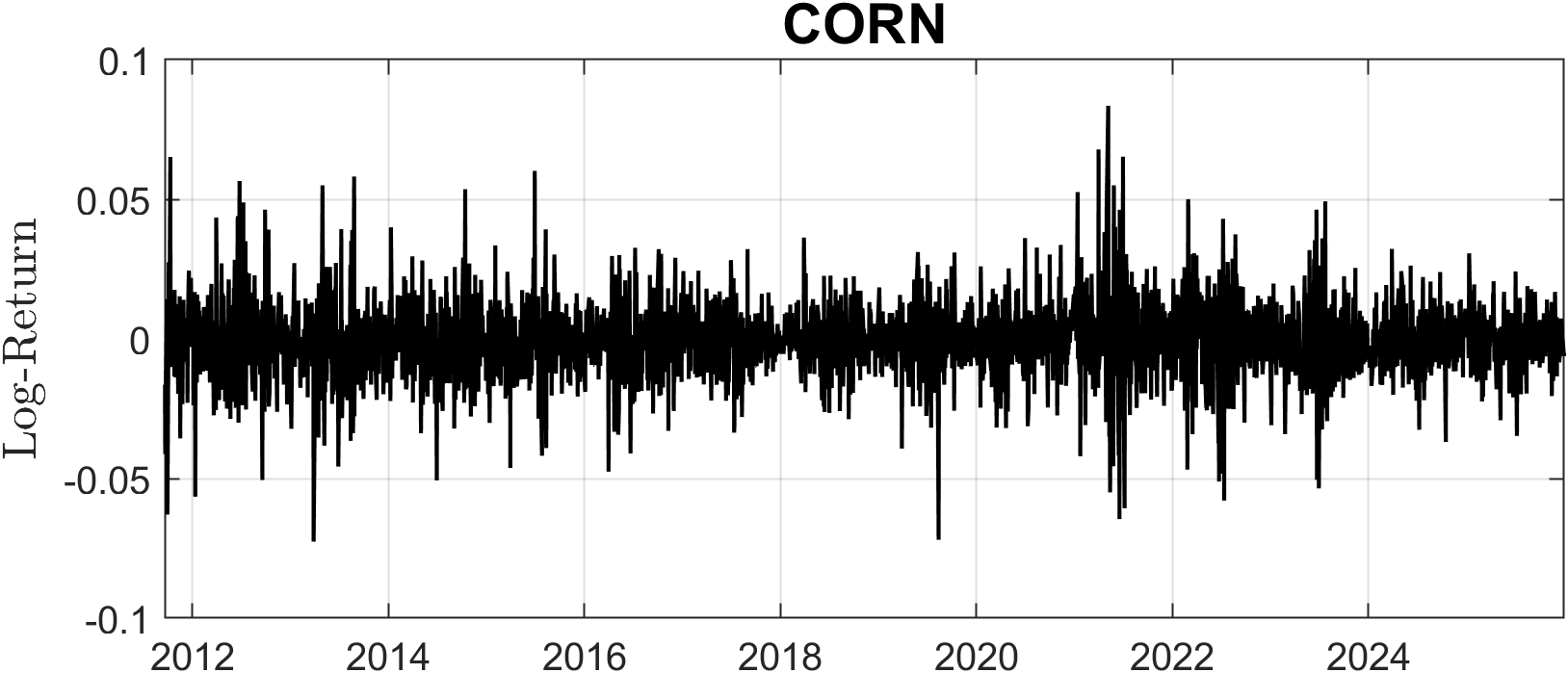}
\caption{CORN}
\end{subfigure}
\hfill
\begin{subfigure}[b]{0.32\textwidth}
\includegraphics[width=\textwidth]{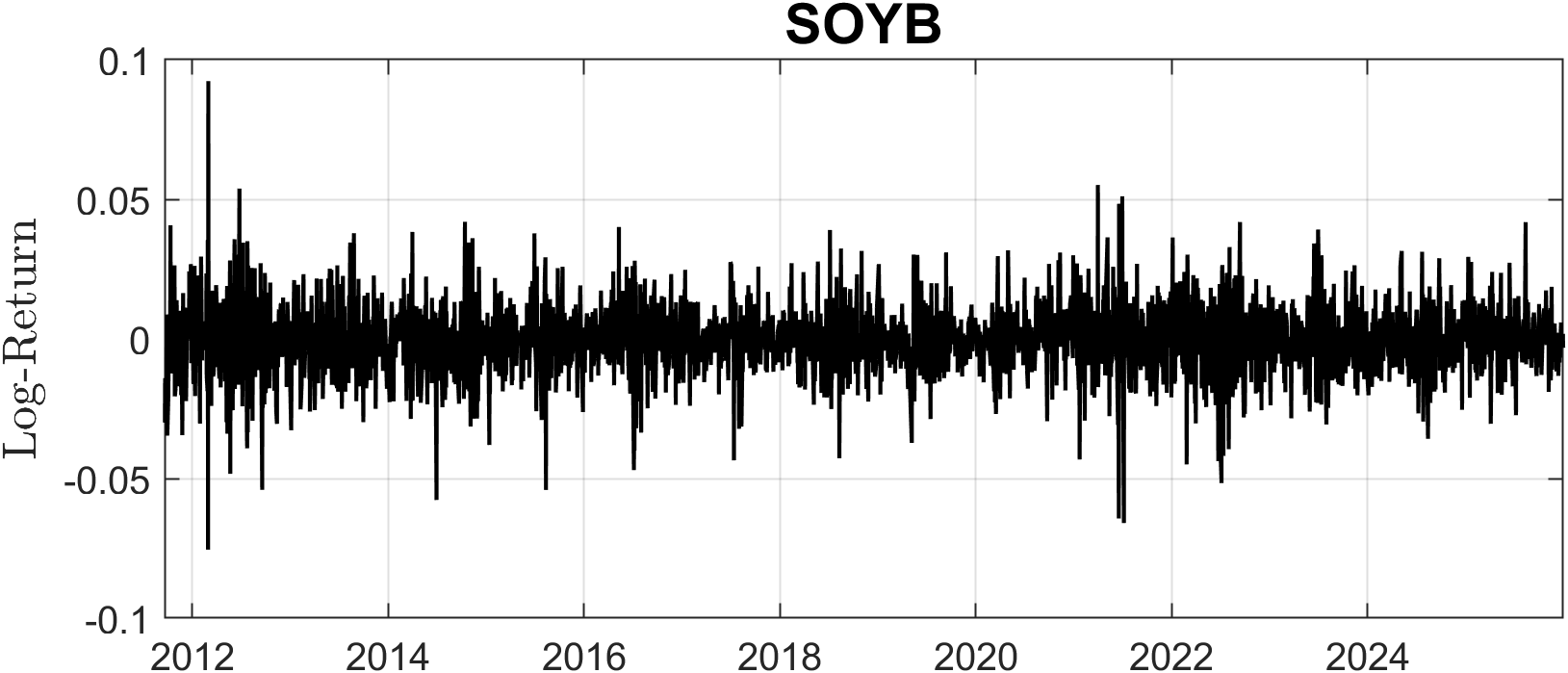}
\caption{SOYB}
\end{subfigure}
\hfill
\begin{subfigure}[b]{0.32\textwidth}
\includegraphics[width=\textwidth]{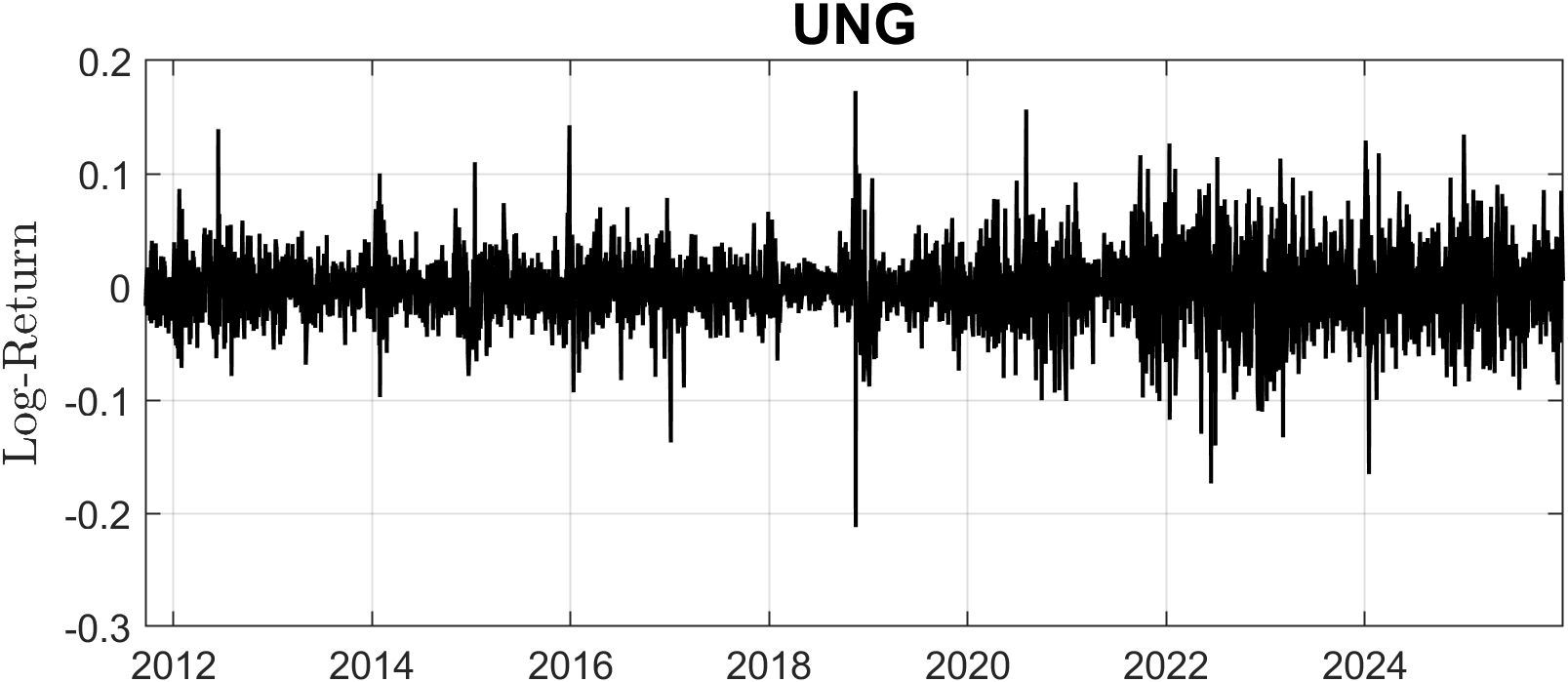}
\caption{UNG}
\end{subfigure}
\hfill
\begin{subfigure}[b]{0.32\textwidth}
\includegraphics[width=\textwidth]{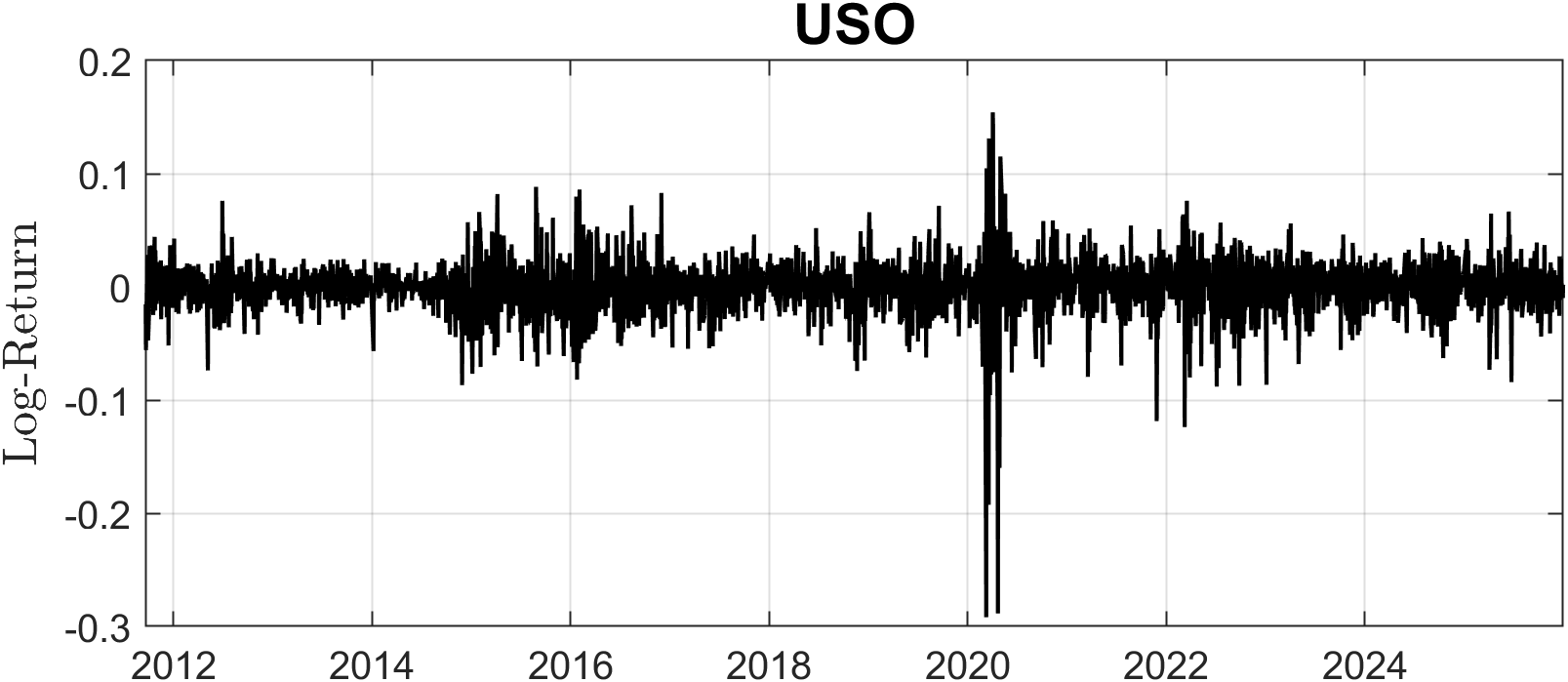}
\caption{USO}
\end{subfigure}
\hfill
\begin{subfigure}[b]{0.32\textwidth}
\includegraphics[width=\textwidth]{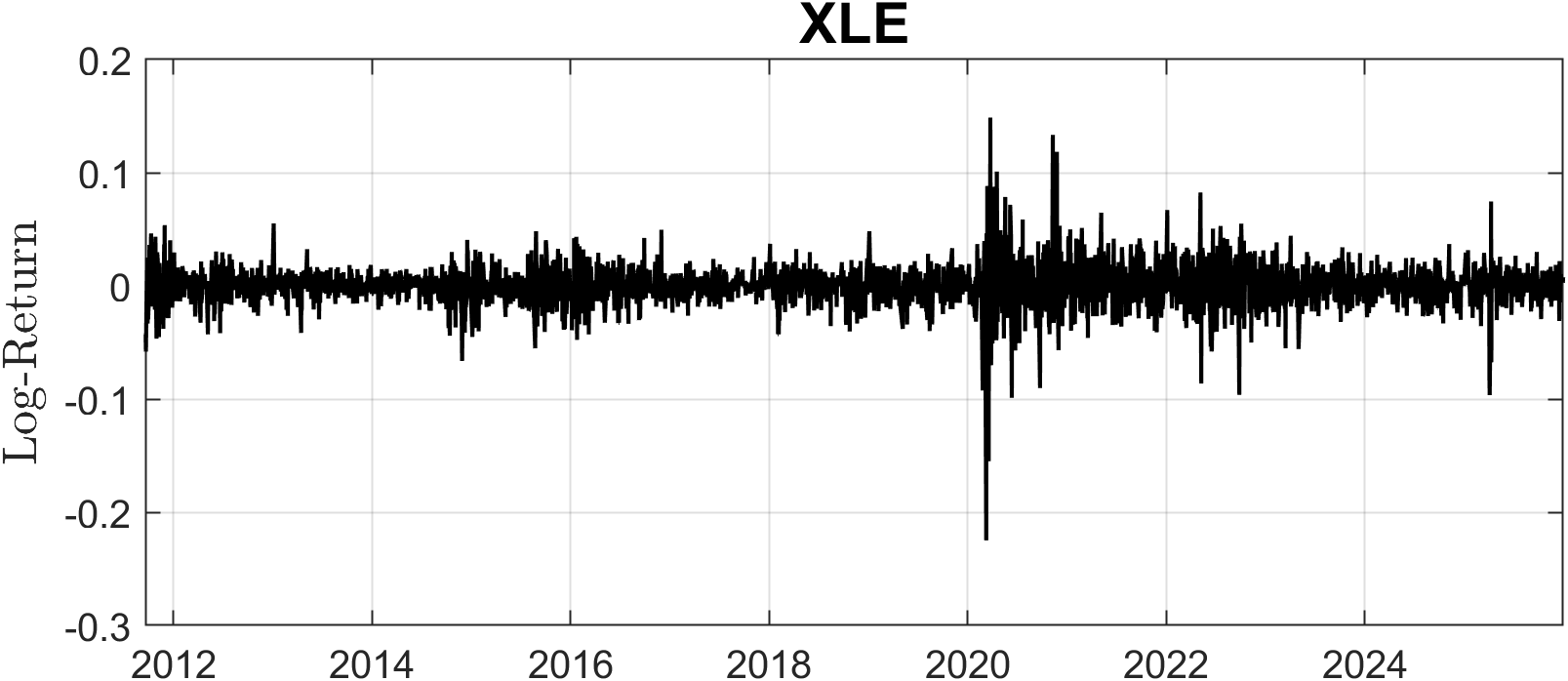}
\caption{XLE}
\end{subfigure}
\caption{Daily returns of the nine assets}
\label{fig:daily_returns}
\end{figure}

Figure~\ref{fig:daily_returns} displays the daily returns of the nine assets. Across the panels, returns fluctuate around zero, with most observations ranging from $-0.15$ to 0.15 for stocks and commodities and from approximately $-0.3$ to 0.2 for energy assets. The series also exhibit volatility clustering: calm periods tend to be followed by further calm periods, while turbulent periods are followed by further turbulence. This phenomenon, first documented by~\cite{mandelbrot1963variation} and later emphasized by~\cite{fama1965behavior}, is a well-known stylized fact of financial time series.

\begin{figure}[htbp]
\centering
\begin{subfigure}[b]{0.32\textwidth}
\includegraphics[width=\textwidth]{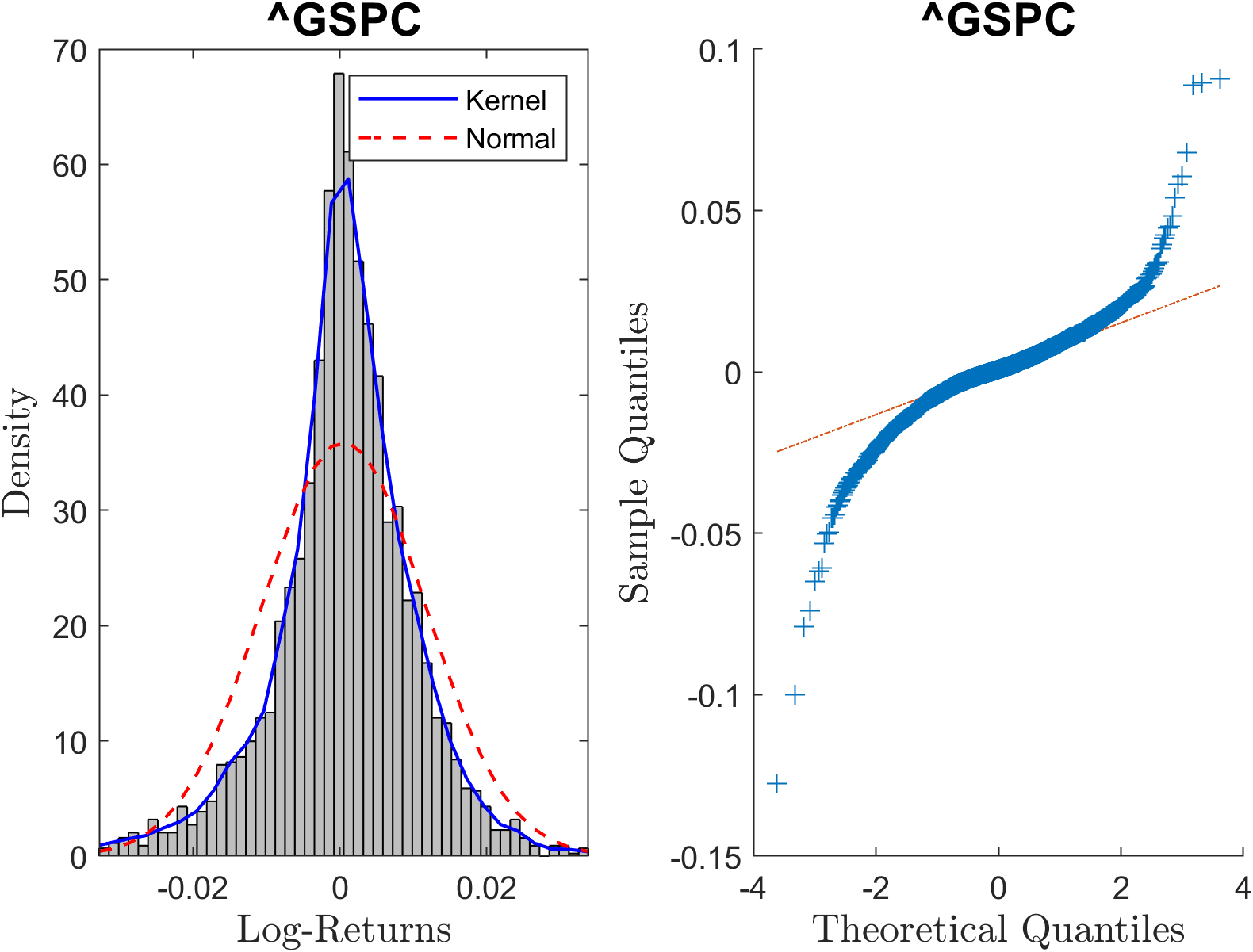}
\caption{\textasciicircum GSPC}
\end{subfigure}
\hfill
\begin{subfigure}[b]{0.32\textwidth}
\includegraphics[width=\textwidth]{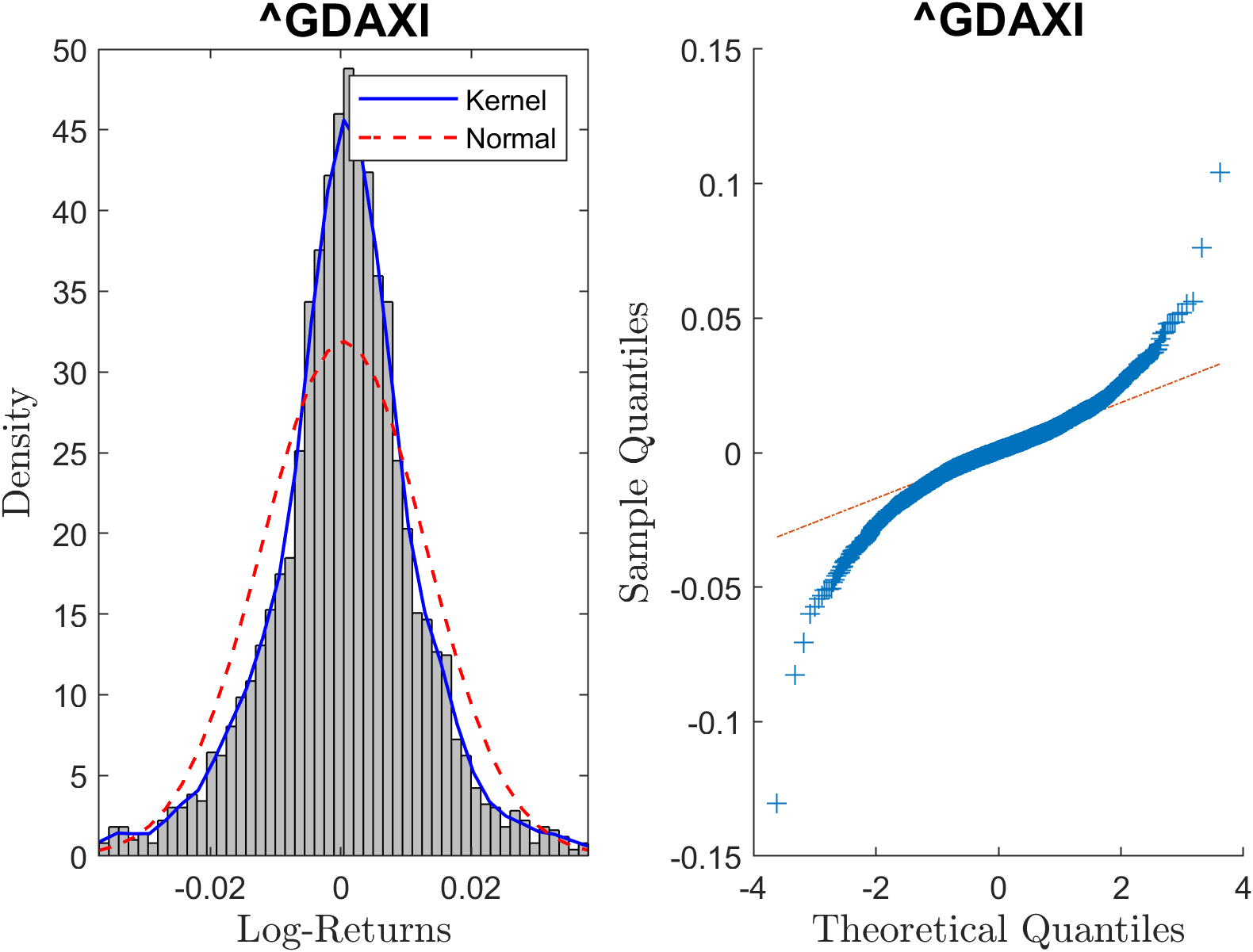}
\caption{\textasciicircum GDAXI}
\end{subfigure}
\hfill
\begin{subfigure}[b]{0.32\textwidth}
\includegraphics[width=\textwidth]{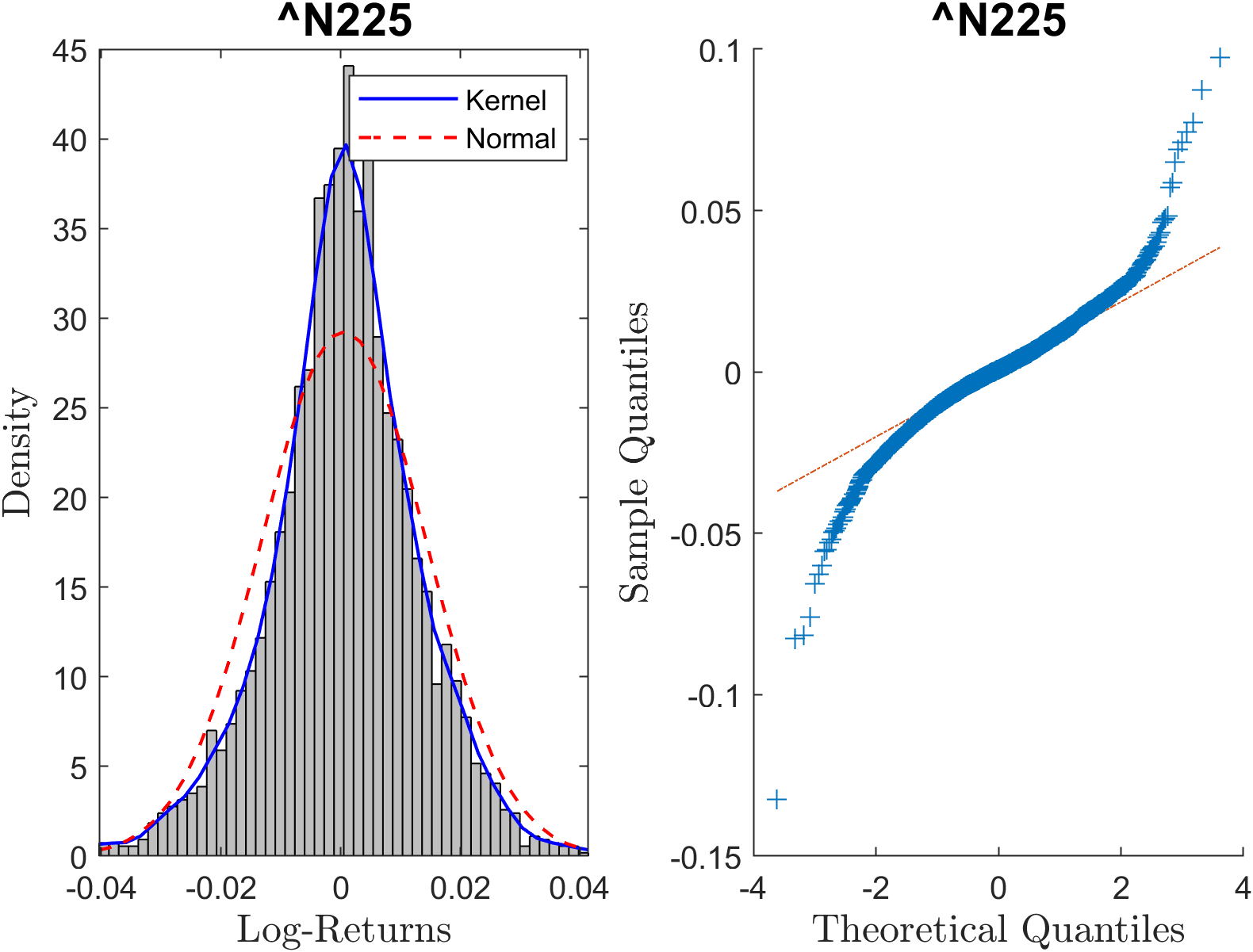}
\caption{\textasciicircum N225}
\end{subfigure}
\hfill
\begin{subfigure}[b]{0.32\textwidth}
\includegraphics[width=\textwidth]{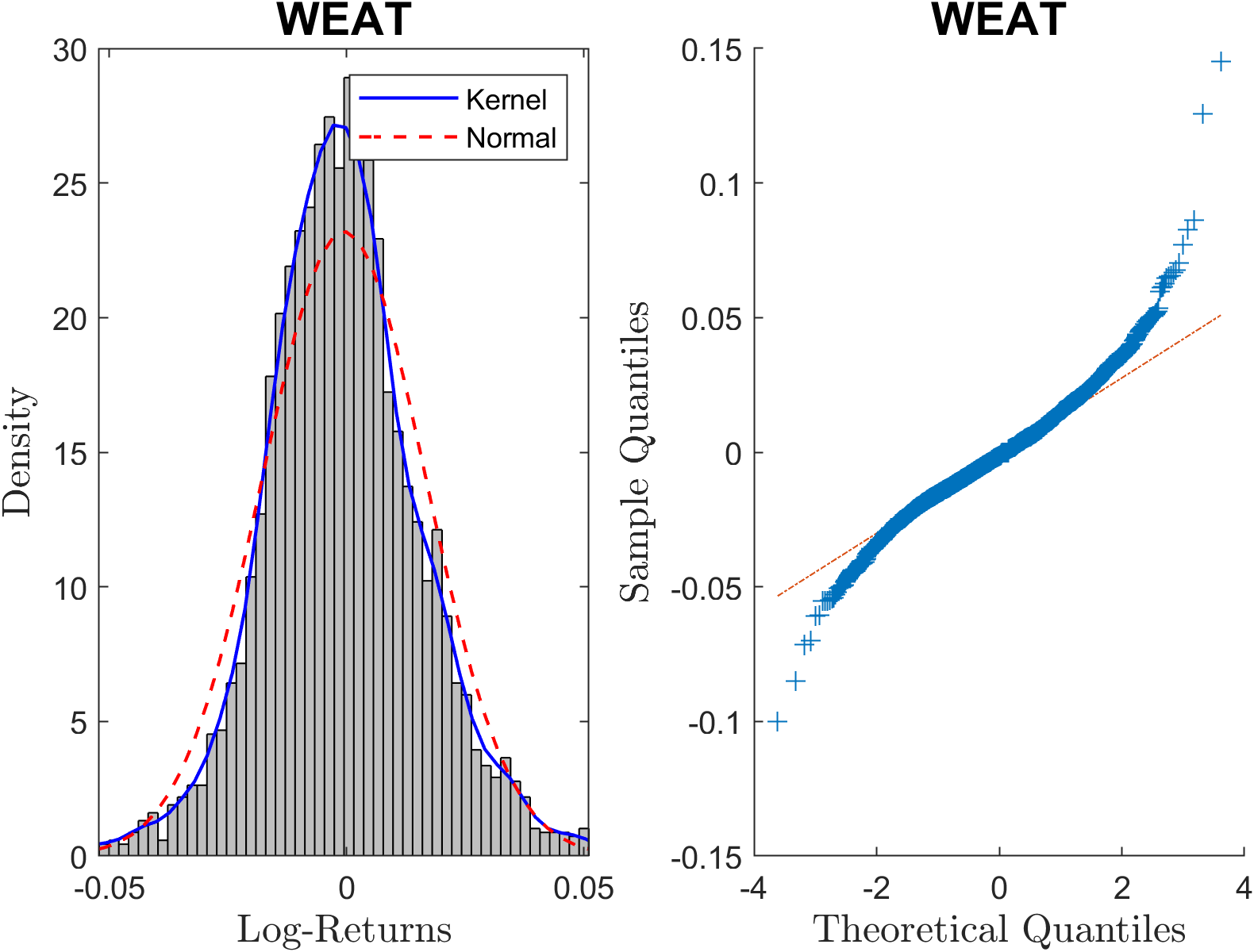}
\caption{WEAT}
\end{subfigure}
\hfill
\begin{subfigure}[b]{0.32\textwidth}
\includegraphics[width=\textwidth]{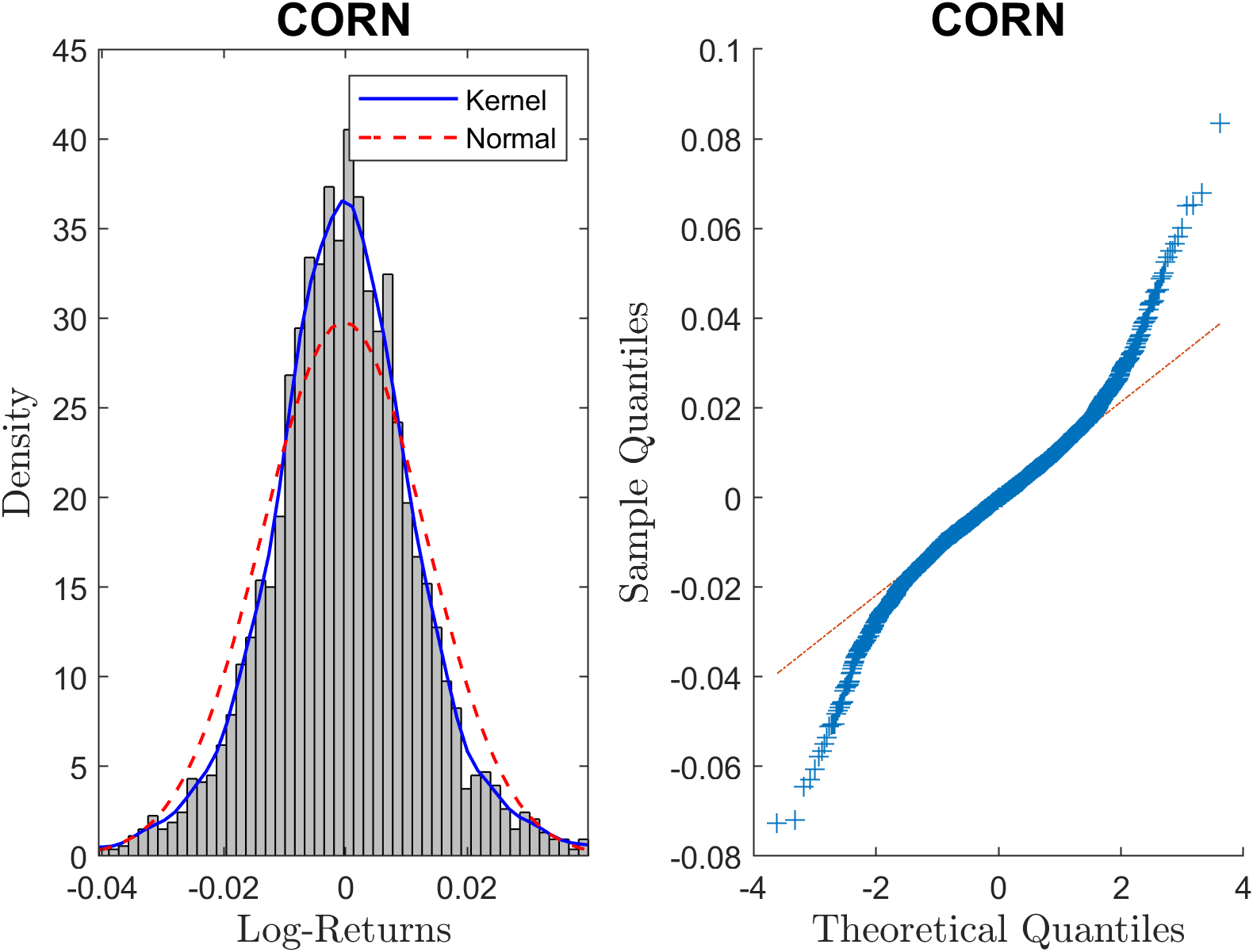}
\caption{CORN}
\end{subfigure}
\hfill
\begin{subfigure}[b]{0.32\textwidth}
\includegraphics[width=\textwidth]{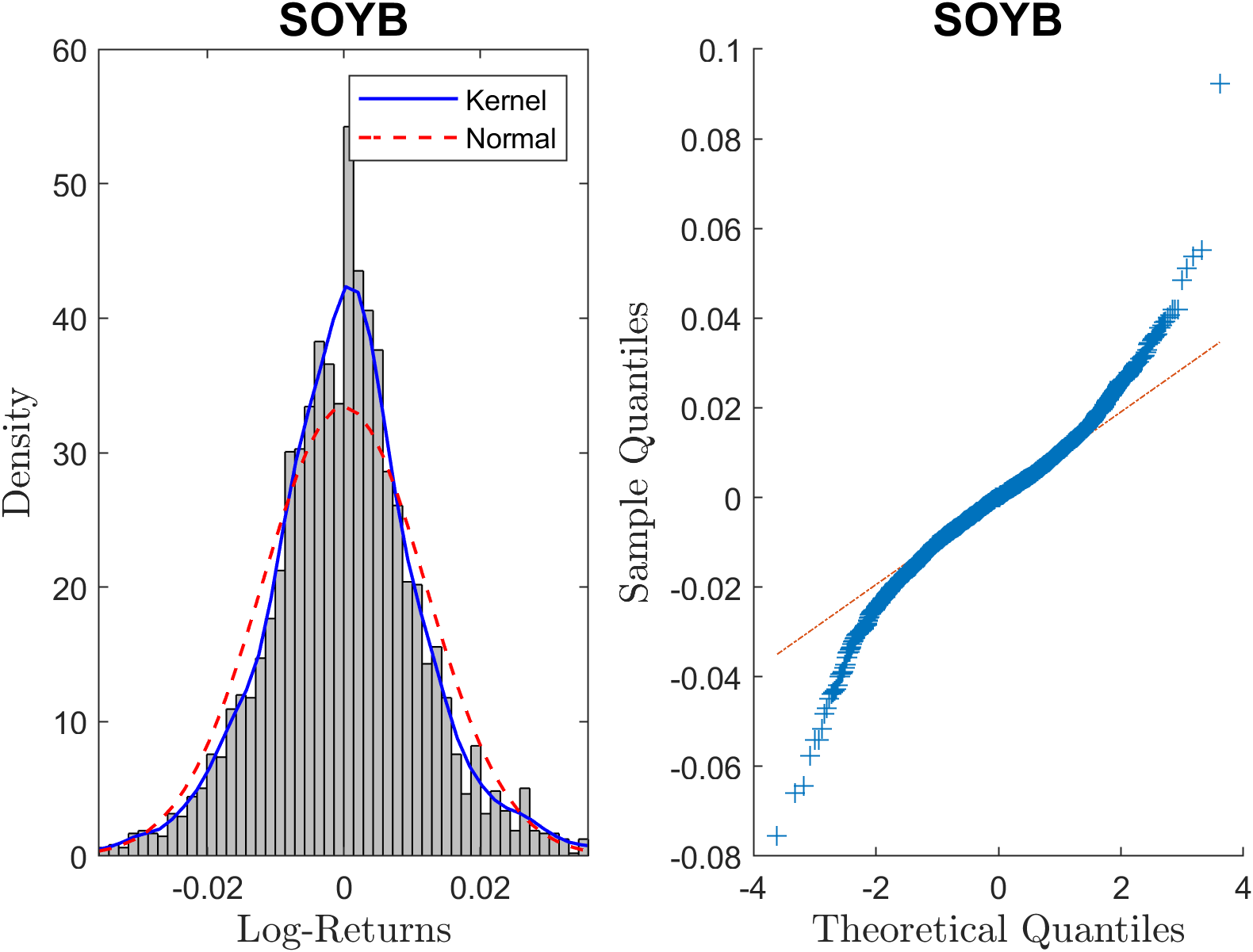}
\caption{SOYB}
\end{subfigure}
\hfill
\begin{subfigure}[b]{0.32\textwidth}
\includegraphics[width=\textwidth]{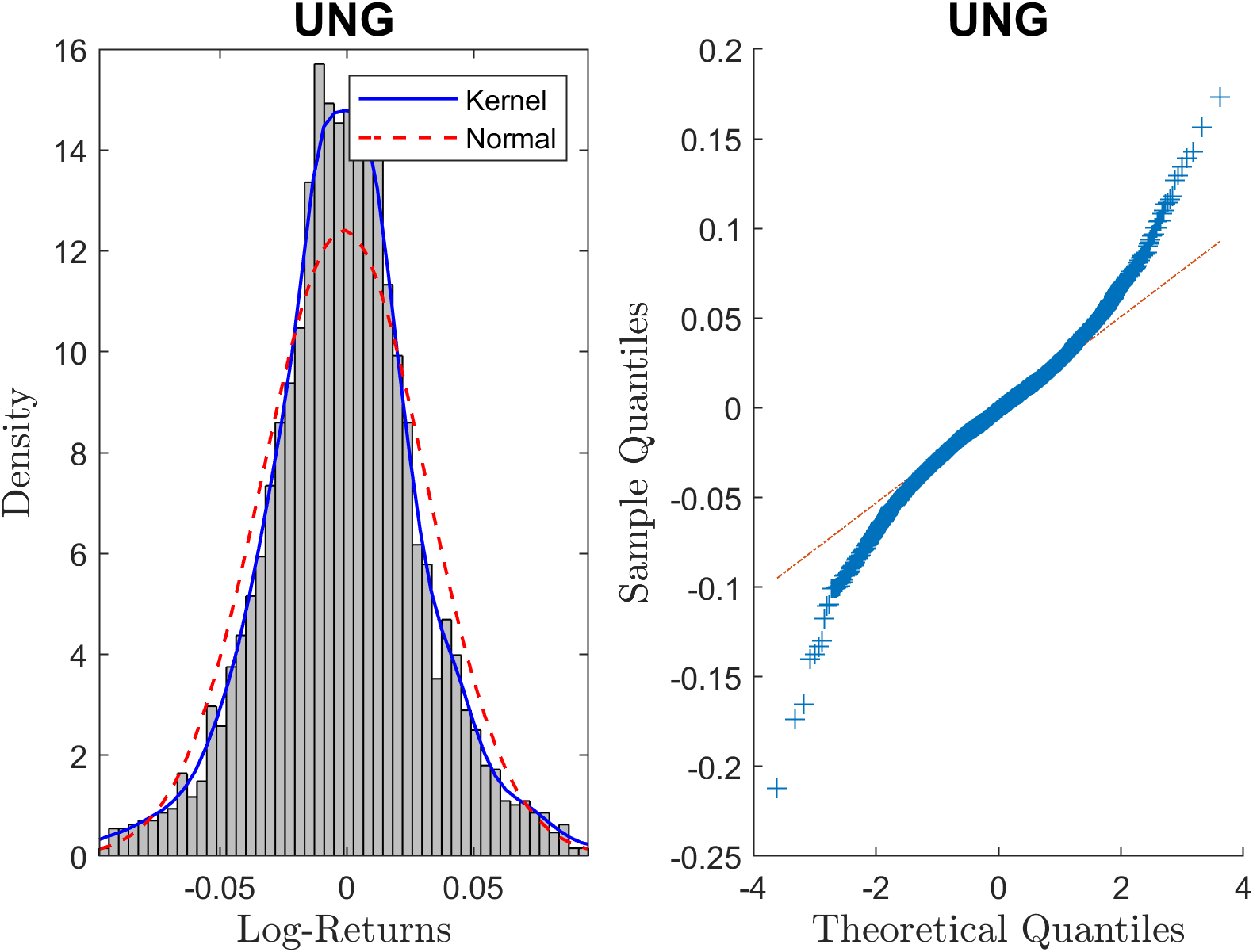}
\caption{UNG}
\end{subfigure}
\hfill
\begin{subfigure}[b]{0.32\textwidth}
\includegraphics[width=\textwidth]{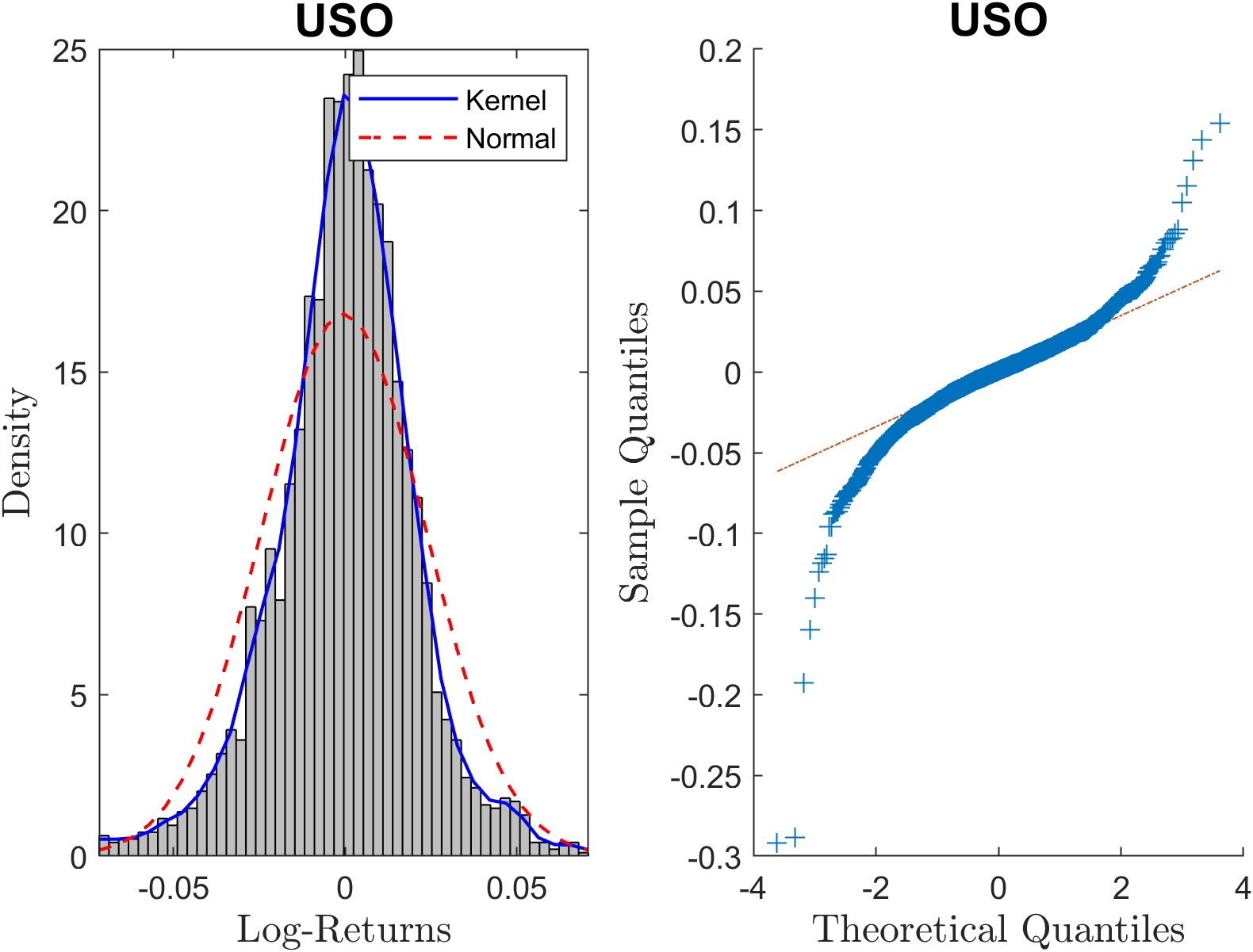}
\caption{USO}
\end{subfigure}
\hfill
\begin{subfigure}[b]{0.32\textwidth}
\includegraphics[width=\textwidth]{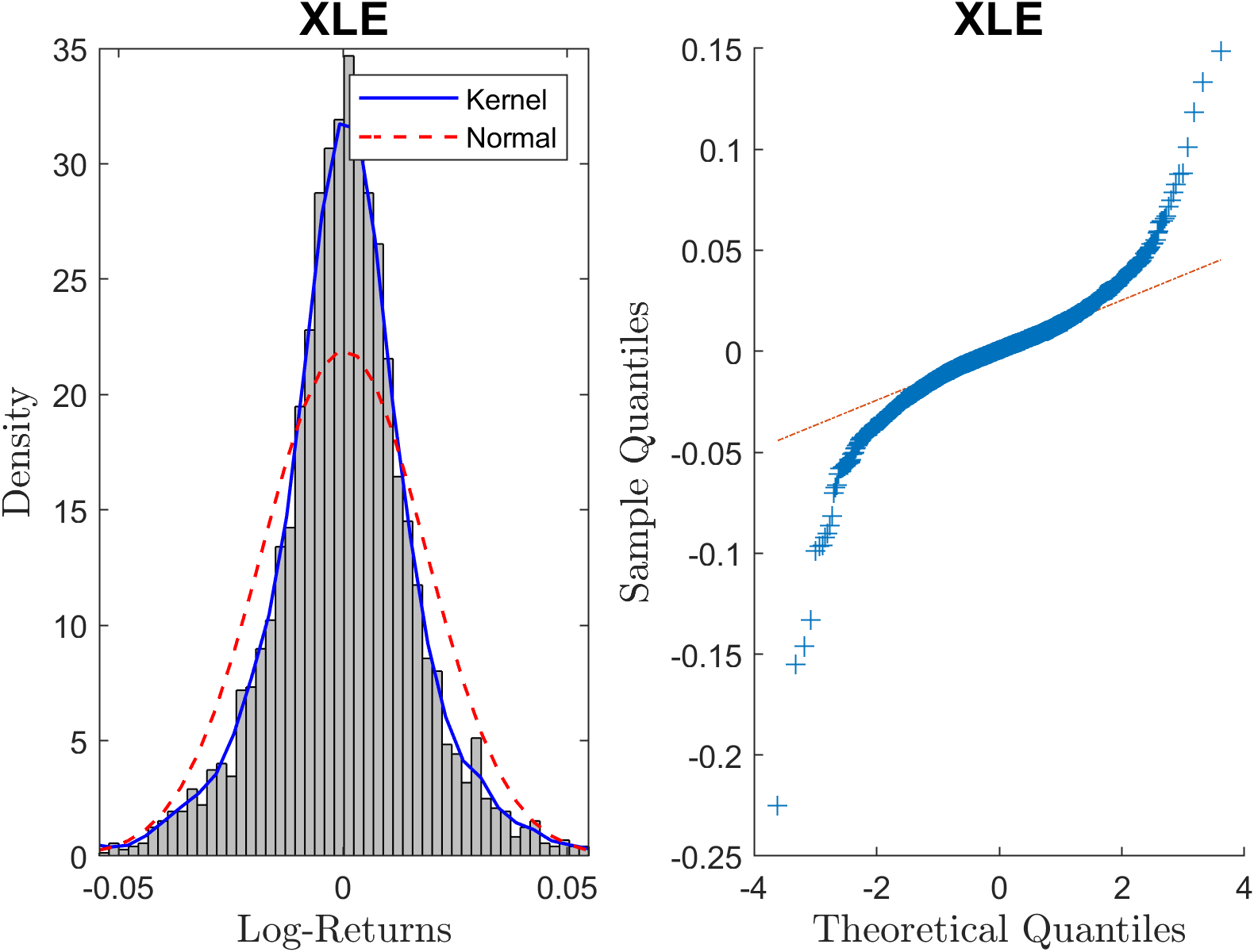}
\caption{XLE}
\end{subfigure}
\caption{Histograms, kernel density estimates, normal density fits, and QQ plots of the daily returns for the nine assets}
\label{fig:hist_qq}
\end{figure}

We next examine whether the return distributions are compatible with Gaussian assumptions. Figure~\ref{fig:hist_qq} presents the distributional diagnostics for each asset. In each panel, the left sub-panel displays the histogram of returns together with a kernel density estimate (blue solid line) and a fitted normal density (red dashed line), while the right sub-panel shows the corresponding quantile--quantile (QQ) plot. The histograms reveal sharper central peaks and heavier tails than the fitted normal density. The QQ plots provide the same evidence from a quantile perspective: sample quantiles deviate substantially from the normal reference line, especially in the tails.\footnote{Fat-tailed distributions exhibit large negative and positive values with greater probability than a normal distribution.}

A normal distribution has skewness $\gamma = 0$ and kurtosis $\kappa = 3$. The difference $\kappa-3$ is known as the \textit{excess kurtosis}. Hypothesis tests are commonly used to assess deviations from normality arising from skewness, excess kurtosis, or both jointly~\citep{d1990suggestion}. In this study, we employ the \texttt{skewtest()}, \texttt{kurtosistest()}, and \texttt{normaltest()} functions from the \texttt{scipy.stats} Python subpackage to conduct these respective tests. In each case, the null hypothesis assumes that the examined moments are consistent with those of a normal distribution.

Table~\ref{tab:normal_test} reports the skewness and excess-kurtosis diagnostics for each asset. The kurtosis tests reject the normal benchmark at the 0.001 level for all assets, indicating substantial tail thickness. The skewness tests also reject normality at the 0.001 level in most cases, with the exceptions of CORN and SOYB, whose $p$-values are 0.115 and 0.182, respectively. Taken jointly, the skewness--kurtosis tests reject the null hypothesis of normality at the 0.001 level for all nine assets.
\begin{table}[htbp]
\centering
\begin{tabular}{c | l r c}
\hline
Category & Ticker & $\gamma$ ($p$-value) & $\kappa - 3^{a}$\\
\hline
 & \textasciicircum GSPC & $-0.711$ (***)$^{b}$ & 15.2\\
Stock & \textasciicircum GDAXI & $-0.464$ (***) & 8.12\\
 & \textasciicircum N225 & $-0.333$ (***) & 6.99 \\ 
\hline
 & WEAT & 0.473 (***) & 4.03 \\
Commodity & CORN & 0.0670 (0.115) & 3.36 \\
 & SOYB & $-0.0566$ (0.182)  & 3.56 \\ 
\hline
 & UNG & $-0.0186$ (***) & 2.63 \\
Energy & USO & $-1.48$ (***) & $18.0$\\
 & XLE & $-0.788$ (***) & $15.7$ \\
\hline
\multicolumn{4}{l}{\footnotesize $^{a}$ All kurtosis $p$-values are less than 0.001.\quad \footnotesize $^{b}$ Indicates a $p$-value less than 0.001.}\\
\hline
\end{tabular}
\caption{The results of the test for normality on the skewness $\gamma$ and excess kurtosis $\kappa - 3$ for each asset}
\label{tab:normal_test}
\end{table}

Combining the evidence from the QQ plots and hypothesis tests on the central moments, we find strong empirical support for rejecting the assumption of normally distributed returns for all nine assets. Given these findings, we adopt heavy-tailed error distributions--such as Student's $t$-distribution--in the ARFIMA-FIGARCH modeling framework discussed in Section~\ref{sec:arfima_figarch}.

\section{Measurement of Long-Range Dependence}\label{sec:measure_LRD}
In this section, we focus on the key concept of this work--LRD, which refers to the slow decay of autocorrelations over time in a stochastic process. The notion of LRD was first introduced by~\cite{hurst1951long} in the context of hydrology, where Hurst observed that the fluctuations of the Nile River levels exhibited a persistent memory effect over long horizons. Since then, the study of LRD has attracted considerable attention in various fields, including geophysics, network traffic modeling, and particularly financial econometrics. In financial markets, the presence of LRD in asset returns or volatility has significant implications for risk management, asset pricing, and forecasting, as it indicates that shocks may have persistent effects over extended time periods~\citep{doukhan2002theory, beran2017statistics}.

In Section~\ref{sec:rs}, we first employ the classical R/S analysis, proposed by~\cite{hurst1951long} and later formalized by~\cite{mandelbrot1968noah}. This technique estimates the Hurst exponent to characterize the degree of persistence in a time series. While R/S analysis has been widely used, it suffers from several limitations, including its sensitivity to nonstationarities, structural breaks, and short-term dependencies~\citep{lo1991long}. These shortcomings often lead to biased estimates of the Hurst exponent, making it less reliable for complex financial data.

To address these issues, Section~\ref{sec:dfa} adopts DFA, originally developed by~\cite{peng1994mosaic} in the context of DNA sequence analysis. DFA improves upon R/S analysis by systematically removing local trends within different time scales before measuring the scaling behavior of fluctuations. This makes DFA particularly robust for nonstationary financial time series and allows a more accurate detection of LRD~\citep{kantelhardt2001detecting}.

Section~\ref{sec:segmented_scaling} then examines whether the equity-market scaling results are stable across the COVID-19 period. Motivated by the visible change in the normalized price paths, we split the equity-index samples into pre- and post-COVID subsamples and complement the R/S and DFA estimates with a segmented MF-DFA spectrum analysis.

Finally, because financial returns typically exhibit not only long memory but also volatility clustering and heavy-tailed distributions, Section~\ref{sec:arfima_figarch} employs a combined modeling framework based on the ARFIMA(1,$d_m$,1)--FIGARCH(1,$d_v$,1) process with Student's $t$-distributed innovations. The ARFIMA component captures LRD in the conditional mean \citep{granger1980introduction}, while the FIGARCH component accounts for long memory in volatility \citep{baillie1996fractionally}. The use of Student's $t$ further accommodates the leptokurtic nature of financial returns~\citep{bollerslev1987conditionally}, making the model suitable for capturing key stylized facts in financial markets, as we discussed in Section~\ref{sec:non_normal_return}.

\subsection{Rescaled Range Analysis}\label{sec:rs}
To measure LRD, we use the Hurst exponent ($H$), which was first proposed by~\cite{hurst1951long}. For a time series, $H = 0.5$ indicates a memoryless or random process, such as white noise or a standard random walk, where past values provide no information about future dynamics. When $H < 0.5$, the series exhibits anti-persistent behavior, meaning that increases are likely to be followed by decreases and vice versa; such processes tend to be mean-reverting, and their autocorrelations decay rapidly, implying no LRD. In contrast, $H > 0.5$ indicates persistent behavior, where positive (or negative) deviations are more likely to be followed by deviations of the same sign. This persistence corresponds to slowly decaying autocorrelations over long horizons, which is the hallmark of LRD.

We first explain how to estimate $H$ by R/S analysis. Consider a time series $\{X_t\}_{t = 1}^N$. To examine scaling behavior across different time horizons, we define a set of subsequence lengths (scales) by
\begin{equation}\label{eq:step1}
n = \frac{N}{4}, \frac{N}{8}, \frac{N}{16},\cdots,
\end{equation}
where the maximum scale is chosen in such a way that at least four non-overlapping subsequences are available for the estimation. In practice, we set
\begin{equation}\label{eq:step2}
n = \biggl\lfloor \frac{N}{2^p} \biggl\rfloor,\quad p = 2,3,\cdots, \biggl\lfloor\log_2\left(\frac{N}{4}\right)\biggl\rfloor,
\end{equation}
which ensures a sequence of evenly spaced scales on a logarithmic axis. For each selected scale $n$, the dataset is divided into $K = \lfloor N/n\rfloor$\footnote{$a = \lfloor b \rfloor$ means $a$ is an integer satisfying $a\le b < a+1$.} non-overlapping blocks of length $n$.

For the $k$-th block, we compute the sample mean $\bar{X}_k$ and the cumulative deviation from the mean, defined as
\begin{equation*}
Y_{k,j} = \sum_{i=1}^j (X_{k,i} - \bar{X}_k),\quad j = 1,\cdots,n.
\end{equation*}
The range $R_k$ of cumulative deviations within the block is given by
\begin{equation*}
R_k = \max(Y_{k,1},\cdots, Y_{k,n}) - \min(Y_{k,1},\cdots, Y_{k,n}),
\end{equation*}
and the corresponding standard deviation is
\begin{equation*}
S_k = \sqrt{\frac{1}{n}\sum_{i=1}^n (X_{k,i} - \bar{X}_k)^2}.
\end{equation*}

The rescaled range for the block is $(R/S)_k = R_k / S_k$. For each scale $n$, the overall statistic $(R/S)_n$ is then obtained by averaging over all blocks:
\begin{equation*}
(R/S)_n = \frac{1}{K}\sum_{k=1}^K \frac{R_k}{S_k}.
\end{equation*}
\cite{hurst1951long} found that the expected value of the rescaled range follows a power-law relationship with the subsequence length $n$:
\begin{equation*}
\mathrm{E}\left[(R/S)_n\right] = c\cdot n^H,
\end{equation*}
where $c$ is a constant and $H$ is the Hurst exponent introduced above. Taking natural logarithms on both sides yields
\begin{equation}\label{eq:rs_fit}
\ln (R/S)_n = \ln c + H\cdot\ln n,
\end{equation}
which implies that $H$ can be estimated as the slope of a linear regression of $\ln (R/S)_n$ on $\ln n$.

\begin{figure}[htbp]
\centering
\begin{subfigure}[b]{0.32\textwidth}
\includegraphics[width=\textwidth]{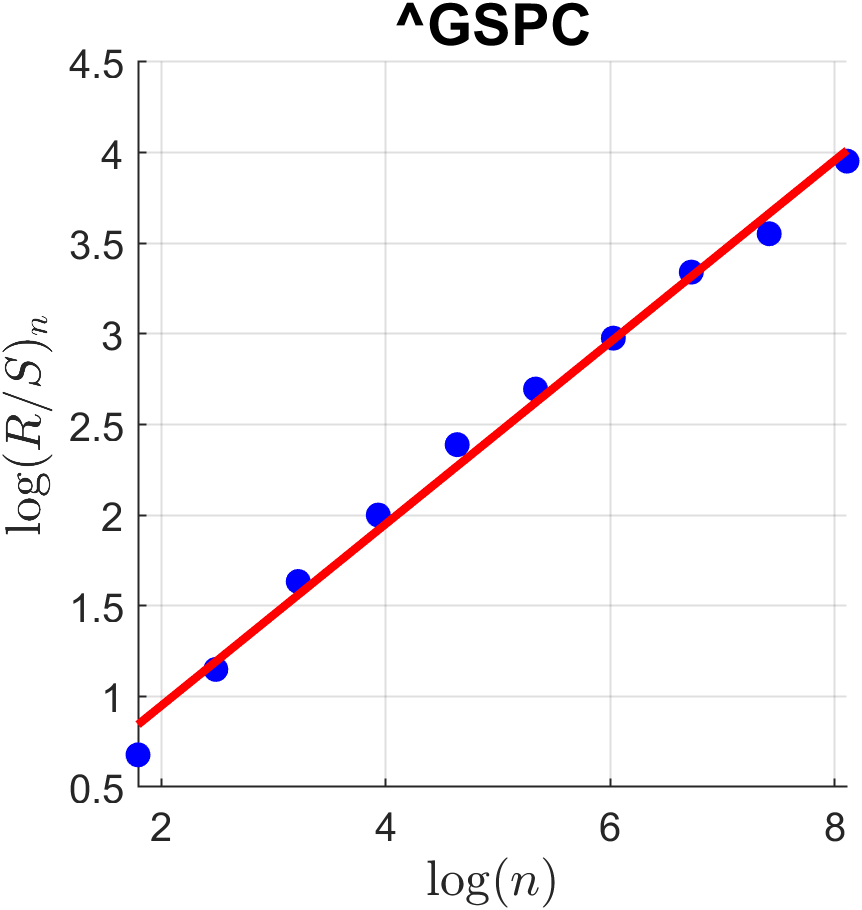}
\caption{\textasciicircum GSPC}
\end{subfigure}
\hfill
\begin{subfigure}[b]{0.32\textwidth}
\includegraphics[width=\textwidth]{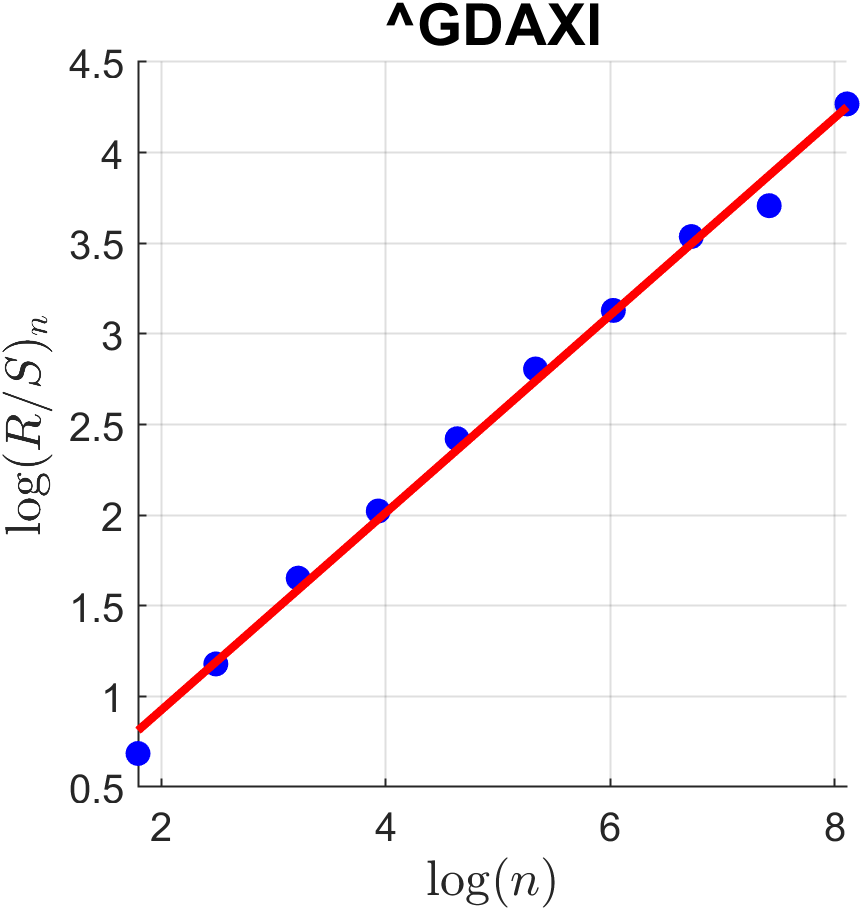}
\caption{\textasciicircum GDAXI}
\end{subfigure}
\hfill
\begin{subfigure}[b]{0.32\textwidth}
\includegraphics[width=\textwidth]{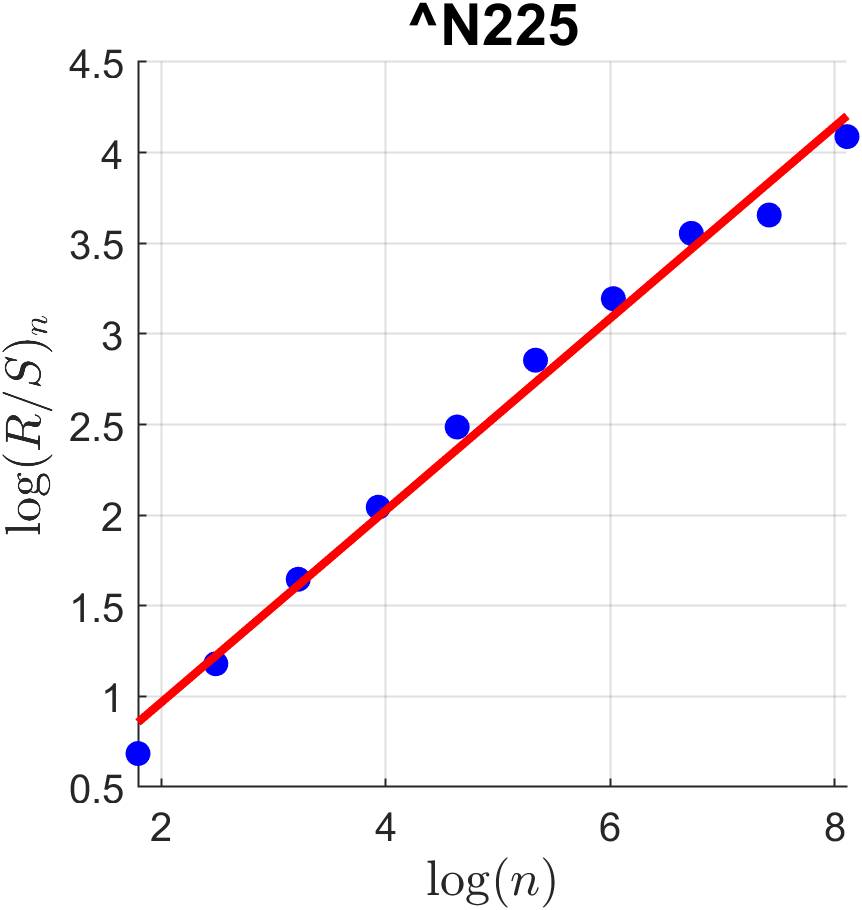}
\caption{\textasciicircum N225}
\end{subfigure}
\hfill
\begin{subfigure}[b]{0.32\textwidth}
\includegraphics[width=\textwidth]{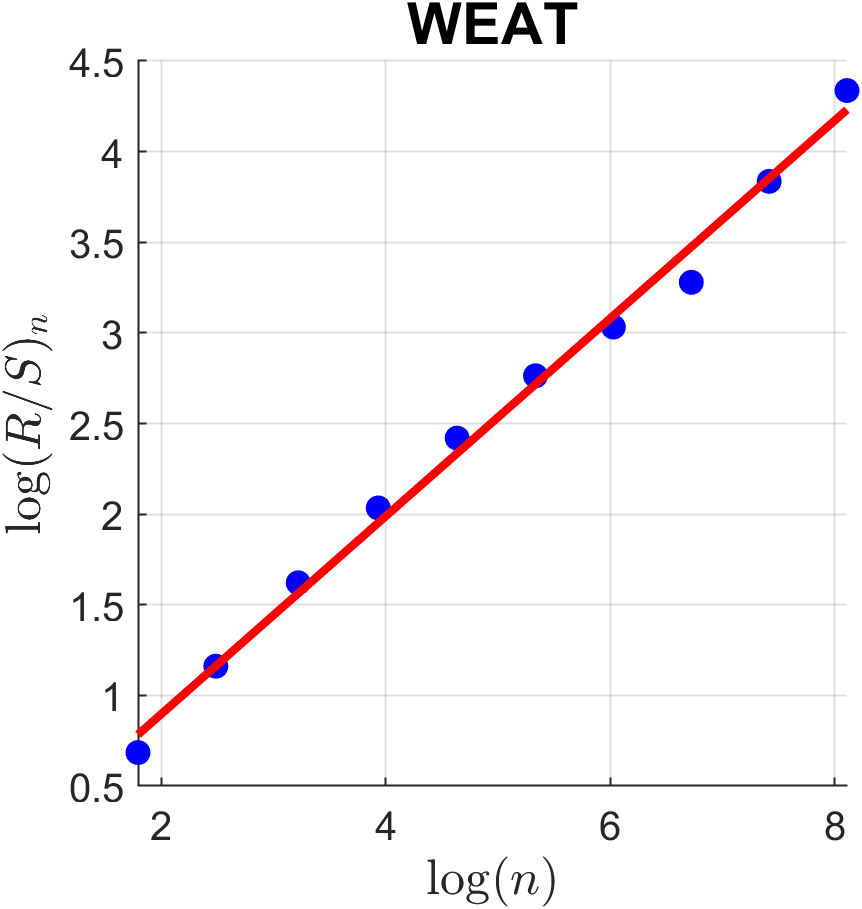}
\caption{WEAT}
\end{subfigure}
\hfill
\begin{subfigure}[b]{0.32\textwidth}
\includegraphics[width=\textwidth]{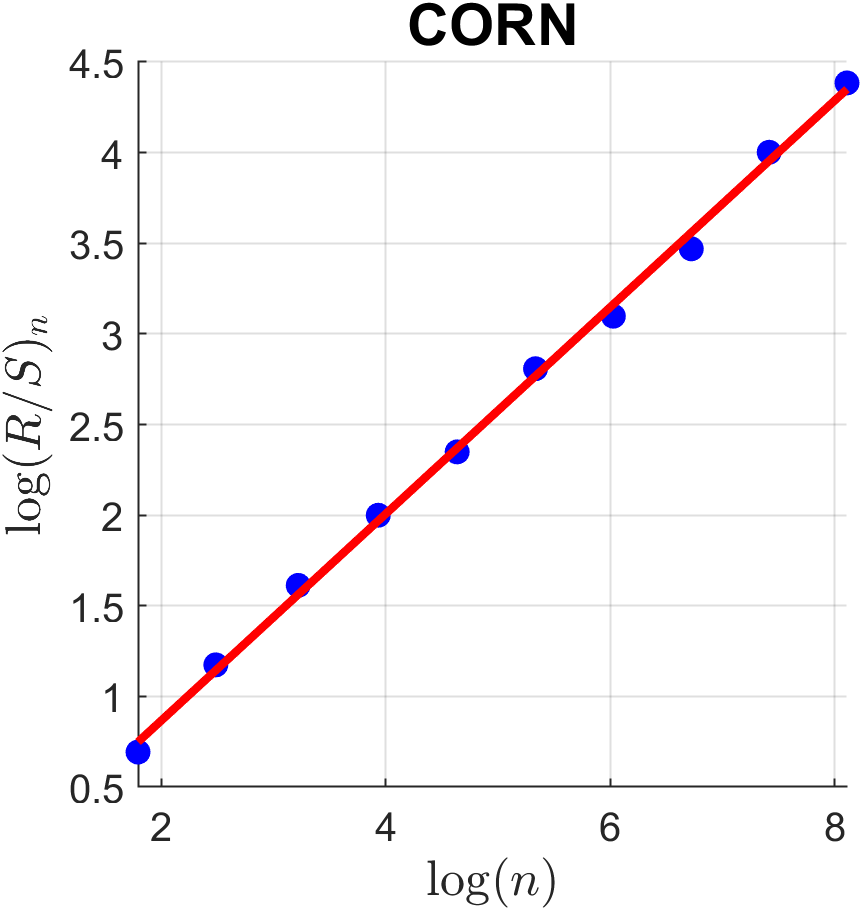}
\caption{CORN}
\end{subfigure}
\hfill
\begin{subfigure}[b]{0.32\textwidth}
\includegraphics[width=\textwidth]{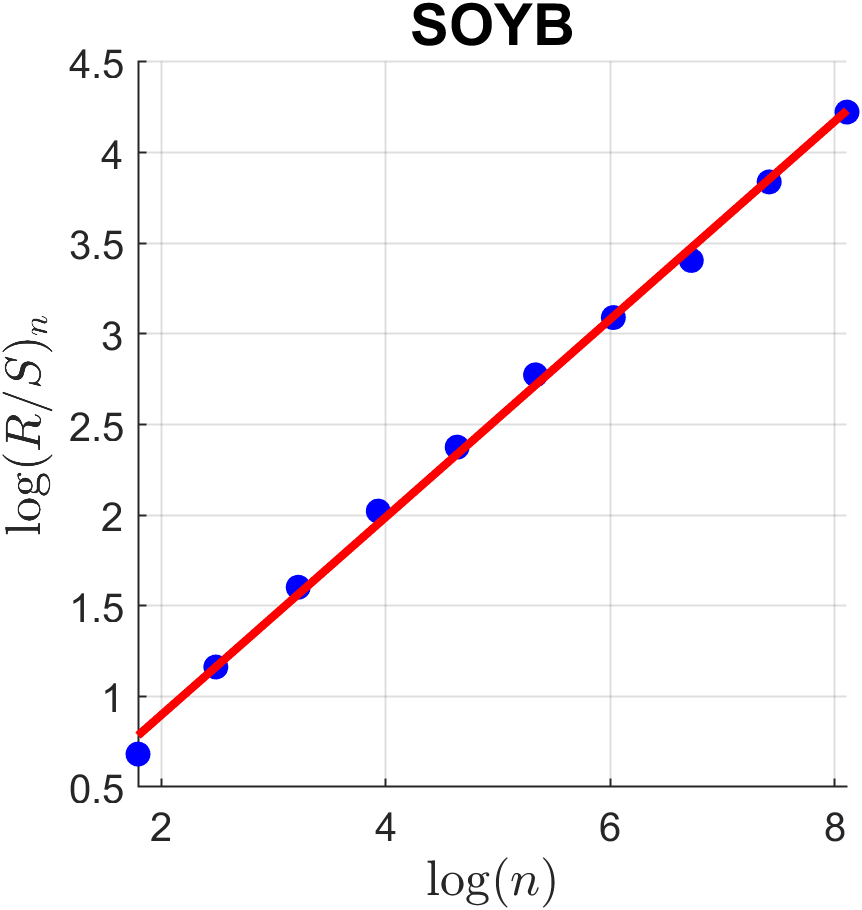}
\caption{SOYB}
\end{subfigure}
\hfill
\begin{subfigure}[b]{0.32\textwidth}
\includegraphics[width=\textwidth]{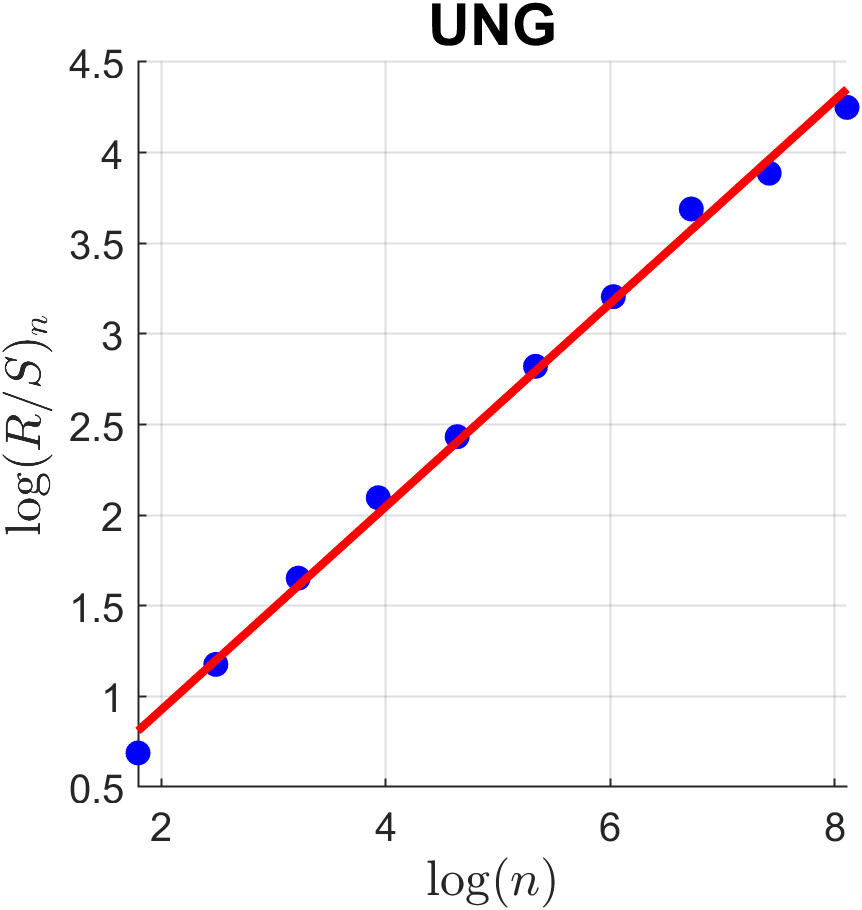}
\caption{UNG}
\end{subfigure}
\hfill
\begin{subfigure}[b]{0.32\textwidth}
\includegraphics[width=\textwidth]{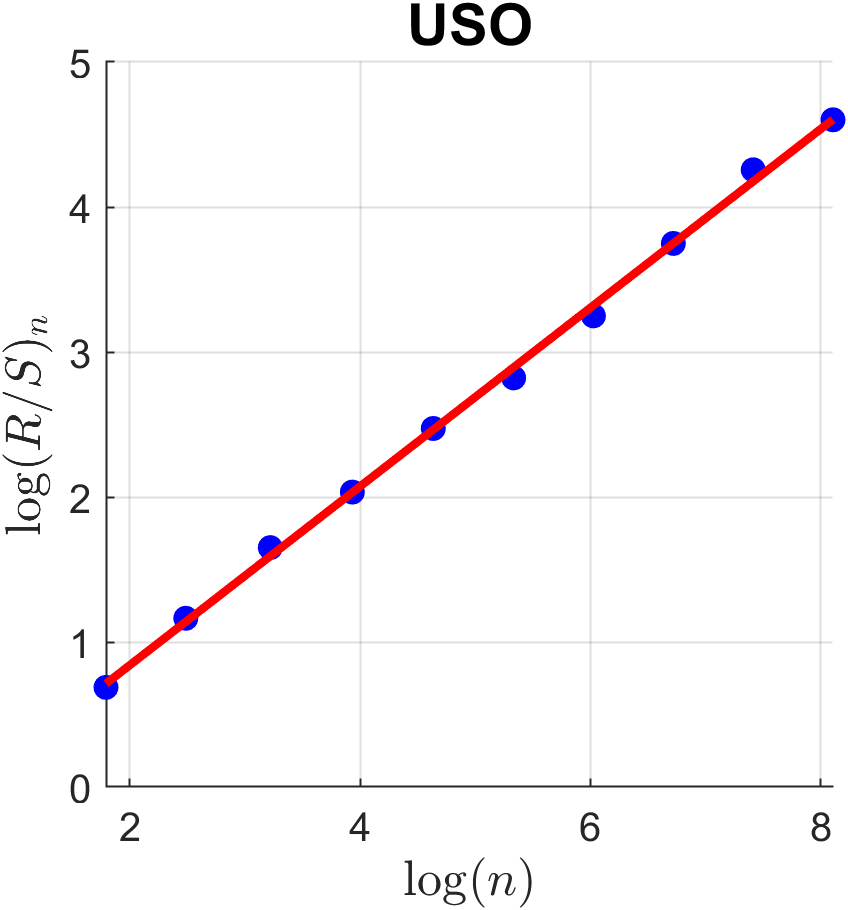}
\caption{USO}
\end{subfigure}
\hfill
\begin{subfigure}[b]{0.32\textwidth}
\includegraphics[width=\textwidth]{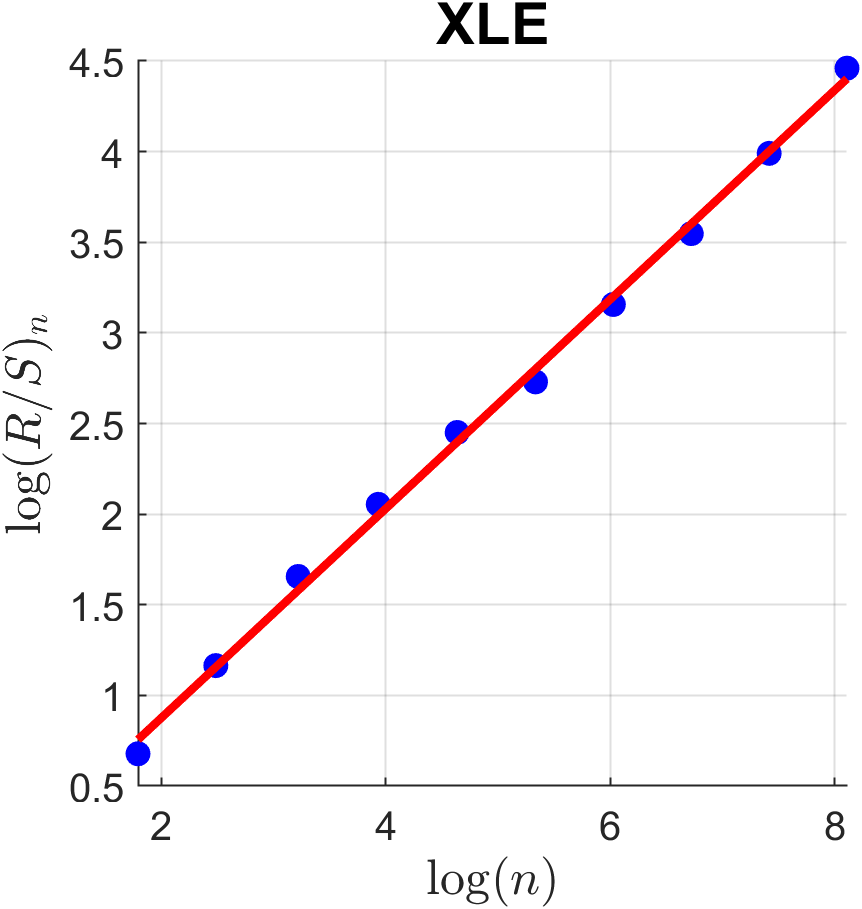}
\caption{XLE}
\end{subfigure}
\caption{Rescaled range analysis for daily returns of the nine assets}
\label{fig:hurst_rs_real}
\end{figure}

Table~\ref{tab:rs} reports the R/S estimates for the daily return series across the three asset classes. The log--log regressions have uniformly high goodness-of-fit, with $R^2$ values exceeding 0.98 for all assets. Figure~\ref{fig:hurst_rs_real} provides the corresponding scaling plots, which show that the estimated linear relationships are visually stable across the selected scales.
\begin{table}[htbp]
\centering
\begin{tabular}{c | l r c c}
\hline
Category & Ticker & $\widehat{H}$ ($p$-value) & 95\% CI & $R^2$\\
\hline
 & \textasciicircum GSPC & 0.506 (0.459) & $[0.466, 0.537]$ & 0.992\\
Stock & \textasciicircum GDAXI & 0.546 (**)$^{a}$ & $[0.513, 0.576]$ & 0.995\\
 & \textasciicircum N225 & 0.530 (0.0854) & $[0.484, 0.575]$ & 0.989 \\ 
 \hline
 & WEAT & 0.546 (*)$^{b}$ & $[0.509, 0.582]$ & 0.993\\
Commodity & CORN & 0.572 (***)$^{c}$ & $[0.550, 0.591]$ & 0.998\\
 & SOYB & 0.544 (***) & $[0.525, 0.567]$ & 0.998\\
\hline
 & UNG & 0.561 (***) & $[0.531, 0.591]$ & 0.996 \\
Energy & USO & 0.616 (***) & $[0.597, 0.635]$ & 0.999\\
 & XLE & 0.578 (***) & $[0.555, 0.600]$ & 0.998\\ 
\hline
\multicolumn{5}{l}{\footnotesize $^{a}$ Indicates a $p$-value less than 0.01.}\\
\multicolumn{5}{l}{\footnotesize $^{b}$ Indicates a $p$-value less than 0.05.}\\
\multicolumn{5}{l}{\footnotesize $^{c}$ Indicates a $p$-value less than 0.001.}\\
\hline
\end{tabular}
\caption{The results of rescaled range analysis}
\label{tab:rs}
\end{table}

Turning to the stock market indices, the estimated Hurst exponents for the S\&P 500 (\textasciicircum GSPC), DAX (\textasciicircum GDAXI), and Nikkei 225 (\textasciicircum N225) are close to the benchmark value of 0.5. Among them, the DAX exhibits a Hurst exponent of 0.546, which is statistically significant at the 1\% level, with a 95\% confidence interval (CI) of $[0.513, 0.576]$, indicating mild but statistically detectable LRD. In contrast, the S\&P 500 and Nikkei 225 display Hurst estimates that are not statistically different from 0.5 at conventional significance levels, suggesting behavior closer to a short-memory or weakly dependent process. 

For the commodity markets, all three series--wheat (WEAT), corn (CORN), and soybeans (SOYB)--exhibit Hurst exponents significantly greater than 0.5. In particular, CORN and SOYB show strong statistical significance, with $p$-values well below 1\%, and their 95\% CIs lying entirely above 0.5. These results provide clear evidence of LRD in agricultural commodity returns, consistent with the presence of persistent dynamics in these markets.

A similar pattern emerges in the energy sector. The Hurst exponents for UNG, USO, and XLE are all significantly greater than 0.5, with USO displaying the strongest degree of persistence ($\widehat{H}=0.616$). The corresponding $p$-values strongly reject the null hypothesis $\textnormal{H}_0: H = 0.5$ in favor of the one-sided alternative $\textnormal{H}_a: H > 0.5$, and the 95\% CIs for all three energy assets lie well above the benchmark value. This indicates pronounced LRD in energy market returns.

Overall, the R/S analysis reveals clear heterogeneity in LRD across asset classes. Persistence is relatively weak for major stock indices, whereas commodity and energy markets exhibit stronger and more statistically robust evidence of long memory.

\subsection{Detrended Fluctuation Analysis}\label{sec:dfa}
In Section~\ref{sec:rs}, we used the R/S analysis to measure the LRD of the return time series. However, this classical method suffers from several limitations. First, the R/S statistic is highly sensitive to short-range correlations and nonstationarities in the data, which can lead to an upward bias in estimating the Hurst exponent~\citep{mandelbrot1968noah}. Moreover, R/S analysis assumes that the underlying process is stationary, making it less robust when dealing with financial time series that often exhibit trends, volatility clustering, and structural breaks~\citep{weron2002estimating}.

To overcome these shortcomings, we now employ DFA, which has been shown to provide a more reliable estimation of LRD by systematically removing local trends before calculating fluctuations, thereby mitigating the influence of nonstationarity~\citep{peng1994mosaic, kantelhardt2001detecting}. In particular, DFA can effectively distinguish between true long-range correlations and apparent scaling behavior caused by underlying trends or heteroskedasticity, which makes it especially suitable for analyzing financial return series~\citep{barunik2010hurst}.

The initial preparations for DFA are the same as Equations~\eqref{eq:step1} and \eqref{eq:step2}, i.e., we select a logarithmically spaced set of scales $n$ and, for each $n$, partition the series $\{X_t\}_{t=1}^N$ into $K = \lfloor N/n\rfloor$ non-overlapping blocks of length $n$. We then form the mean-adjusted cumulative profile $Y(i) = \sum_{t=1}^{i} (X_t - \bar{X})$ for $i=1,\cdots, N$. Within each block $v = 1,\cdots, K$, with local index $i = 1,\cdots, n$, we fit a least-squares linear trend to the profile, $\hat{Y}_v(i) = a_v i + b_v$, where the $(a_v, b_v)$ satisfy
\begin{equation*}
(a_v, b_v) = \arg\min_{(a, b)}\frac{1}{n}\sum_{i=1}^n \left[Y((v-1)n + i) - (ai + b)\right]
^2.
\end{equation*}
Let the corresponding detrended deviations be $e_v(i) = Y((v-1)n + i) - \hat{Y}_v(i)$. The detrended root-mean-square fluctuation for block $v$ at scale $n$ is
\begin{equation*}
F_v(n) = \sqrt{\frac{1}{n}\sum_{i=1}^n e_v(i)^2},
\end{equation*}
and the fluctuation function at scale $n$ used in our estimation aggregates blocks by simple averaging,
\begin{equation*}
F(n) = \frac{1}{K}\sum_{v=1}^{K}F_v(n).
\end{equation*}
Repeating this construction over all selected scales $n$ yields pairs $(\ln n, \ln F(n))$. The Hurst exponent $H$ is then obtained by ordinary least squares from the log--log scaling relation
\begin{equation}\label{eq:dfa_fit}
\ln F(n) = H\cdot\ln n + C,
\end{equation}
where $C$ is a constant.

\begin{figure}[htbp]
\centering
\begin{subfigure}[b]{0.32\textwidth}
\includegraphics[width=\textwidth]{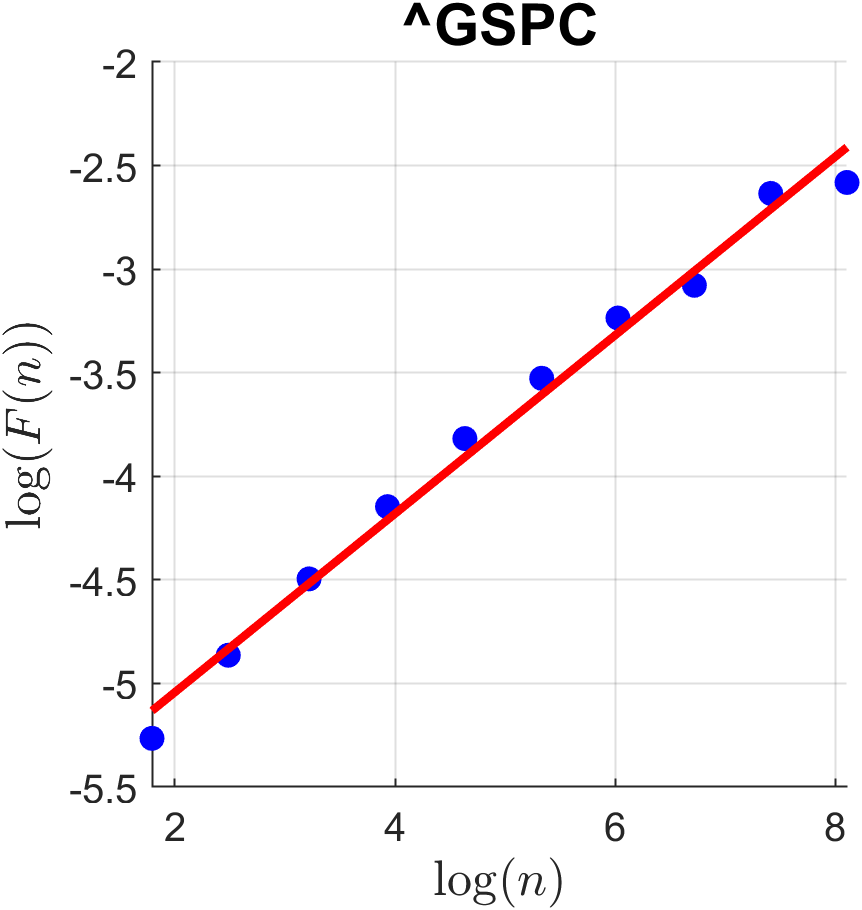}
\caption{\textasciicircum GSPC}
\end{subfigure}
\hfill
\begin{subfigure}[b]{0.32\textwidth}
\includegraphics[width=\textwidth]{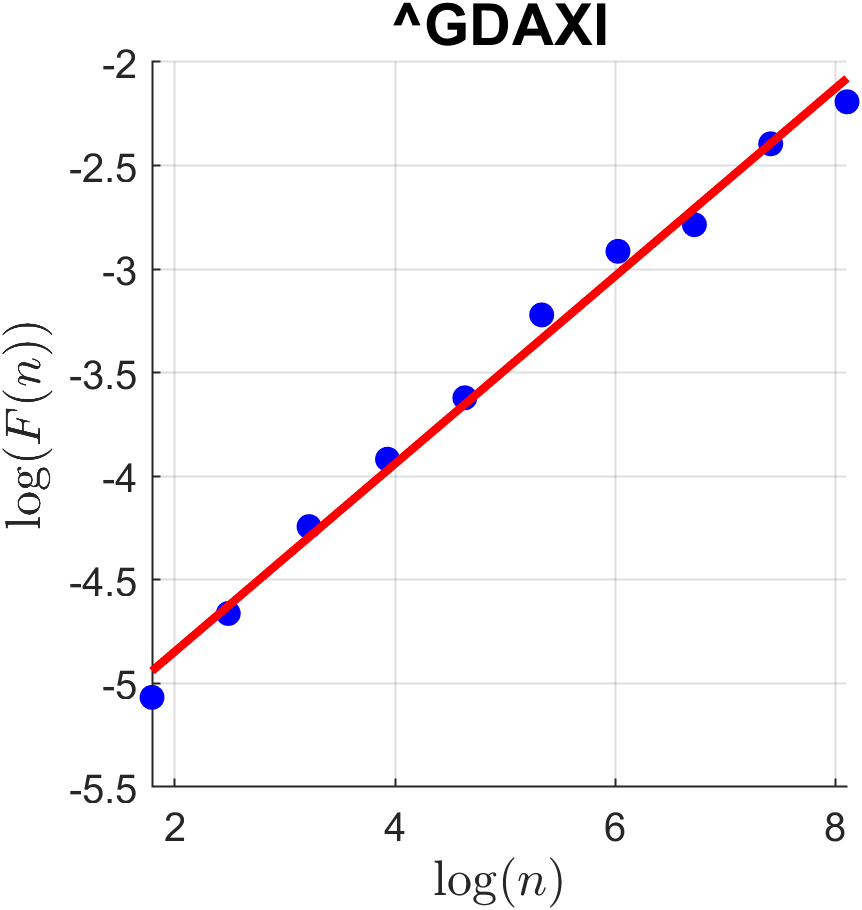}
\caption{\textasciicircum GDAXI}
\end{subfigure}
\hfill
\begin{subfigure}[b]{0.32\textwidth}
\includegraphics[width=\textwidth]{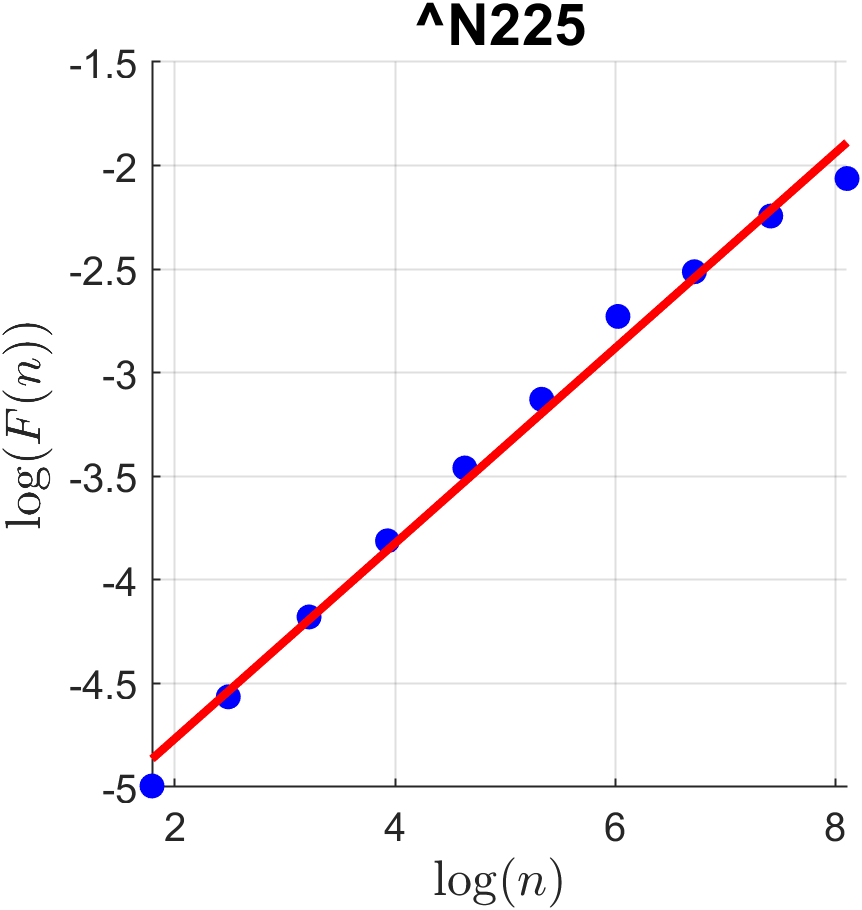}
\caption{\textasciicircum N225}
\end{subfigure}
\hfill
\begin{subfigure}[b]{0.32\textwidth}
\includegraphics[width=\textwidth]{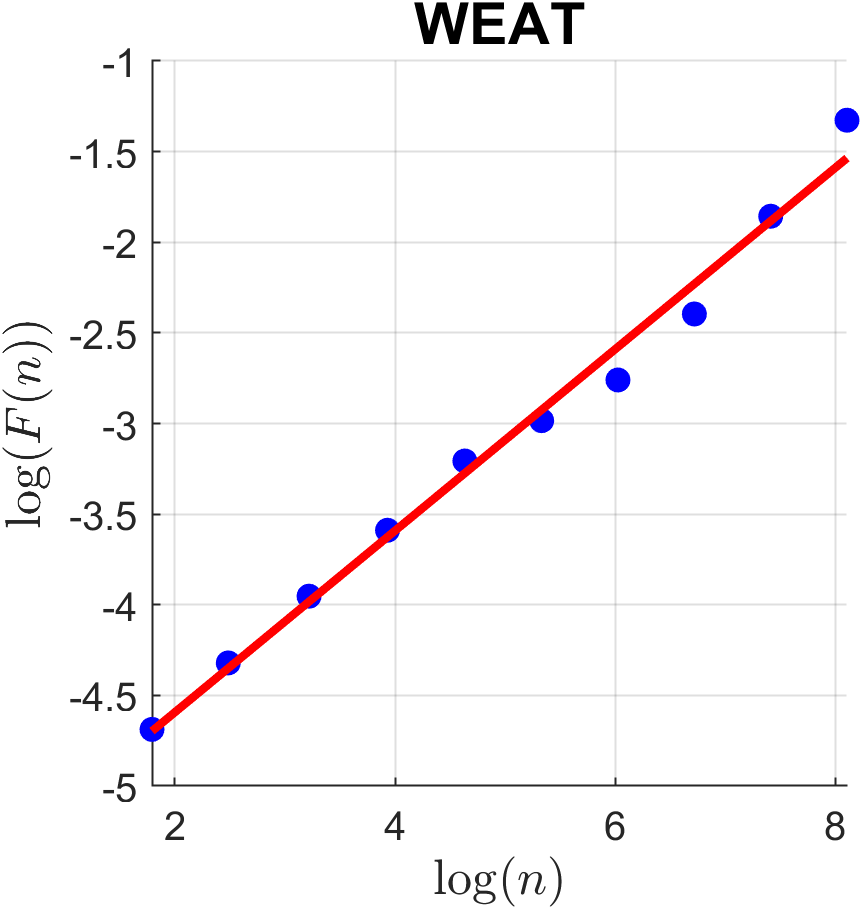}
\caption{WEAT}
\end{subfigure}
\hfill
\begin{subfigure}[b]{0.32\textwidth}
\includegraphics[width=\textwidth]{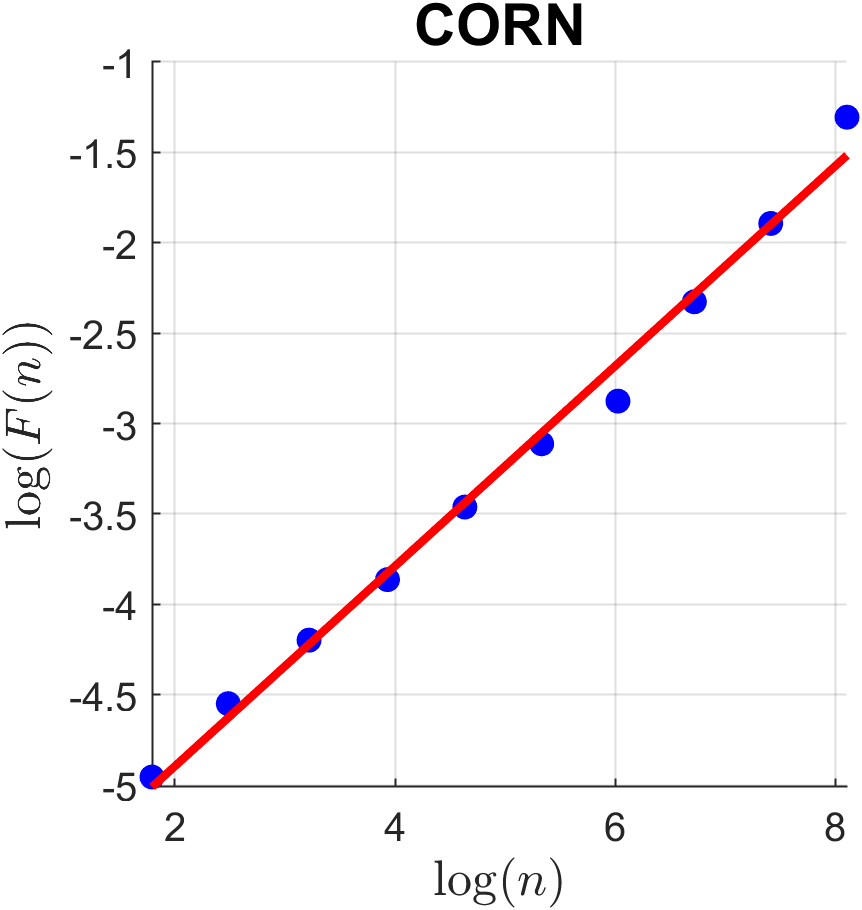}
\caption{CORN}
\end{subfigure}
\hfill
\begin{subfigure}[b]{0.32\textwidth}
\includegraphics[width=\textwidth]{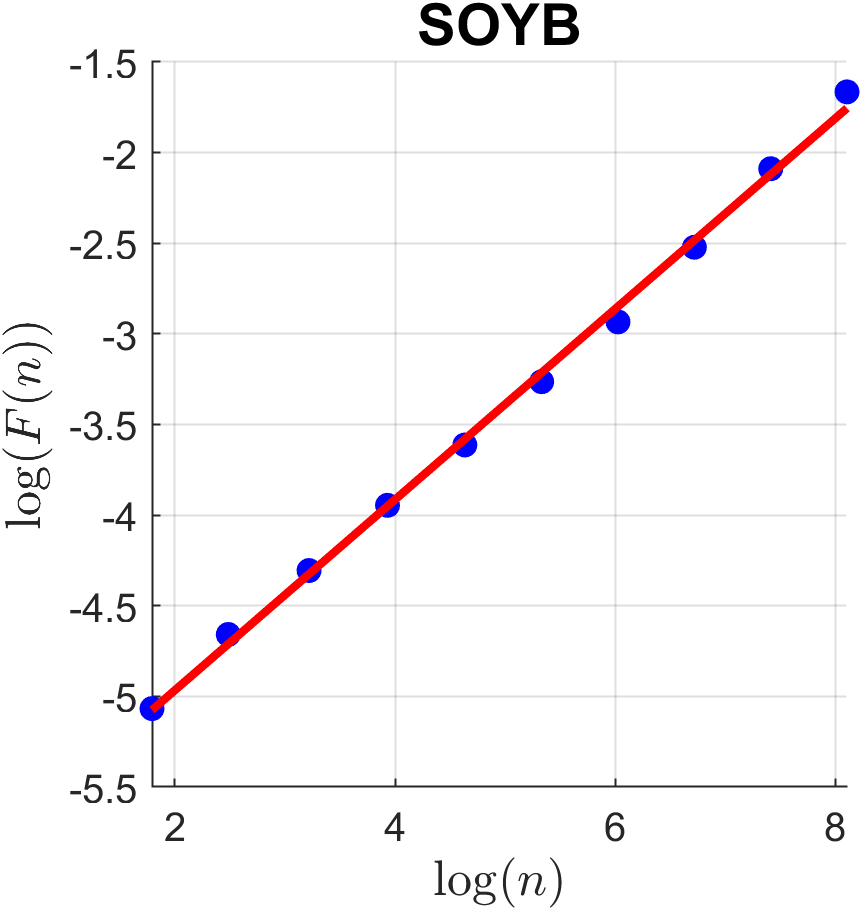}
\caption{SOYB}
\end{subfigure}
\hfill
\begin{subfigure}[b]{0.32\textwidth}
\includegraphics[width=\textwidth]{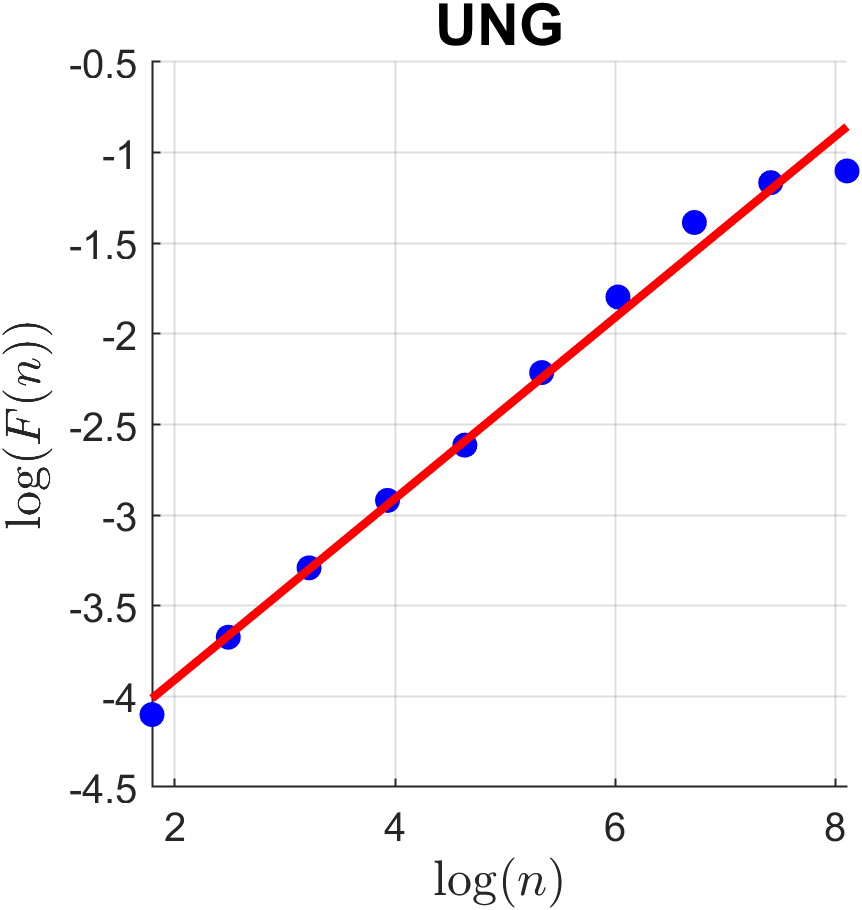}
\caption{UNG}
\end{subfigure}
\hfill
\begin{subfigure}[b]{0.32\textwidth}
\includegraphics[width=\textwidth]{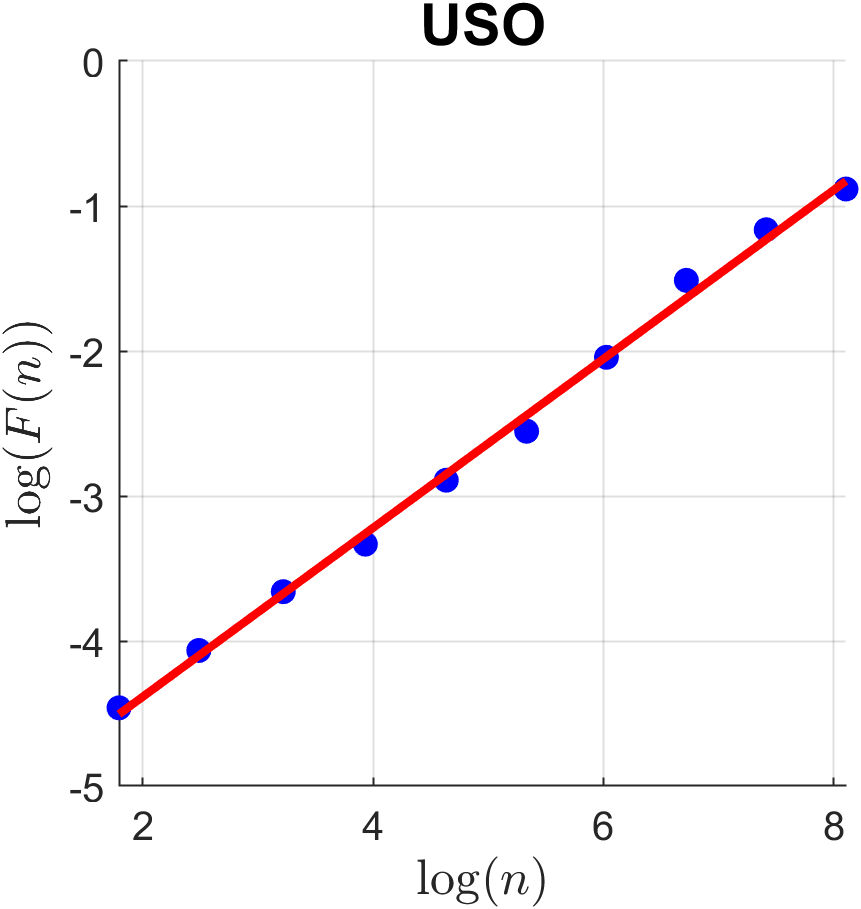}
\caption{USO}
\end{subfigure}
\hfill
\begin{subfigure}[b]{0.32\textwidth}
\includegraphics[width=\textwidth]{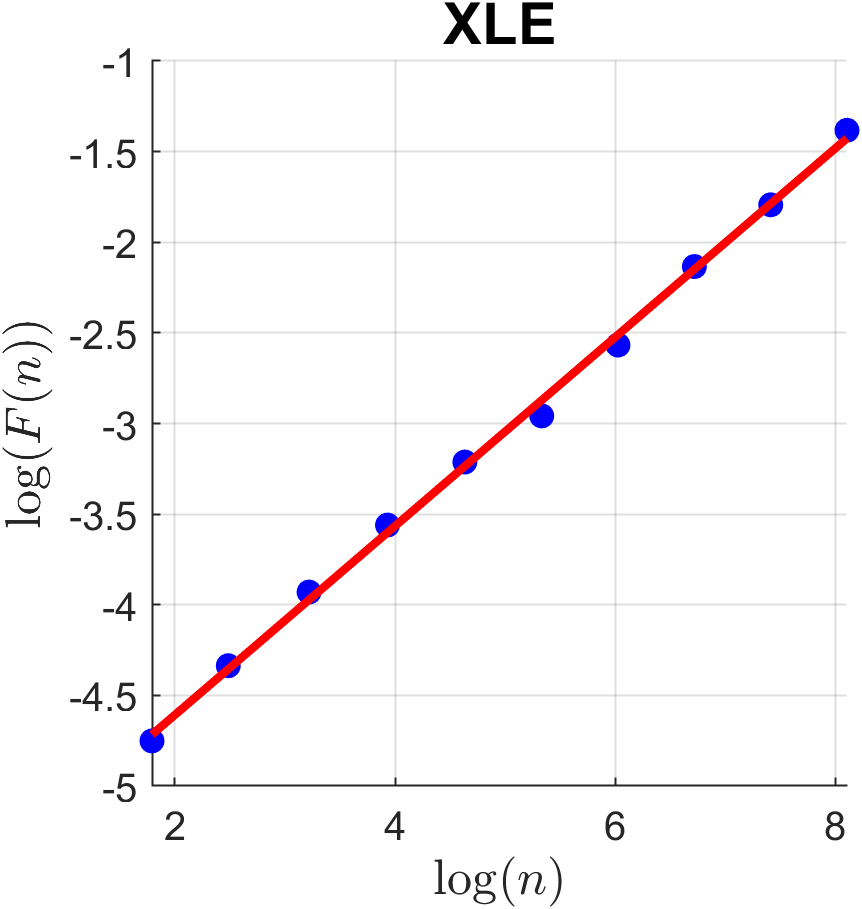}
\caption{XLE}
\end{subfigure}
\caption{Detrended fluctuation analysis for daily returns of the nine assets}
\label{fig:hurst_dfa_real}
\end{figure}

Table~\ref{tab:dfa} presents the DFA estimates for the daily return series across the three asset classes. Similar to the R/S analysis, the DFA regressions exhibit uniformly high goodness-of-fit, with $R^2$ values above 0.98 for all assets. Figure~\ref{fig:hurst_dfa_real} displays the associated log--log scaling relationships and provides a visual check on the stability of the fitted slopes.
\begin{table}[htbp]
\centering
\begin{tabular}{c | l r c c}
\hline
Category & Ticker & $\widehat{H}$ ($p$-value) & 95\% CI & $R^2$\\
\hline
 & \textasciicircum GSPC & 0.432 (0.999) & $[0.394, 0.467]$ & 0.989\\
Stock & \textasciicircum GDAXI & 0.454 (0.994) & $[0.420, 0.486]$ & 0.992\\
 & \textasciicircum N225 & 0.470 (0.947) & $[0.435, 0.508]$ & 0.991 \\ 
 \hline
 & WEAT & 0.501 (0.488) & $[0.457, 0.544]$ & 0.989\\
Commodity & CORN & 0.556 (***)$^{a}$ & $[0.512, 0.595]$ & 0.992\\
 & SOYB & 0.527 (***) & $[0.506, 0.546]$ & 0.998\\
\hline
 & UNG & 0.500 (0.509) & $[0.458, 0.542]$ & 0.990 \\
Energy & USO & 0.582 (***) & $[0.555, 0.609]$ & 0.997\\
 & XLE & 0.522 (*)$^{b}$ & $[0.504, 0.538]$ & 0.998\\ 
\hline
\multicolumn{5}{l}{\footnotesize $^{a}$ Indicates a $p$-value less than 0.001.\quad \footnotesize $^{b}$ Indicates a $p$-value less than 0.05.}\\
\hline
\end{tabular}
\caption{The results of detrended fluctuation analysis}
\label{tab:dfa}
\end{table}

For the stock market indices, all three estimated Hurst exponents are below 0.5: S\&P 500 (\textasciicircum GSPC), DAX (\textasciicircum GDAXI), and Nikkei 225 (\textasciicircum N225). The associated $p$-values are large, and the 95\% CIs either include or lie entirely below 0.5, indicating no statistically significant evidence of LRD in stock returns under DFA. These results suggest that stock market returns are largely consistent with short-memory dynamics once deterministic trends are removed.

In the commodity markets, mixed evidence of LRD is observed. While wheat (WEAT) exhibits a Hurst exponent very close to 0.5 and fails to reject the null hypothesis $\textnormal{H}_0: H = 0.5$, both corn (CORN) and soybeans (SOYB) display statistically significant Hurst exponents greater than 0.5. In particular, their $p$-values strongly reject the null in favor of the one-sided alternative $\textnormal{H}_a: H > 0.5$, and the corresponding 95\% CIs lie entirely above 0.5, providing evidence of persistent behavior in these commodity returns.

For the energy sector, the DFA results indicate heterogeneous persistence. While UNG does not exhibit statistically significant LRD, both USO and XLE display Hurst exponents greater than 0.5. The persistence is especially pronounced for USO, whose 95\% CI $[0.555, 0.609]$ lies well above 0.5 and whose $p$-value strongly rejects the null hypothesis. XLE shows weaker but still statistically significant evidence of LRD at the 5\% level.

Overall, the DFA results provide a more conservative picture than the R/S estimates. They still support LRD in selected commodity and energy markets, but they suggest that stock market returns exhibit little to no persistent behavior once local trends are removed.

\subsection{Segmented Scaling Analysis Around the COVID-19 Period}\label{sec:segmented_scaling}
The normalized equity price paths in Figure~\ref{fig:price_evolution} display visible changes in slope and fluctuation amplitude around the COVID-19 period. To examine whether this visual pattern is associated with changes in scaling behavior, we conduct a segmented analysis for the three equity indices. Following the change-point motivation in recent work on skewed multifractal scaling during COVID-19~\citep{saadaoui2023skewed}, the sample is divided into a pre-COVID subsample ending on December 31, 2019 and a post-COVID subsample starting on January 1, 2020. For each segment, we re-estimate the Hurst exponent using both R/S and DFA.

We also compute the width of the multifractal spectrum using linear-detrending MF-DFA. For a return segment $\{x_t\}_{t=1}^{N}$, we first construct the profile
\begin{equation*}\label{eq:mfdfa_profile}
Y(i)=\sum_{t=1}^{i}\left(x_t-\bar{x}\right), \qquad i=1,\ldots,N.
\end{equation*}
For each scale $s$, the profile is divided into $N_s=\lfloor N/s\rfloor$ non-overlapping windows from both the beginning and the end of the sample. A linear trend $\widehat{Y}_{\nu,s}(i)$ is fitted in each window, and the local detrended variance is computed as
\begin{equation*}\label{eq:mfdfa_variance}
F^2(\nu,s)=\frac{1}{s}\sum_{i=1}^{s}\left[Y\left((\nu-1)s+i\right)-\widehat{Y}_{\nu,s}(i)\right]^2 .
\end{equation*}
The $q$th-order fluctuation function is then
\begin{equation*}\label{eq:mfdfa_fq}
F_q(s)=
\begin{cases}
\left\{\dfrac{1}{2N_s}\sum_{\nu=1}^{2N_s}\left[F^2(\nu,s)\right]^{q/2}\right\}^{1/q}, & q\neq 0,\\[1.1em]
\exp\left\{\dfrac{1}{4N_s}\sum_{\nu=1}^{2N_s}\log\left[F^2(\nu,s)\right]\right\}, & q=0.
\end{cases}
\end{equation*}
Thus, $q=0$ is included in $q\in\{-4,-3,\ldots,4\}$ and is handled by logarithmic averaging. The generalized Hurst exponent $h(q)$ is obtained from the scaling relation $F_q(s)\sim s^{h(q)}$. The multifractal scaling exponent is $\tau(q)=qh(q)-1$, and the singularity spectrum is obtained through the Legendre transform $\alpha(q)=d\tau(q)/dq$ and $f(\alpha)=q\alpha-\tau(q)$. We report the spectrum width
\begin{equation*}\label{eq:mfdfa_width}
\Delta\alpha=\max_q\alpha(q)-\min_q\alpha(q),
\end{equation*}
where a larger $\Delta\alpha$ indicates stronger multifractality and broader scaling heterogeneity.

\begin{table}[htbp]
\centering
\begin{tabular}{l l r c c c}
\hline
Ticker & Segment & $N$ & R/S $\widehat{H}$ & DFA $\widehat{H}$ & $\Delta\alpha$\\
\hline
\textasciicircum GSPC & Pre-COVID & 1923 & 0.504 & 0.417 & 0.348\\
 & Post-COVID & 1395 & 0.563 & 0.509 & 0.304\\ \hline
\textasciicircum GDAXI & Pre-COVID & 1923 & 0.522 & 0.464 & 0.341\\ 
 & Post-COVID & 1395 & 0.559 & 0.514 & 0.424\\ \hline
\textasciicircum N225 & Pre-COVID & 1923 & 0.533 & 0.503 & 0.314\\ 
 & Post-COVID & 1395 & 0.564 & 0.488 & 0.252\\
\hline
\end{tabular}
\caption{Segmented scaling analysis for equity returns around the COVID-19 period}
\label{tab:segmented_scaling}
\end{table}

\begin{figure}[htbp]
\centering
\includegraphics[width=\textwidth]{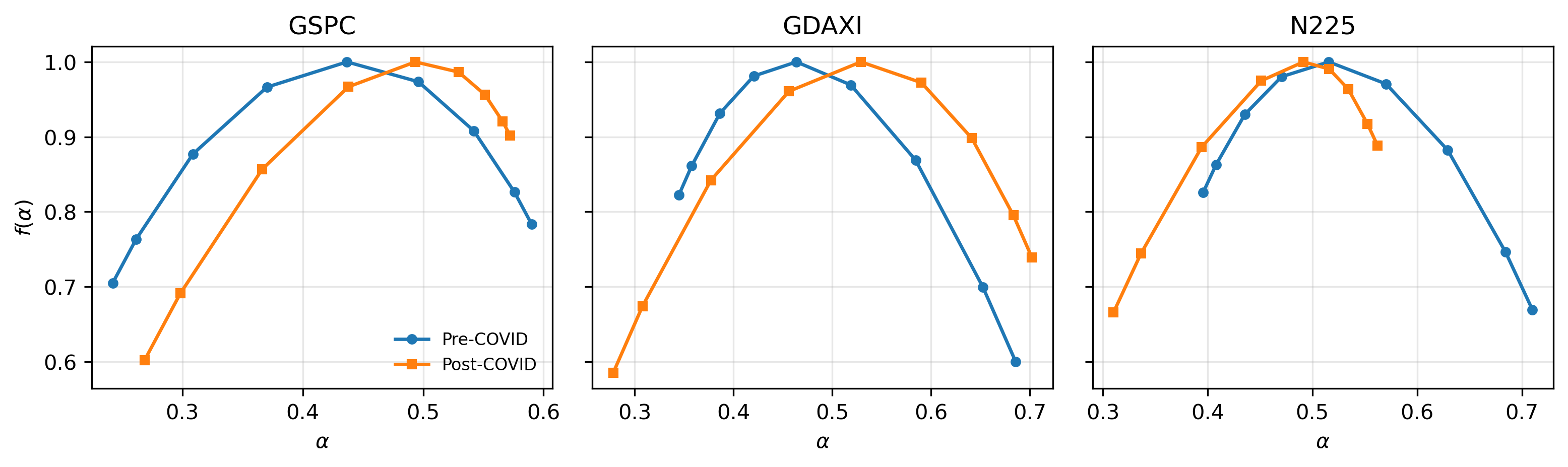}
\caption{Pre- and post-COVID MF-DFA spectra for equity returns}
\label{fig:mfdfa_segmented}
\end{figure}

Table~\ref{tab:segmented_scaling} and Figure~\ref{fig:mfdfa_segmented} show that the R/S-based Hurst exponent increases for all three equity indices after the COVID-19 split, from 0.504 to 0.563 for the S\&P 500, from 0.522 to 0.559 for the DAX, and from 0.533 to 0.564 for the Nikkei 225. The DFA estimates also rise for the S\&P 500 and DAX, moving closer to or above the persistence benchmark of 0.5, although the Nikkei 225 does not show the same DFA increase. The multifractal spectrum width increases for the DAX, suggesting stronger scaling heterogeneity after 2020, while it narrows for the S\&P 500 and Nikkei 225. Hence, the COVID-19 period is associated with a non-negligible shift in scaling properties, but the direction and strength of the change are market-specific. Full-sample estimates should therefore be interpreted as averages over distinct regimes rather than as time-invariant descriptions of equity-market memory.

\subsection[ARFIMA-FIGARCH with Student's t-Distribution]{ARFIMA(1,$d_m$,1)-FIGARCH(1,$d_v$,1) with Student's $t$-Distribution}\label{sec:arfima_figarch}
In Sections~\ref{sec:rs} and \ref{sec:dfa}, we analyzed LRD using R/S analysis and DFA, respectively. While these non-parametric approaches effectively measure the presence of LRD through the Hurst exponent, they do not explicitly model the dynamic structure of the returns or account for additional empirical features commonly observed in financial time series, such as volatility clustering and heavy-tailed distributions. 

To address these limitations, this subsection employs a combined long-memory mean--volatility framework with Student's $t$-distributed innovations. Specifically, the mean component is an ARFIMA model with orders $(1,d_m,1)$, and the volatility component is a FIGARCH model with orders $(1,d_v,1)$. In this setup, the ARFIMA component captures LRD in the conditional mean~\citep{granger1980introduction}, while the FIGARCH component accounts for long memory in conditional volatility~\citep{baillie1996fractionally}. Student's $t$ distribution further accommodates the leptokurtic nature of financial returns~\citep{bollerslev1987conditionally}, making the model suitable for capturing the key stylized facts in financial markets, as discussed in Section~\ref{sec:non_normal_return}.

Given a time series $\{y_t\}$ (e.g., asset returns), suppose we have
\begin{equation}\label{eq:decomp_y_t}
\begin{aligned}
y_t &= \mu_t + \epsilon_t,\\
\epsilon_t &= \sigma_t z_t,
\end{aligned}
\end{equation}
where
\begin{itemize}[noitemsep, topsep=0pt]
\item $\mu_t = \mathrm{E}\left[y_t\mid\mathcal{F}_{t-1}\right]$ is the conditional mean ($\mathcal{F}_{t-1}$ is the information we have up to time $t-1$),
\item $\sigma_t^2 = \mathrm{Var}\left[y_t \mid \mathcal{F}_{t-1}\right]$ is the conditional variance, and
\item the $z_t$ are independent and identically distributed (iid) residuals. In our case, we assume that the $z_t$ follow a Student's $t$ distribution.
\end{itemize}
The structure of ARFIMA($p, d_m, q$) is
\begin{equation}\label{eq:arfima}
\Phi_M(L)(1-L)^{d_m}(y_t - \mu) = \Theta_M(L)\epsilon_t,
\end{equation}
where
\begin{itemize}[noitemsep, topsep=0pt]
\item $\mu = \mathrm{E}[y_t]$ is the unconditional mean,
\item $L$ is the lag operator (i.e., $Ly_t = y_{t-1}$),
\item $\Phi_M(L) = 1 - \sum_{i=1}^{p}\phi_i L^i$,
\item $\Theta_M(L) = 1 + \sum_{i=1}^{q}\theta_i L^i$, and
\item $d_m\in(-0.5, 0.5)$ is the fractional differencing parameter for the mean.
\end{itemize}
The structure of FIGARCH($m,d_v,s$) is
\begin{equation}\label{eq:figarch}
\Phi_V(L)(1-L)^{d_v}\epsilon_t^2 = \omega + [1 - B_V(L)]\sigma_t^2,
\end{equation}
where
\begin{itemize}[noitemsep, topsep=0pt]
\item $\Phi_V(L) = 1 - \sum_{i=1}^{m}\alpha_i L^i$,
\item $B_V(L) = \sum_{i=1}^{s}\beta_i L^i$,
\item $d_v\in[0, 1)$ is the fractional differencing parameter for the volatility.
\end{itemize}
In our study, we combine Equations~\eqref{eq:decomp_y_t} through~\eqref{eq:figarch} and set $p = q = m = s = 1$. This yields a parsimonious ARFIMA--FIGARCH framework with first-order autoregressive and moving-average terms in both the mean and volatility equations.

Table~\ref{tab:arfima_figarch} reports the estimation results of the ARFIMA--FIGARCH model with Student's $t$-distributed innovations for all nine assets. This framework allows us to jointly examine LRD in both the conditional mean and conditional variance processes.
\begin{table}[htbp]
\centering
\begin{tabular}{c | l r l r c}
\hline
Category & Ticker & $\widehat{d}_m$ ($p$-value) & 95\% CI for $\widehat{d}_m$ & $\widehat{d}_v$ $^{a}$ & 95\% CI for $\widehat{d}_v$ \\ \hline
 & \textasciicircum GSPC & $4.95\cdot 10^{-3}$ (***) $^{b}$ & $[3.42\cdot 10^{-3}, 6.48\cdot 10^{-3}]$ & 0.392 & $[0.391,0.394]$\\
Stock & \textasciicircum GDAXI & $1.00\cdot 10^{-8}$ (0.500) & $[-3.56\cdot 10^{-2}, 3.56\cdot 10^{-2}]$ & 0.494 & $[0.347, 0.641]$ \\ 
 & \textasciicircum N225 & $1.00\cdot 10^{-8}$ (0.500) & $[-6.38\cdot 10^{-4}, 1.00\cdot 10^{-3}]$ & 0.403 & $[0.401, 0.404]$ \\  \hline
 & WEAT & 0.0357 (***) & $[0.0337, 0.0377]$ & 0.380 & $[0.378, 0.382]$ \\ 
Commodity & CORN & 0.0753 (0.0671) & $[-0.0233, 0.173]$ & 0.462 & $[0.154, 0.772]$ \\
 & SOYB & $1.00\cdot 10^{-8}$ (0.500) & $[-3.35\cdot 10^{-3}, 3.35\cdot 10^{-3}]$ & 0.393 & $[0.392, 0.394]$ \\ \hline
 & UNG & $1.31\cdot 10^{-6}$ (0.500) & $[-0.0427, 0.0427]$ & 0.885 & $[0.844, 0.926]$ \\
Energy & USO & $2.90\cdot 10^{-6}$ (0.500) & $[-0.0966, 0.0966]$ & 0.892 & $[0.844, 0.940]$ \\ 
 & XLE & $1.00\cdot 10^{-8}$ (0.500) & $[-0.0381, 0.0381]$ & 0.578 & $[0.239, 0.918]$ \\ \hline
\multicolumn{5}{l}{\footnotesize $^{a}$ All $p$-values for hypothesis test $\textnormal{H}_0: d_v = 0$ vs. $\textnormal{H}_a: d_v > 0$ are less than 0.001.}\\
\multicolumn{5}{l}{\footnotesize $^{b}$ Indicates a $p$-value less than 0.001.}\\
\hline
\end{tabular}
\caption{The results of ARFIMA--FIGARCH fit}
\label{tab:arfima_figarch}
\end{table}
The parameter $\widehat{d}_m$ captures the degree of fractional integration in returns, while $\widehat{d}_v$ measures LRD in volatility. Since the Hurst exponent is related to the fractional differencing parameter through $H = d + 0.5$, testing the null hypothesis $\textnormal{H}_0: H = 0.5$ is equivalent to testing $\textnormal{H}_0: d = 0$. Accordingly, all reported $p$-values correspond to the one-sided hypothesis test $\textnormal{H}_0: d = 0$ vs. $\textnormal{H}_a: d > 0$.

For the stock market indices, the estimated mean-memory parameters $\widehat{d}_m$ are generally close to zero. While the S\&P 500 (\textasciicircum GSPC) exhibits a statistically significant but economically small estimate of $\widehat{d}_m$, both the DAX (\textasciicircum GDAXI) and Nikkei 225 (\textasciicircum N225) fail to reject the null hypothesis of no LRD in returns, as their 95\% CIs include zero. These findings are consistent with the R/S and DFA results, which indicate weak or absent persistence in stock returns.

In the commodity markets, heterogeneous LRD in returns is observed. Wheat (WEAT) displays a statistically significant and positive $\widehat{d}_m$, with a narrow CI entirely above zero, providing strong evidence of persistent behavior. Corn (CORN) shows marginal evidence of LRD, whereas soybeans (SOYB) do not exhibit statistically significant fractional integration in the mean process.

For the energy sector, none of the three assets--UNG, USO, and XLE--exhibit statistically significant LRD in returns, as indicated by the insignificance of $\widehat{d}_m$.

In contrast, the estimates of the volatility memory parameter $\widehat{d}_v$ are uniformly positive and highly significant across all assets, with all $p$-values below 0.001. This provides strong evidence of LRD in volatility dynamics, regardless of asset class. Notably, energy assets such as UNG and USO exhibit particularly high values of $\widehat{d}_v$, suggesting pronounced persistence in volatility even when return-level LRD is absent.

Overall, the ARFIMA--FIGARCH results confirm that LRD manifests differently across markets and model components: while persistence in returns is asset- and method-dependent, LRD in volatility appears to be a robust and universal feature across stocks, commodities, and energy markets.

\section{Learning Long-Range Dependence through GANs}\label{sec:learn_LRD}
Section~\ref{sec:measure_LRD} introduced several complementary approaches for quantifying LRD, enabling us to determine whether a return time series exhibits such dependence. Building upon these findings, we now turn to a related but distinct question: can the intrinsic LRD structure of financial returns be learned by modern data-driven approaches, particularly those based on machine learning and deep learning?

In this section, we investigate the potential of Quant GANs to replicate the LRD observed in empirical return series. Section~\ref{sec:quantgans} provides a concise overview of the Quant GAN framework, originally developed by~\cite{wiese2020quant}. As the primary objective of this work is not to conduct an in-depth algorithmic analysis, we limit our discussion to an intuitive explanation of the framework. Readers seeking a more comprehensive treatment are referred to the original paper and the corresponding implementation repository at~\url{https://github.com/KseniaKingsep/quantgan}.

Subsequently, Section~\ref{sec:emp} evaluates the ability of Quant GANs to learn from return time series across the nine selected assets. The empirical analysis contrasts synthetic data generated by Quant GANs with real financial data from multiple perspectives, including price dynamics, return distributions, benchmark accuracy, full-ensemble Hurst robustness, and computational workload.

\subsection{Quant GANs: Theoretical Framework}\label{sec:quantgans}
GANs were first introduced by~\cite{goodfellow2014generative} as a novel framework for generative modeling. A GAN consists of two neural networks, a \textit{generator} $G$ and a \textit{discriminator} $D$, which are trained simultaneously in a two-player minimax game. The objective of the generator is to produce synthetic data samples that resemble the real data, while the discriminator aims to distinguish between the real and the generated samples. Formally, the original GAN optimization problem is expressed as
\begin{equation}\label{eq:gans}
\min_{G}\max_{D}V(D, G) = \mathrm{E}_{x\sim p_{\rm data}}[\log D(x)] + \mathrm{E}_{z\sim p_z}[\log(1 - D(G(z)))].
\end{equation}
In Equation~\eqref{eq:gans}, $x\sim p_{\rm data}$ means that $x$ is sampled from the \textit{true data distribution} $p_{\rm data}$---for example, in a financial context, $x$ could represent an observed historical return or price sequence---while $z\sim p_z$ means that $z$ is sampled from a \textit{prior noise distribution} $p_z$, typically chosen to be a simple Gaussian or uniform distribution. The generator $G$ maps this noise $z$ into a synthetic sample $G(z)$ intended to resemble real data. The discriminator $D$ takes an input sample and outputs a scalar value $D(x)\in(0,1)$, representing the estimated probability that the sample is real rather than generated.

The training process forms an adversarial game:
\begin{itemize}[noitemsep, topsep=0pt]
\item The discriminator $D$ maximizes $V(D,G)$ by improving its ability to correctly discriminate between real and generated data.
\item The generator $G$ minimizes $V(D,G)$ by learning to produce data $G(z)$ that the discriminator cannot distinguish from true data.
\end{itemize}
At equilibrium, the generator learns to approximate the true data distribution, i.e., $p_g\approx p_{\rm data}$, resulting in synthetic samples that are statistically indistinguishable from real ones.

Building upon this framework, Quant GANs were proposed by~\cite{wiese2020quant} to address the challenges of generating synthetic financial time series. Unlike standard GANs, which focus purely on matching the overall data distribution, Quant GANs are designed to capture the domain-specific features observed in financial markets. Both the generator and discriminator are implemented using a TCN architecture, which is well suited for modeling sequential data.

Quant GANs serve as a simulation engine for financial markets. The generator $G$ produces synthetic price or return sequences that replicate important statistical properties, such as volatility clustering, heavy-tailed distributions, and serial dependencies. The discriminator $D$ acts as a quality controller, ensuring that the generated series are indistinguishable from real historical data. This makes Quant GANs valuable in applications such as stress testing, portfolio optimization, and derivative pricing, where large amounts of realistic yet synthetic financial data are needed.

\subsection{Empirical Evaluation of Quant GANs}\label{sec:emp}

\subsubsection{Visualization of Simulated Price Paths}
To evaluate the ability of Quant GANs to reproduce the statistical properties of financial time series, we adopt the implementation provided in the official repository:~\url{https://github.com/KseniaKingsep/quantgan}. The model is trained on historical return series using the default hyperparameter configuration described in the repository, including the TCN architecture for both the generator and the discriminator. After training, we generate 10,000 synthetic \textit{log-price paths}, $\ln P_t$, for each asset under consideration. The following analysis focuses exclusively on evaluating these simulated paths and contrasts them with the empirical data in terms of price dynamics, return distributions, and LRD, without delving into the algorithmic details of the underlying model.

\begin{figure}[htbp]
\centering
\begin{subfigure}[b]{0.32\textwidth}
\includegraphics[width=\textwidth]{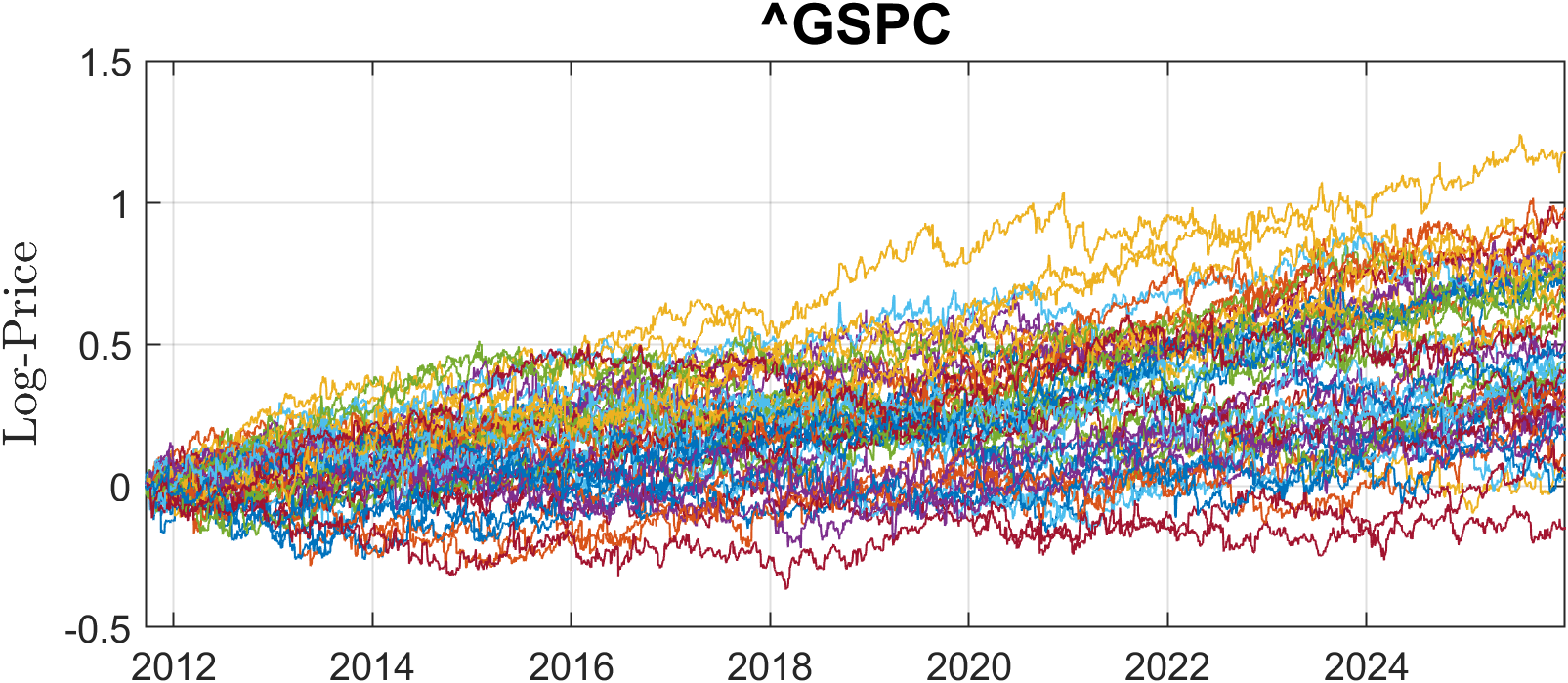}
\caption{\textasciicircum GSPC}
\end{subfigure}
\hfill
\begin{subfigure}[b]{0.32\textwidth}
\includegraphics[width=\textwidth]{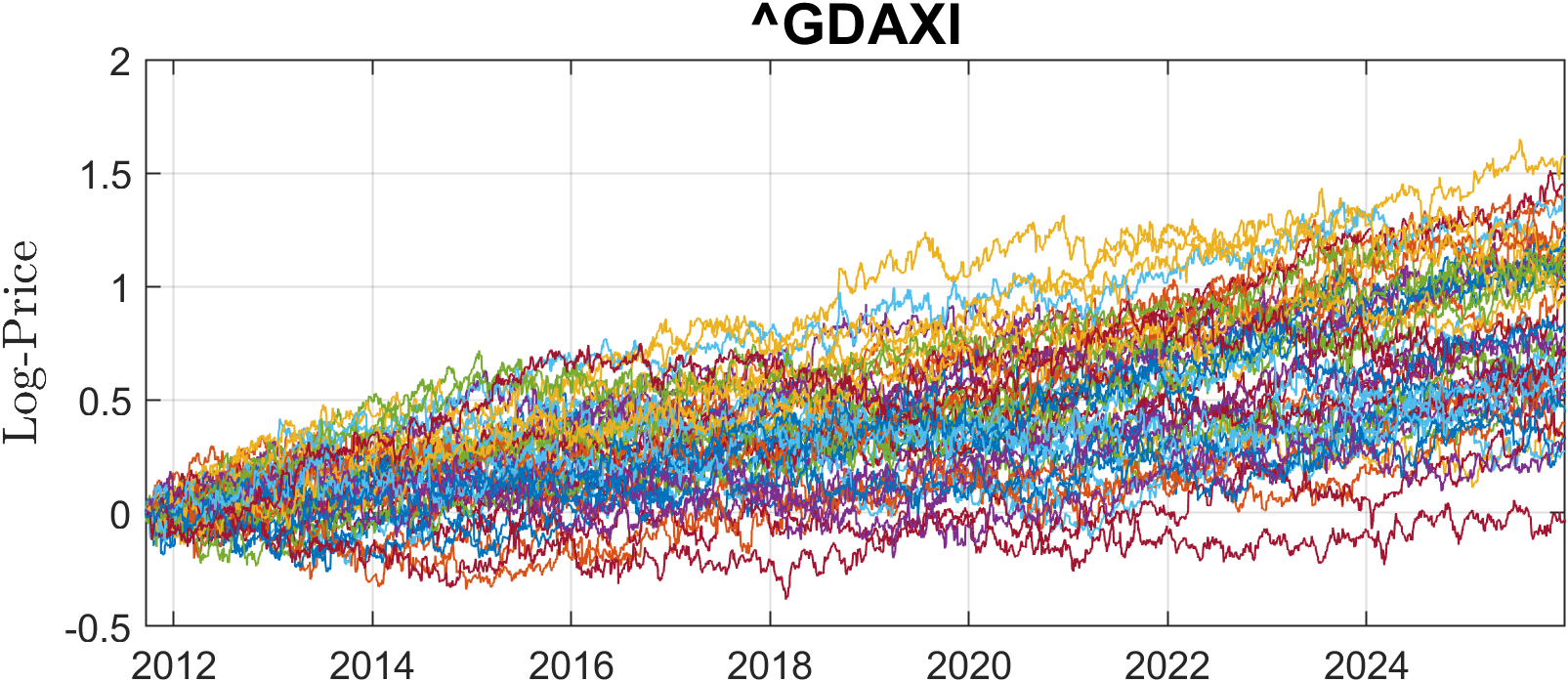}
\caption{\textasciicircum GDAXI}
\end{subfigure}
\hfill
\begin{subfigure}[b]{0.32\textwidth}
\includegraphics[width=\textwidth]{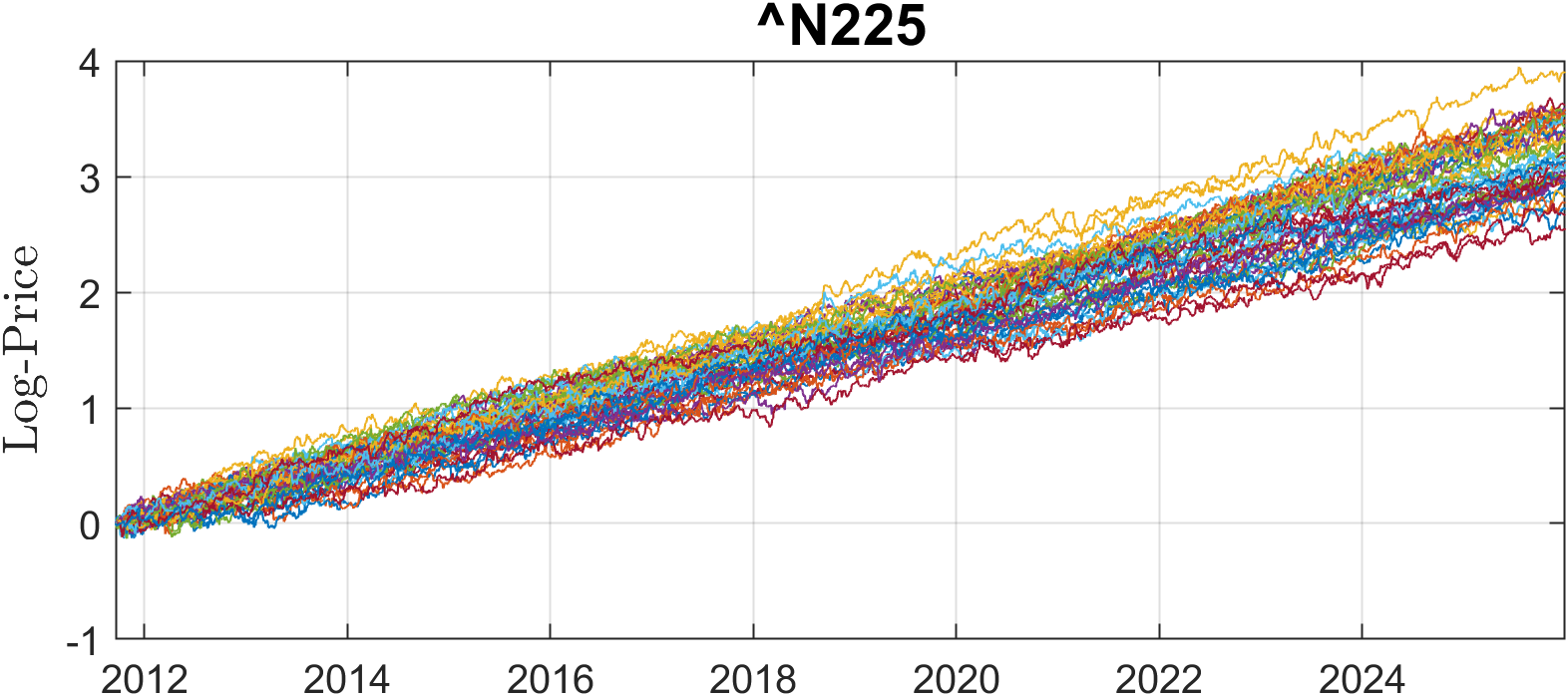}
\caption{\textasciicircum N225}
\end{subfigure}
\hfill
\begin{subfigure}[b]{0.32\textwidth}
\includegraphics[width=\textwidth]{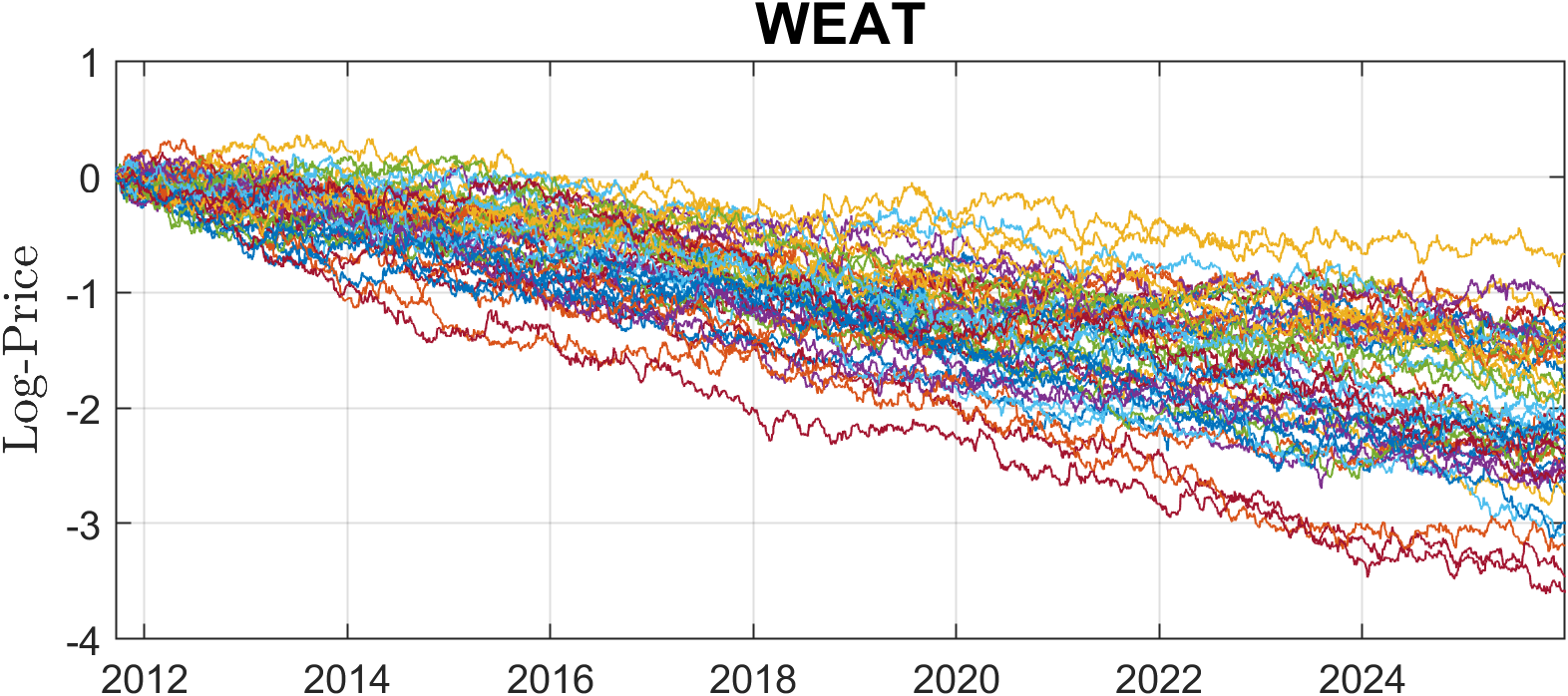}
\caption{WEAT}
\end{subfigure}
\hfill
\begin{subfigure}[b]{0.32\textwidth}
\includegraphics[width=\textwidth]{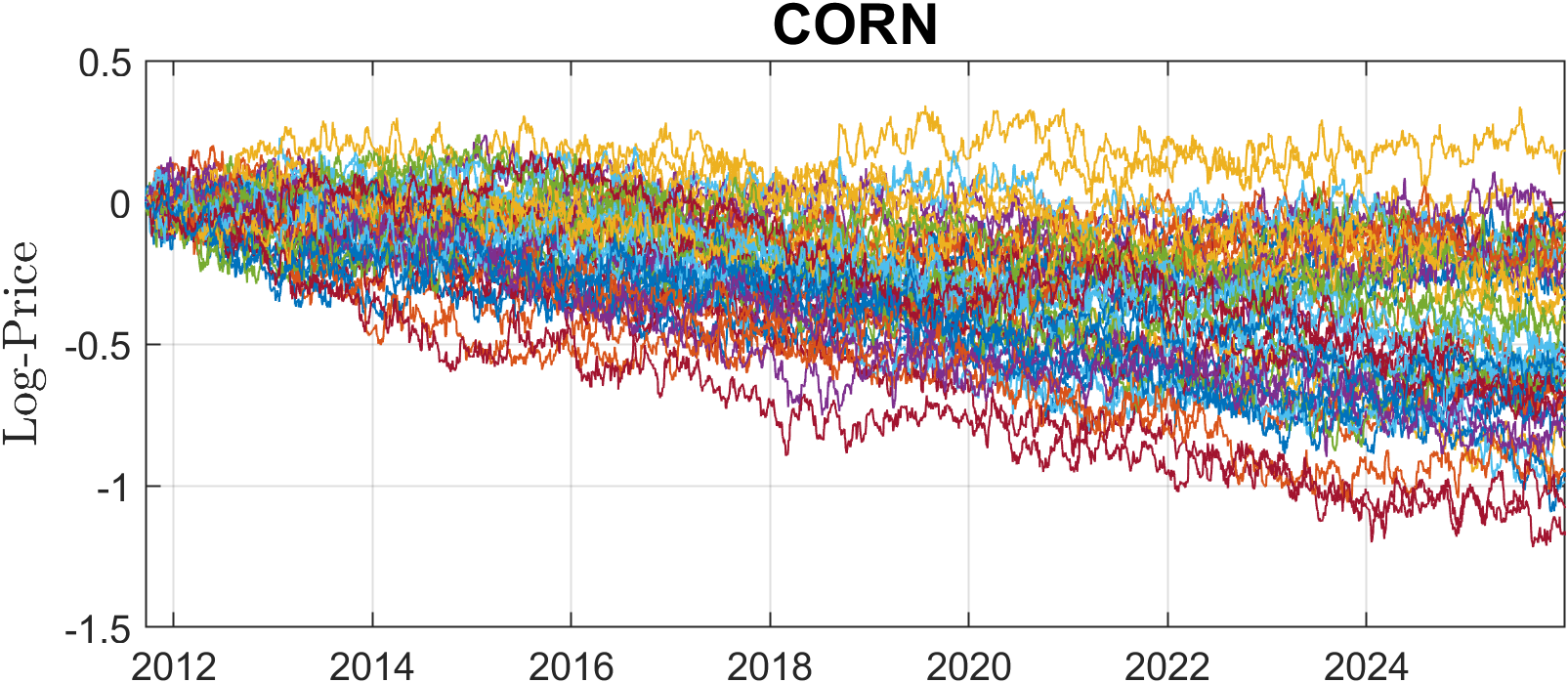}
\caption{CORN}
\end{subfigure}
\hfill
\begin{subfigure}[b]{0.32\textwidth}
\includegraphics[width=\textwidth]{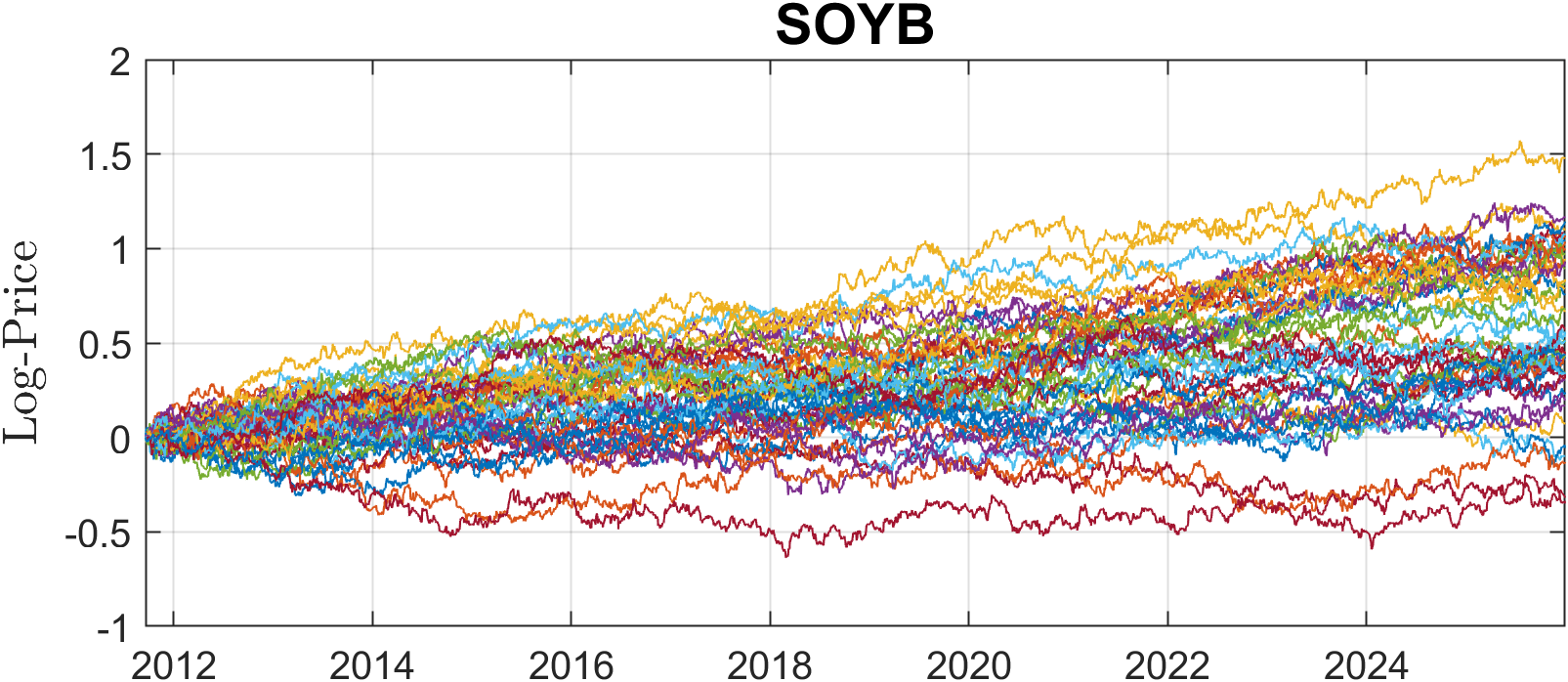}
\caption{SOYB}
\end{subfigure}
\hfill
\begin{subfigure}[b]{0.32\textwidth}
\includegraphics[width=\textwidth]{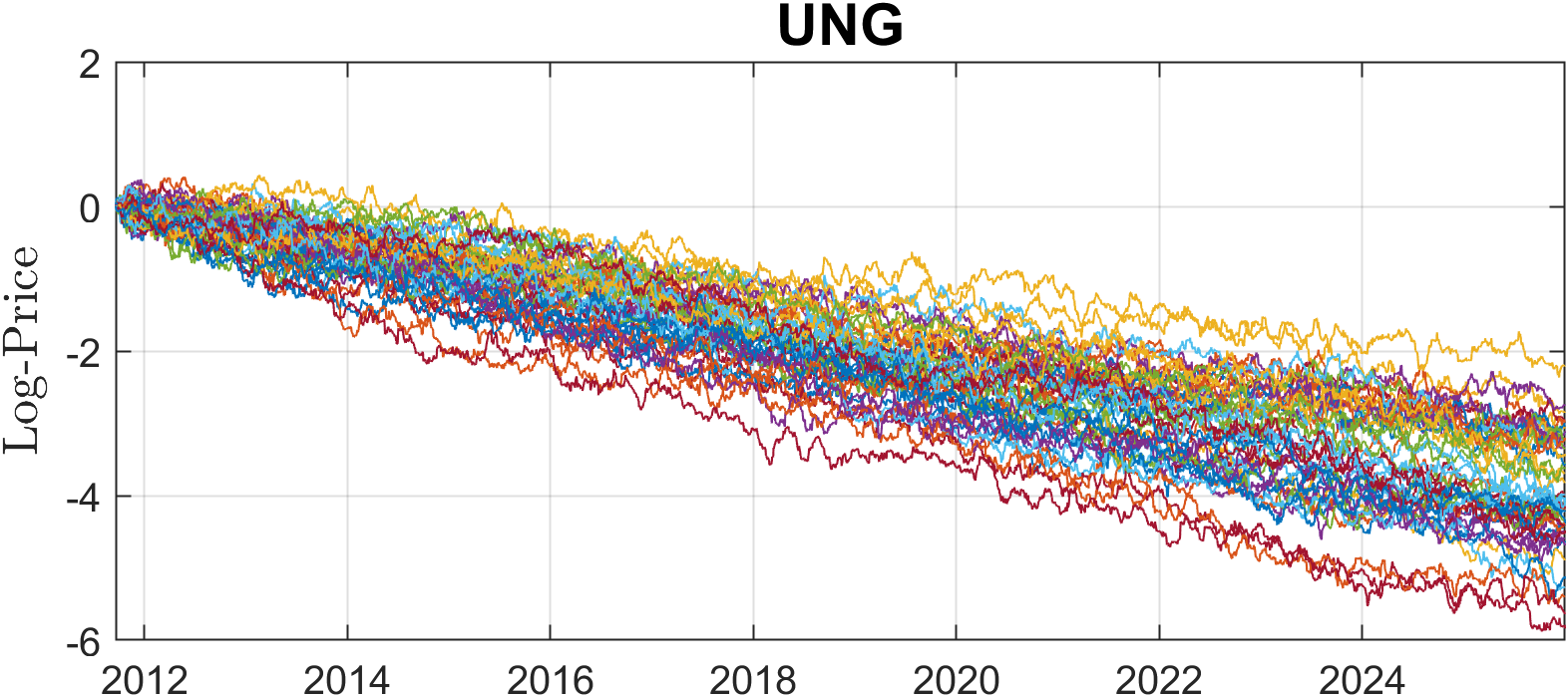}
\caption{UNG}
\end{subfigure}
\hfill
\begin{subfigure}[b]{0.32\textwidth}
\includegraphics[width=\textwidth]{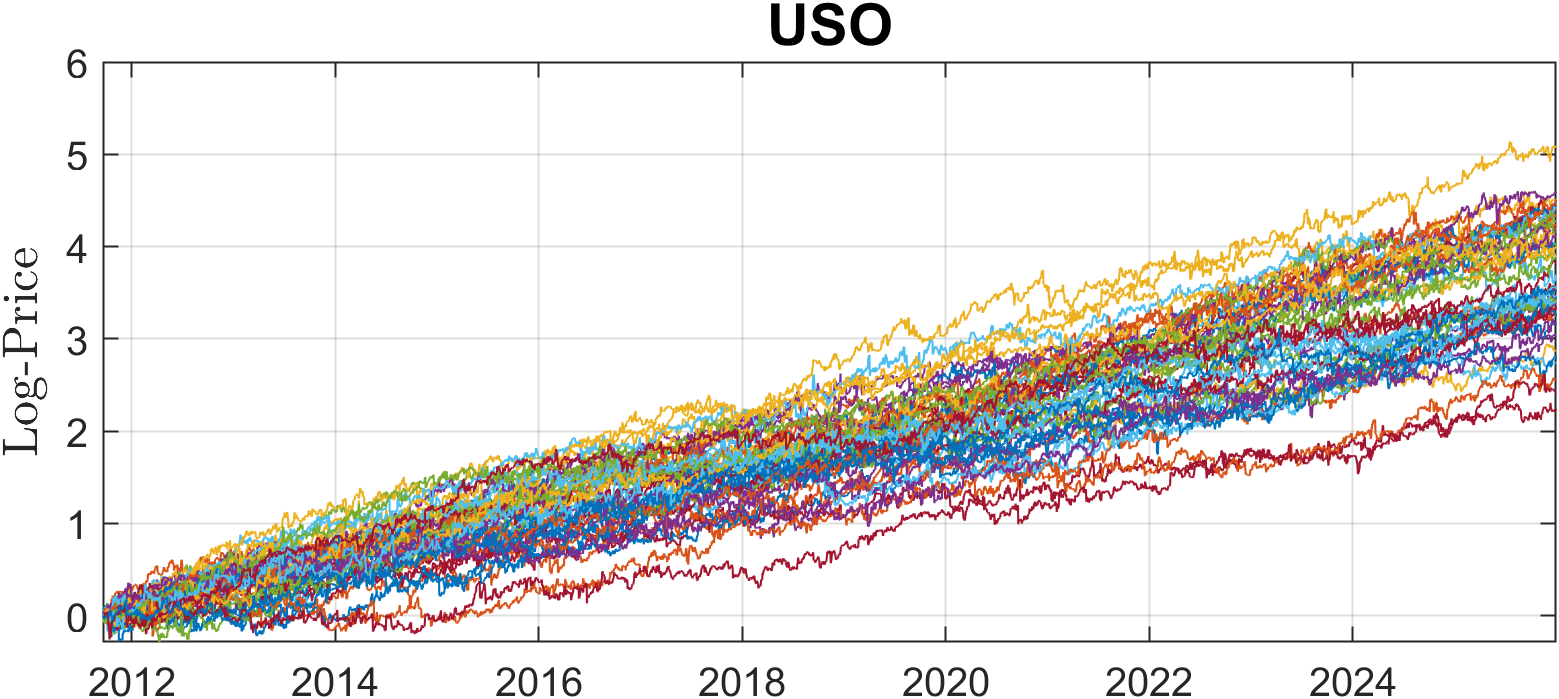}
\caption{USO}
\end{subfigure}
\hfill
\begin{subfigure}[b]{0.32\textwidth}
\includegraphics[width=\textwidth]{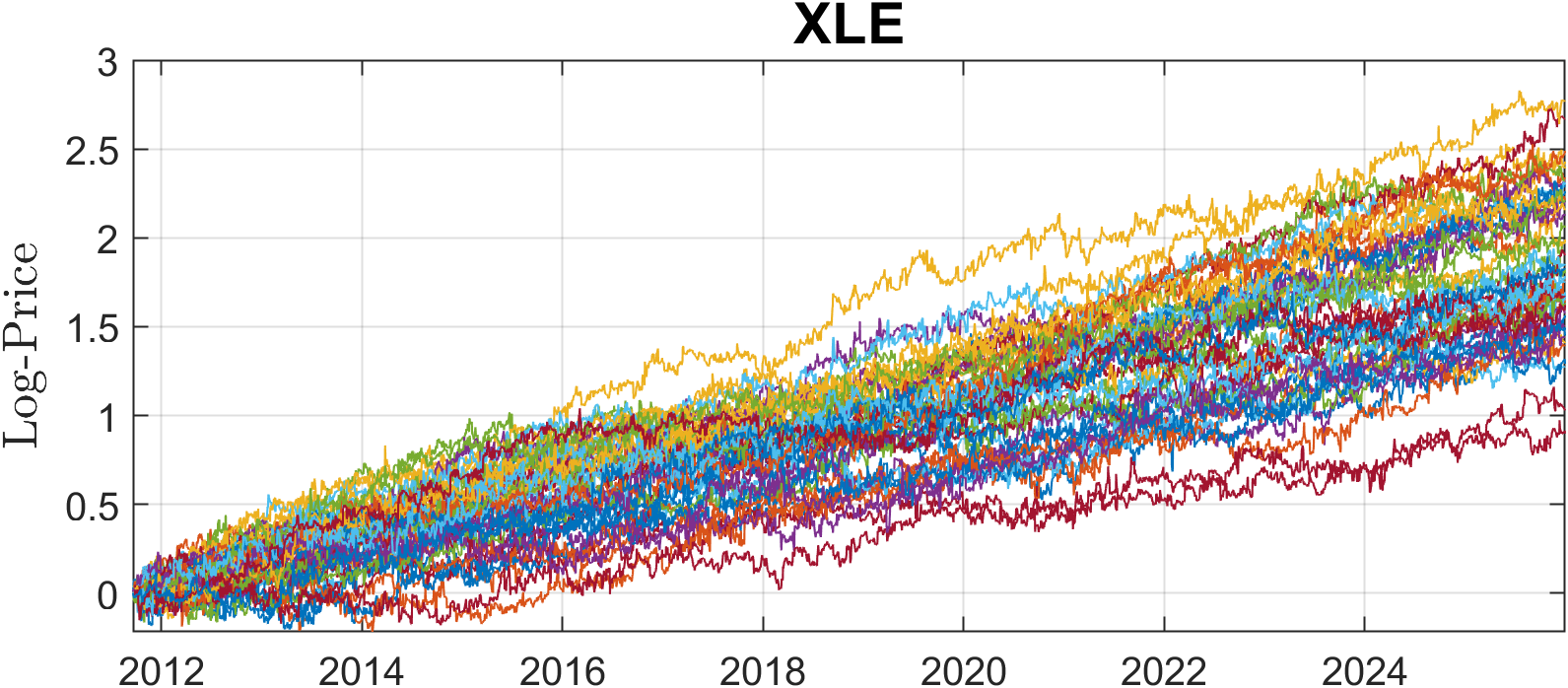}
\caption{XLE}
\end{subfigure}
\caption{Fifty selected simulated daily log-price paths generated by Quant GANs for the nine assets}
\label{fig:sim_log_paths}
\end{figure}

Displaying all 10,000 simulated paths is impractical. For visualization, Figure~\ref{fig:sim_log_paths} reports 50 simulated daily log-price paths for each asset. These paths are used to inspect the range of trajectories produced by the trained generator before turning to path-level and distributional diagnostics.

Across equity markets, the simulated log-price paths for the S\&P 500 (\textasciicircum GSPC), DAX (\textasciicircum GDAXI), and Nikkei 225 (\textasciicircum N225) exhibit broadly similar upward trends. The paths are relatively smooth and display moderate dispersion, indicating stable long-run growth with persistent fluctuations around the trend. This behavior closely resembles the empirical characteristics of major equity indices, where gradual appreciation dominates despite short-term volatility.

In the commodity market, more heterogeneous dynamics are observed. The simulated paths for Wheat (WEAT) and Corn (CORN) tend to follow a downward trajectory, whereas Soybeans (SOYB) exhibit a general upward trend over the simulation horizon. Compared with equities, agricultural commodities display substantially higher path-wise variability, with frequent reversals and wider dispersion across simulated paths. These features suggest weaker trend persistence and stronger influence of short- and medium-term shocks, which are consistent with the well-documented volatility and episodic behavior of agricultural commodity prices.

The energy market exhibits a distinct pattern. UNG shows a pronounced downward trend in the simulated log-price paths, while USO and XLE display upward trends. At the same time, all three energy assets demonstrate elevated volatility, with substantial fluctuations around their respective trends. In particular, the dispersion of simulated paths is markedly larger than that observed in equity markets, reflecting the sensitivity of energy prices to geopolitical events, supply constraints, and macroeconomic conditions.

Overall, the simulated log-price paths generated by Quant GANs capture several broad commonalities and cross-asset differences observed in real financial markets. Equities are characterized by smoother upward trends, commodities exhibit mixed directional movements with high variability, and energy assets combine strong trends with pronounced volatility. These qualitative observations provide an initial visual assessment of the model's ability to reproduce realistic price dynamics across different asset classes and motivate the subsequent quantitative analysis of return distributions and LRD.

\subsubsection{Selection of Representative Paths}
\begin{figure}[htbp]
\centering
\begin{subfigure}[b]{0.32\textwidth}
\includegraphics[width=\textwidth]{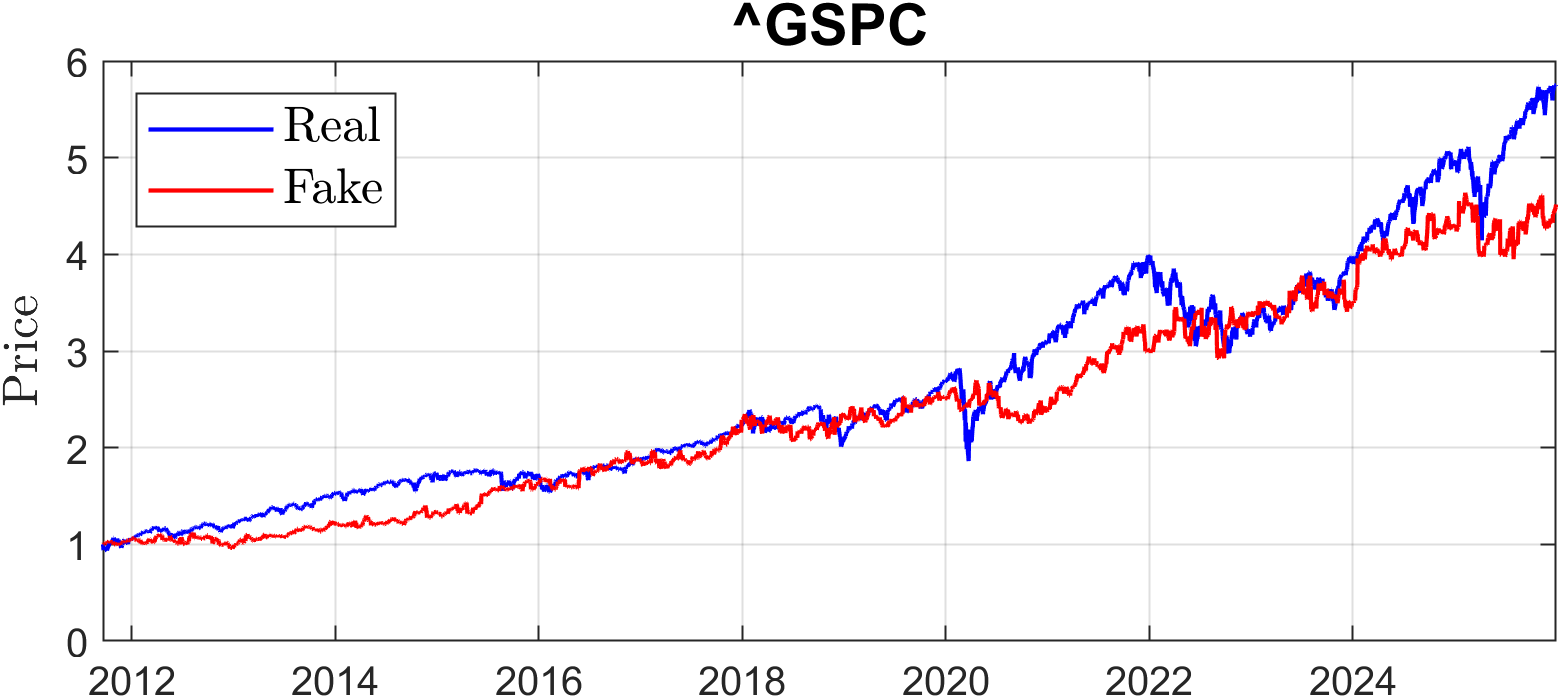}
\caption{\textasciicircum GSPC}
\end{subfigure}
\hfill
\begin{subfigure}[b]{0.32\textwidth}
\includegraphics[width=\textwidth]{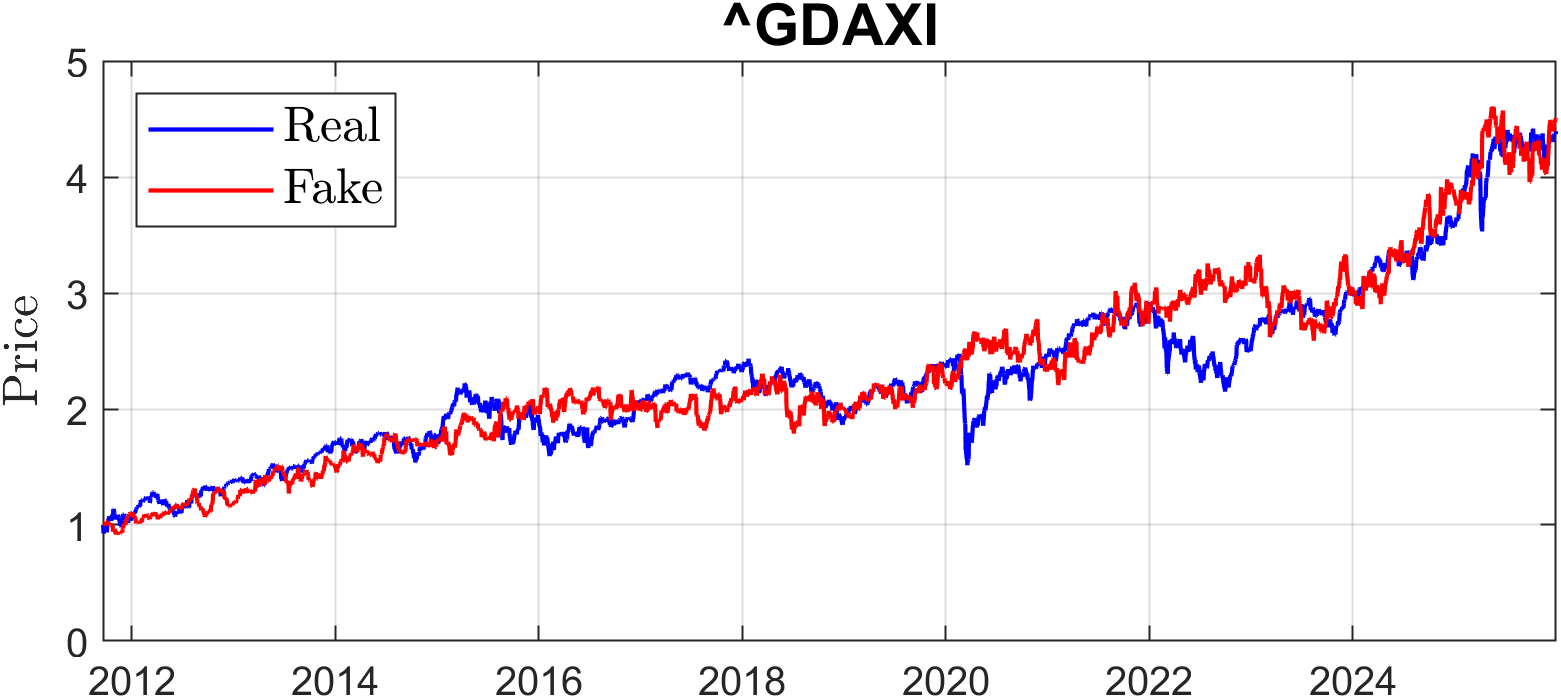}
\caption{\textasciicircum GDAXI}
\end{subfigure}
\hfill
\begin{subfigure}[b]{0.32\textwidth}
\includegraphics[width=\textwidth]{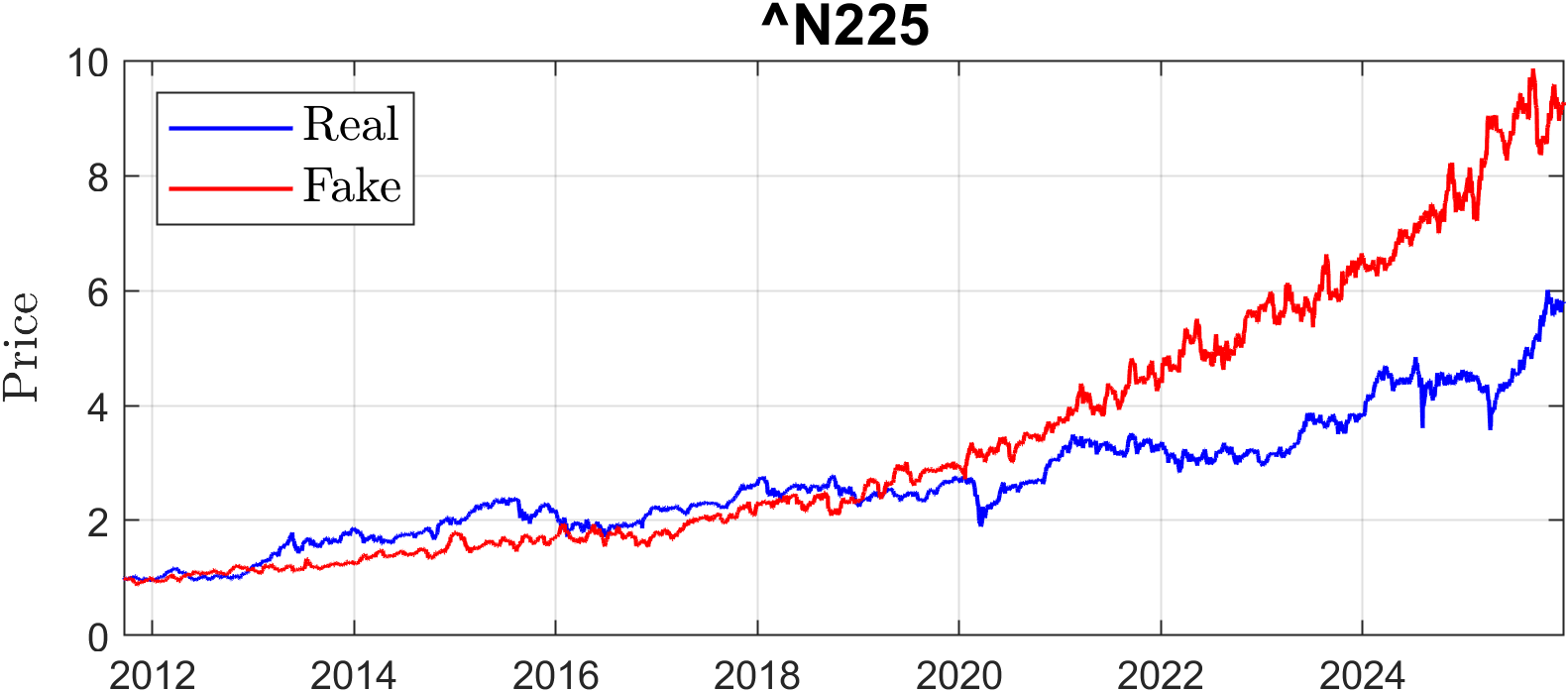}
\caption{\textasciicircum N225}
\end{subfigure}
\hfill
\begin{subfigure}[b]{0.32\textwidth}
\includegraphics[width=\textwidth]{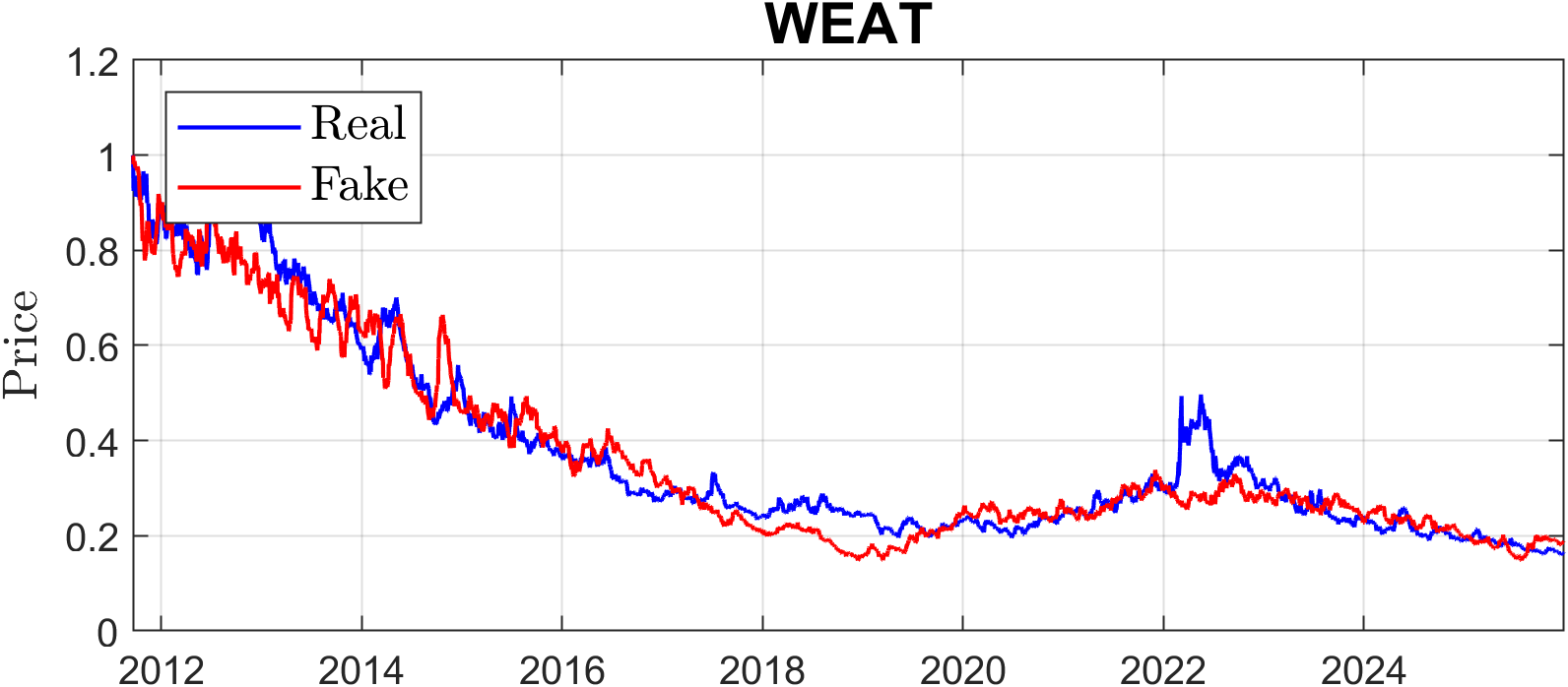}
\caption{WEAT}
\end{subfigure}
\hfill
\begin{subfigure}[b]{0.32\textwidth}
\includegraphics[width=\textwidth]{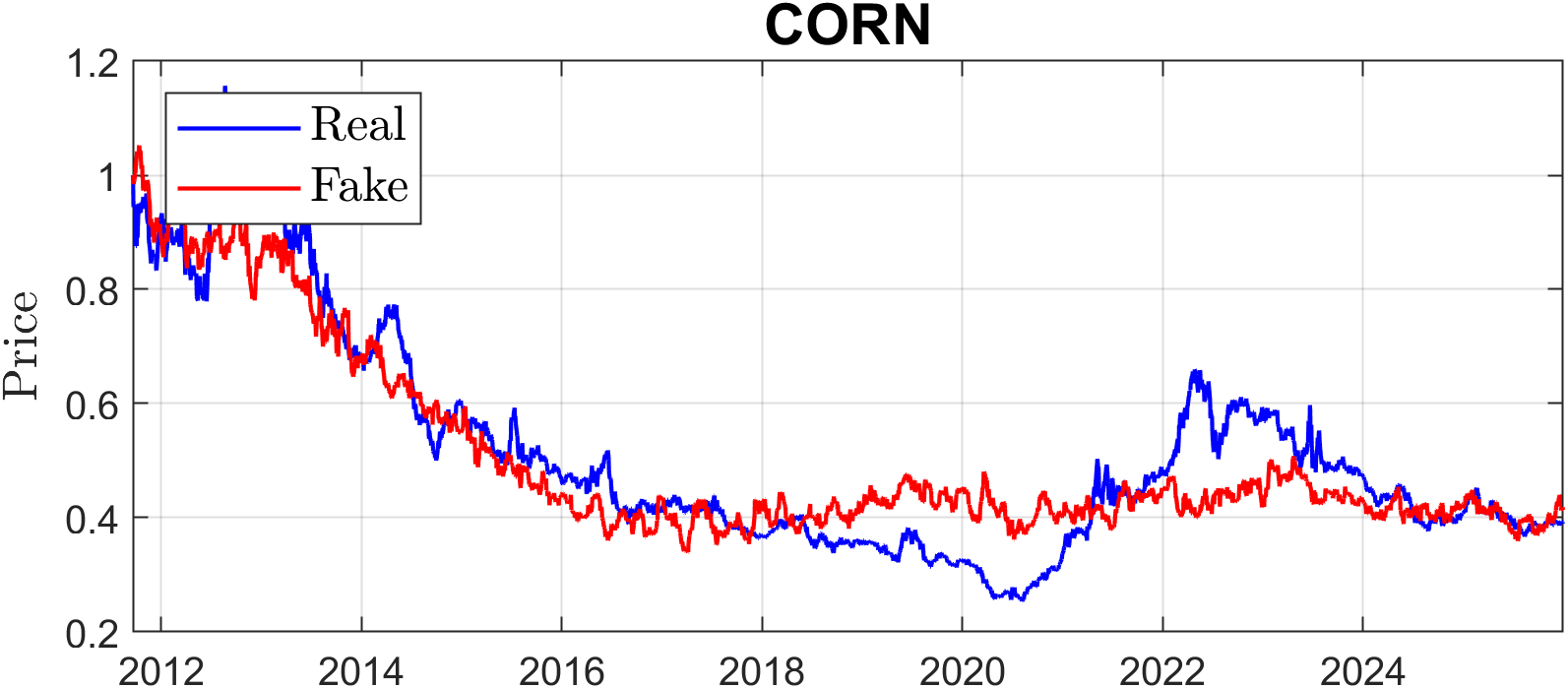}
\caption{CORN}
\end{subfigure}
\hfill
\begin{subfigure}[b]{0.32\textwidth}
\includegraphics[width=\textwidth]{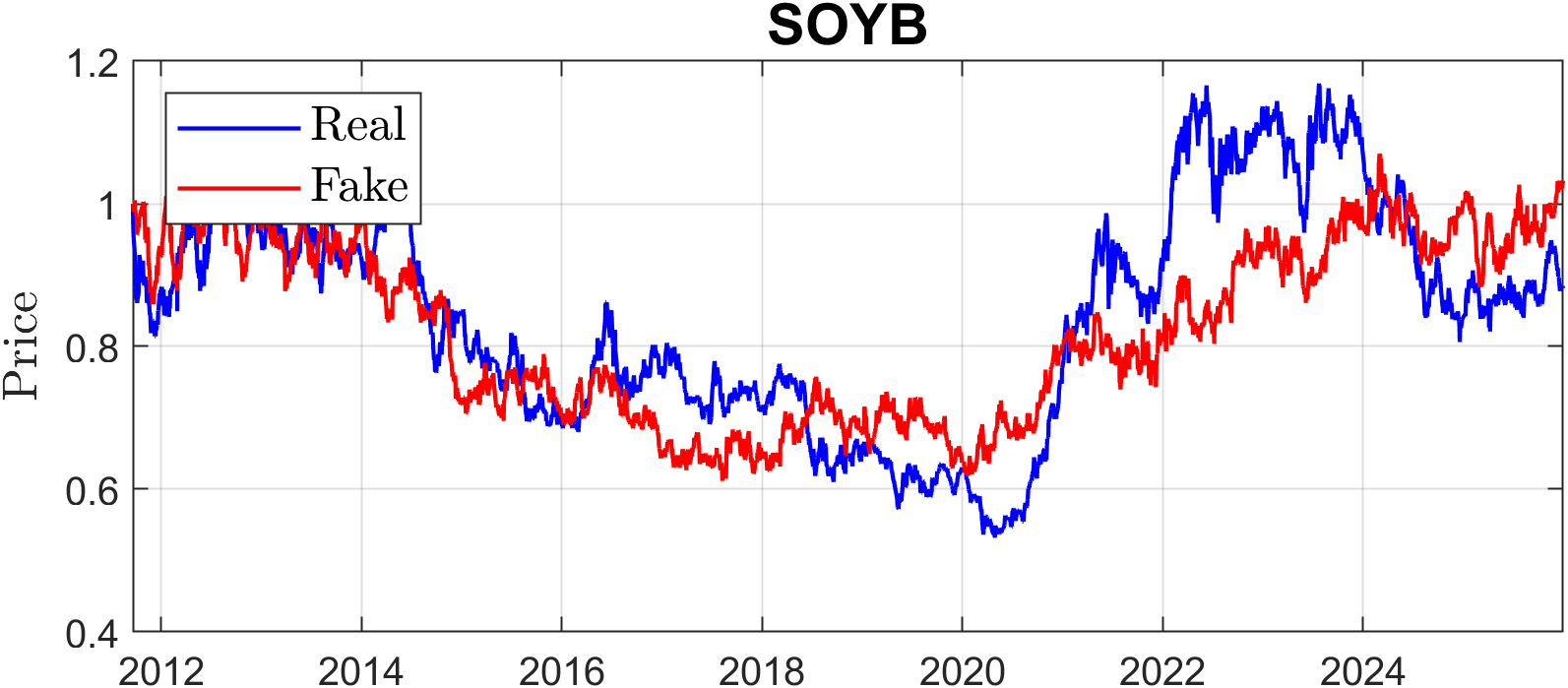}
\caption{SOYB}
\end{subfigure}
\hfill
\begin{subfigure}[b]{0.32\textwidth}
\includegraphics[width=\textwidth]{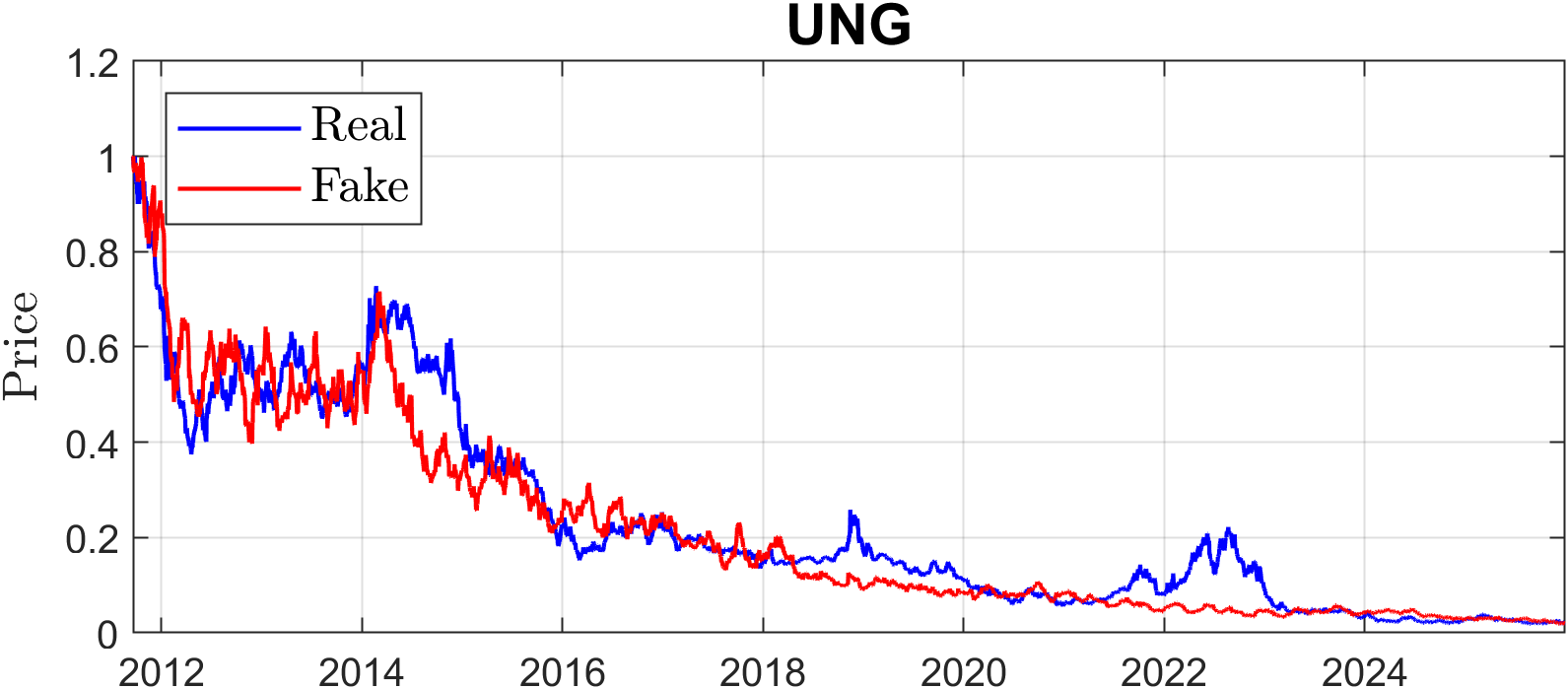}
\caption{UNG}
\end{subfigure}
\hfill
\begin{subfigure}[b]{0.32\textwidth}
\includegraphics[width=\textwidth]{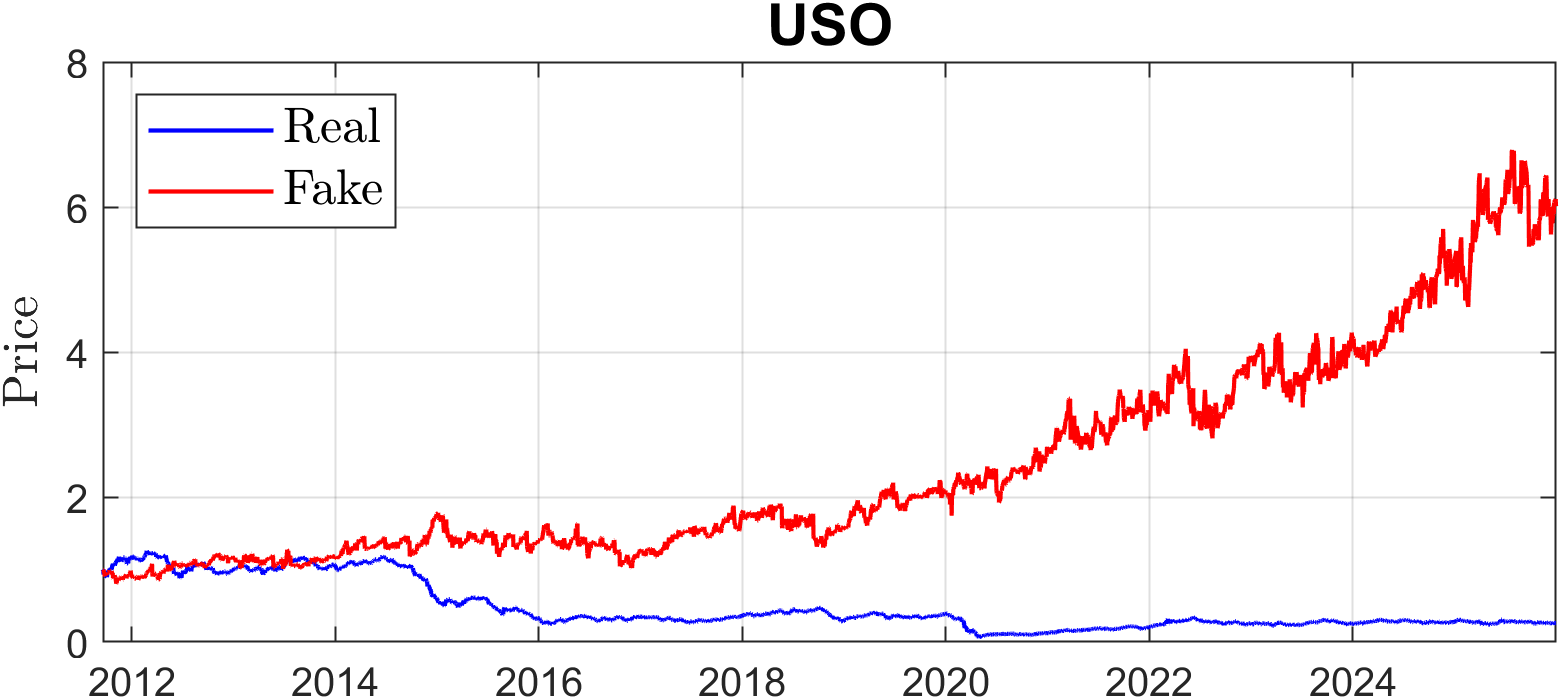}
\caption{USO}
\end{subfigure}
\hfill
\begin{subfigure}[b]{0.32\textwidth}
\includegraphics[width=\textwidth]{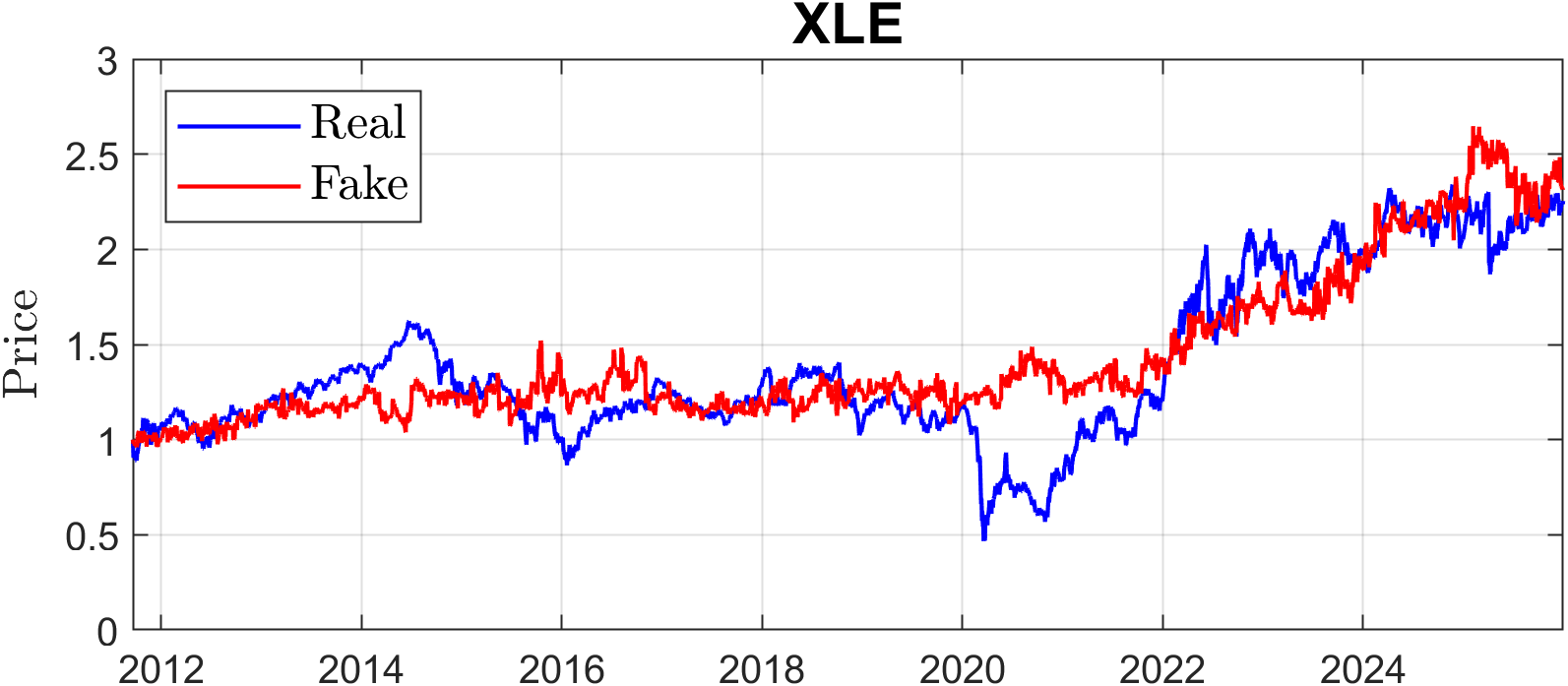}
\caption{XLE}
\end{subfigure}
\caption{Empirical (blue) and simulated (red) daily price series for the nine assets}
\label{fig:real_fake}
\end{figure}

To compare the empirical daily price series $P_t^{\rm real}$ with the Quant GAN-simulated price series $P_t^{\rm fake}$, we select, among the 10,000 generated paths, the one that minimizes the Euclidean distance to the empirical series:
\begin{equation*}
\arg\min_{i} \lVert P_t^{\rm real} - P_{t}^{\textrm{fake}, i} \rVert = \arg\min_{i}\sqrt{\sum_{j=t_0}^{t_T}(P_j^{\rm real} - P_{j}^{\textrm{fake}, i})^2},
\end{equation*}
where $P_{t}^{\textrm{fake}, i}$ denotes the $i$-th simulated path, and $t_0$ and $t_T$ correspond to the start and end dates defined in Section~\ref{sec:data_preprocessing}, namely September 20, 2011, and December 30, 2025, respectively. The optimal index $i$ naturally varies across the nine assets, as the path closest to the empirical series differs for each asset. We use this selected path as a representative diagnostic path for visual comparison, return-distribution comparison, and memory estimation. This choice should be understood as a path-level assessment of the best-aligned generated trajectory, while the broader benchmark analysis below evaluates Quant GAN performance against alternative generators at the return level. Figure~\ref{fig:real_fake} compares the empirical (real) and simulated (fake) price series for the nine assets.

For the equity market, the simulated price series generally reproduces the long-term upward tendency observed in the empirical data, indicating that Quant GANs learn part of the broad growth pattern of major stock indices. Nevertheless, the generated paths smooth several abrupt rises and sharp declines observed in real markets. For example, in the S\&P 500, the pronounced downturns around 2019--2020 and June 2024, as well as the rapid increases during 2020--2022 and again near mid-2024, are largely dampened in the simulated series. A similar pattern is observed for the DAX, where the sudden drops in early 2020 and the first half of 2022 are not fully reflected. In the case of the Nikkei 225, the discrepancy is more visible after 2020, when the simulated path exhibits a steeper and more persistent upward movement than the empirical series.

In the commodity market, a comparable pattern emerges. The simulated paths generally follow broad directional trends, but abrupt spikes and collapses are attenuated. For Wheat (WEAT), the simulated series aligns with the long-term evolution of the real series but does not reproduce the sharp price surge at the beginning of 2022. For Corn (CORN), the empirical data show a more pronounced decline starting around 2018, whereas the simulated series displays a milder downward movement; the rapid increase between roughly 2020 and early 2022 is also insufficiently represented. Soybeans (SOYB) present a clearer pattern characterized by an initial decline followed by a sustained rise. Although the simulated series reflects this general shape, the magnitude of these movements is dampened, and the renewed downward tendency beginning in early 2024 is largely absent.

In the energy market, the model's performance is more heterogeneous. The most visible discrepancy is USO, where the empirical series exhibits an overall downward trend, while the corresponding simulated series follows an upward trajectory and therefore misses the long-term direction. This contrast suggests that the intrinsic characteristics and noise structure of the training data can materially influence the model's learning outcomes. For UNG and XLE, the behavior is closer to the patterns observed in equities and commodities: the simulated series captures the broad trend but does not reproduce short-lived regime shifts or extreme movements at specific time intervals.

Overall, these observations indicate that Quant GANs can learn smooth long-horizon dynamics and general directional behavior, but they tend to underrepresent abrupt shocks and sharp turning points. This smoothing effect is visible across all three asset classes and becomes important when the generated data are used for risk-sensitive applications.

\subsubsection{Return Distribution Comparison}
\begin{figure}[htbp]
\centering
\begin{subfigure}[b]{0.28\textwidth}
\includegraphics[width=\textwidth]{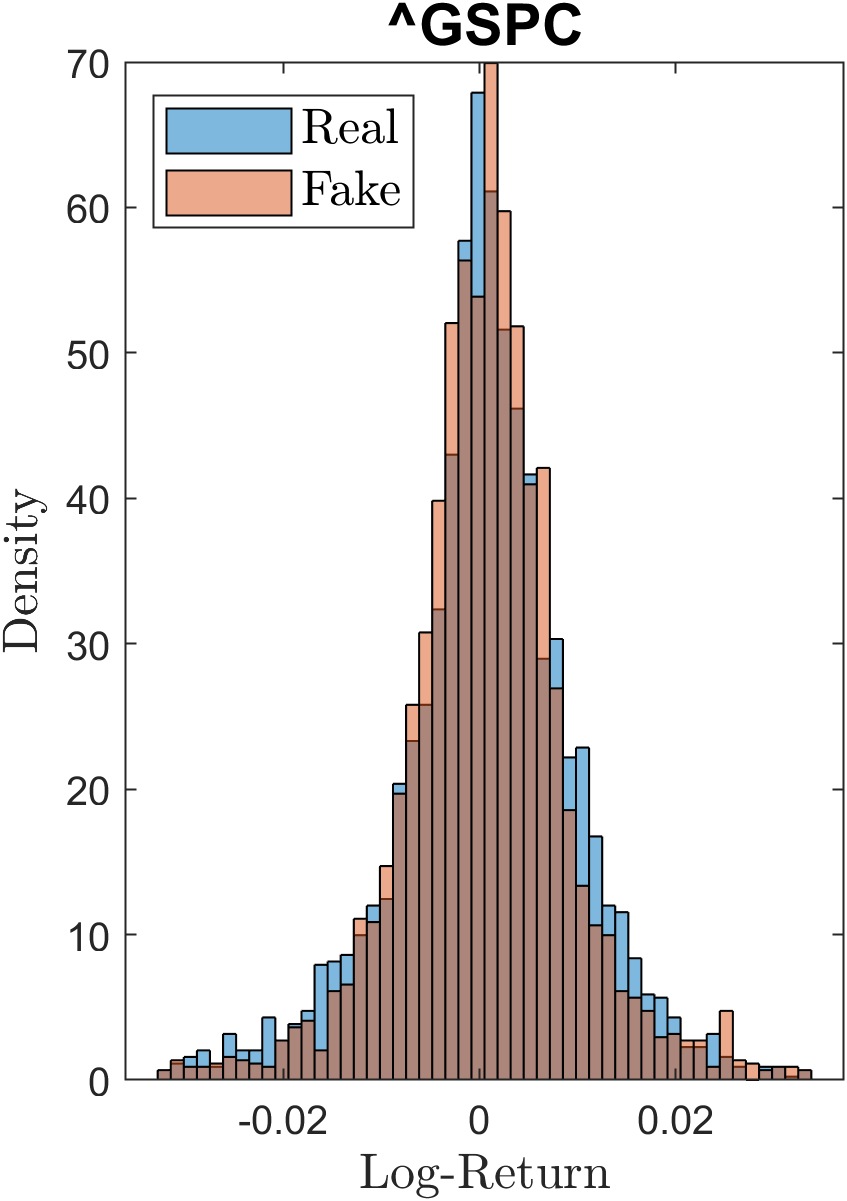}
\caption{\textasciicircum GSPC}
\end{subfigure}
\hfill
\begin{subfigure}[b]{0.28\textwidth}
\includegraphics[width=\textwidth]{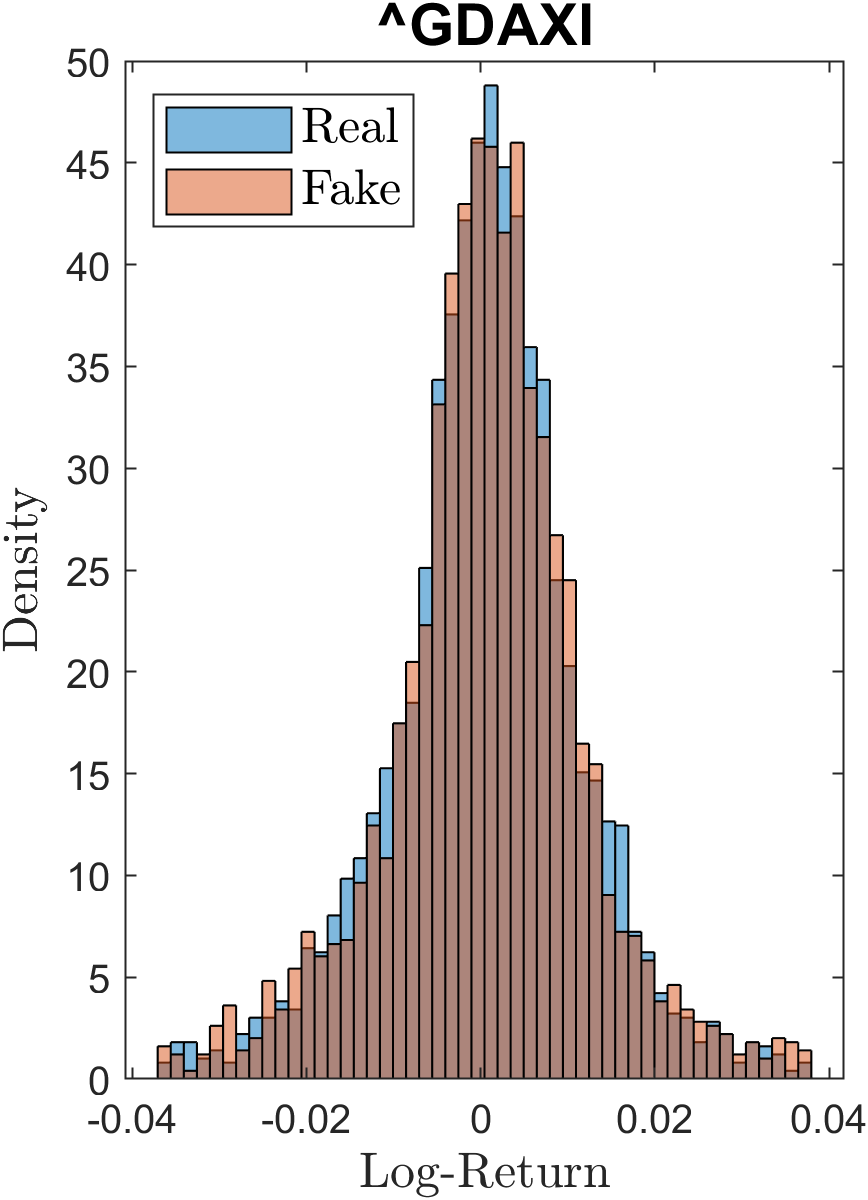}
\caption{\textasciicircum GDAXI}
\end{subfigure}
\hfill
\begin{subfigure}[b]{0.28\textwidth}
\includegraphics[width=\textwidth]{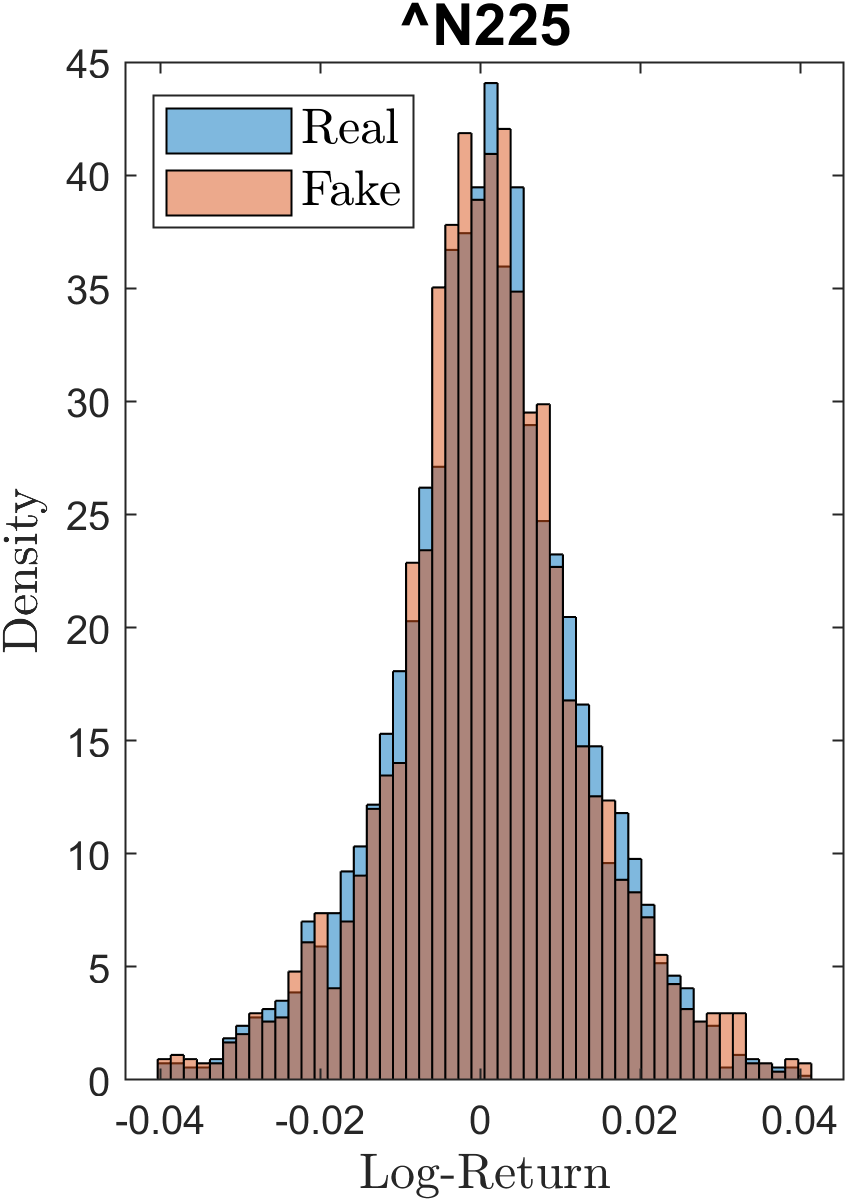}
\caption{\textasciicircum N225}
\end{subfigure}
\hfill
\begin{subfigure}[b]{0.28\textwidth}
\includegraphics[width=\textwidth]{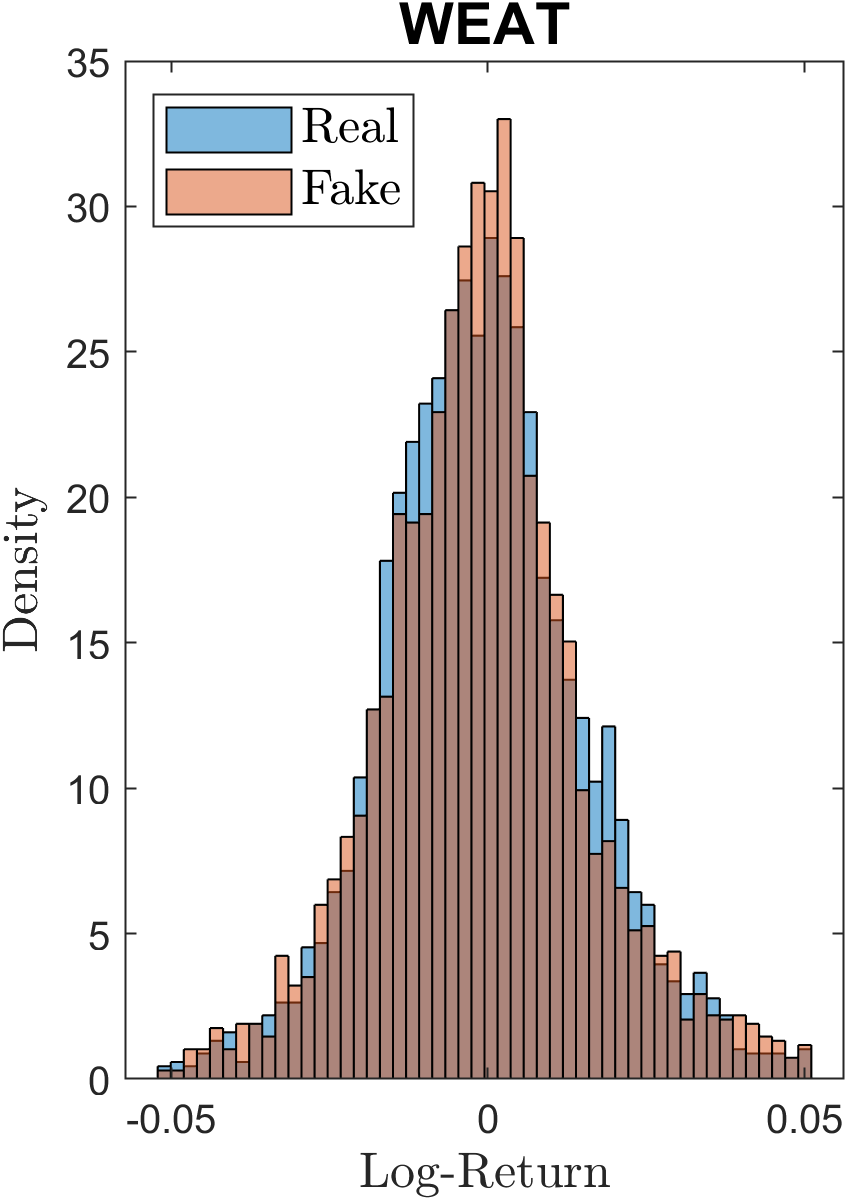}
\caption{WEAT}
\end{subfigure}
\hfill
\begin{subfigure}[b]{0.28\textwidth}
\includegraphics[width=\textwidth]{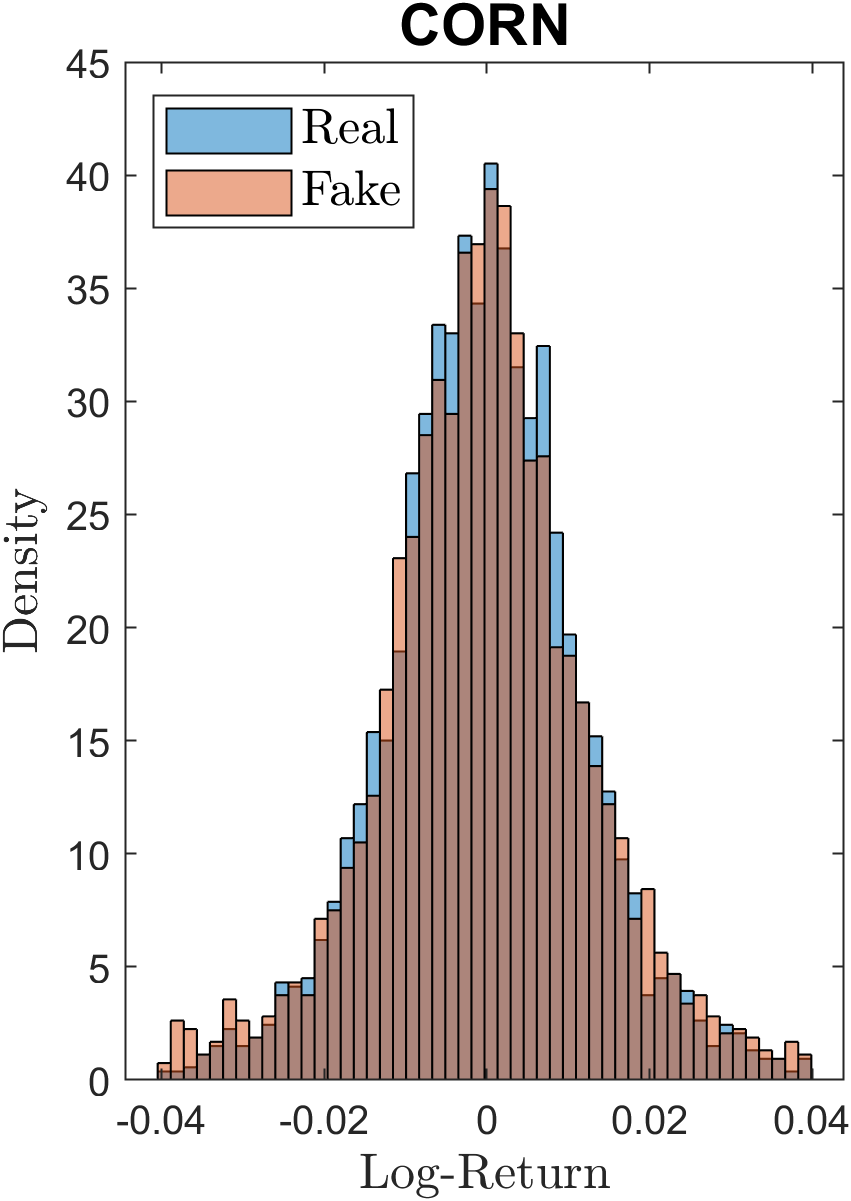}
\caption{CORN}
\end{subfigure}
\hfill
\begin{subfigure}[b]{0.28\textwidth}
\includegraphics[width=\textwidth]{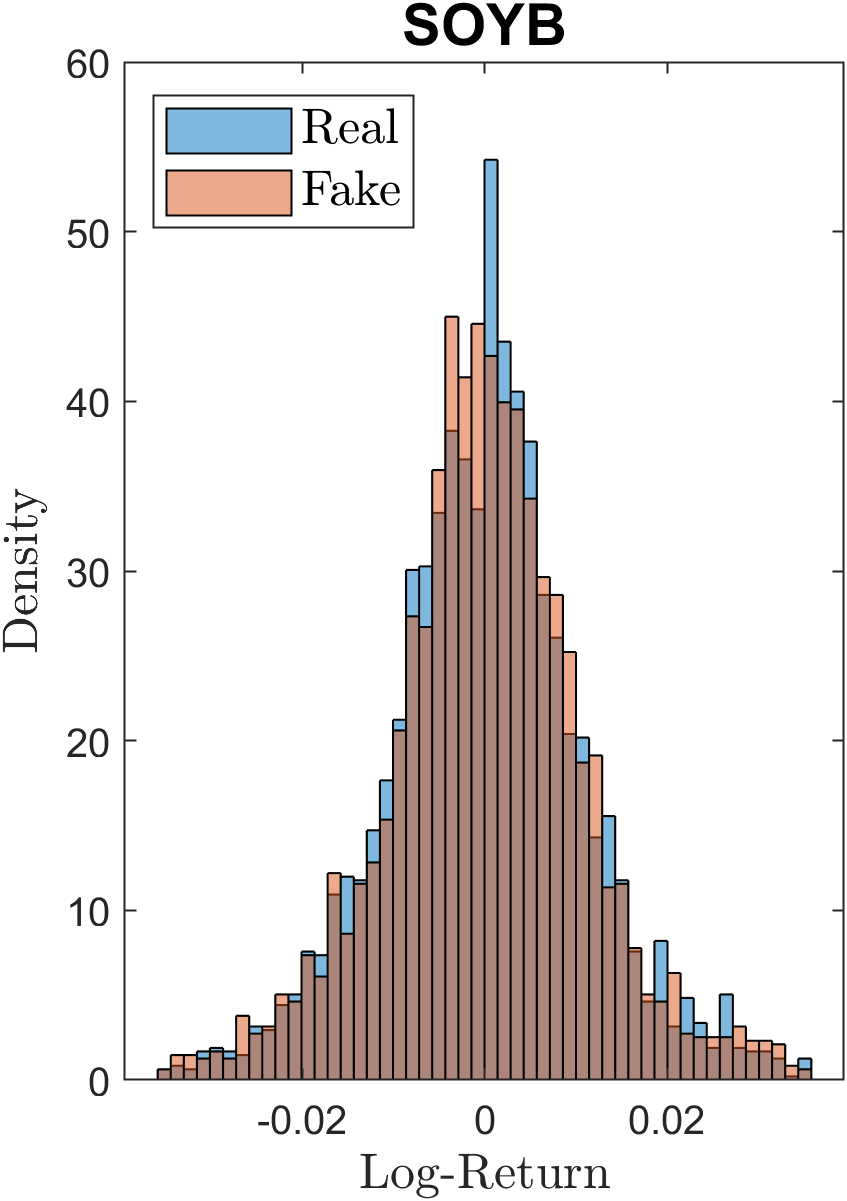}
\caption{SOYB}
\end{subfigure}
\hfill
\begin{subfigure}[b]{0.28\textwidth}
\includegraphics[width=\textwidth]{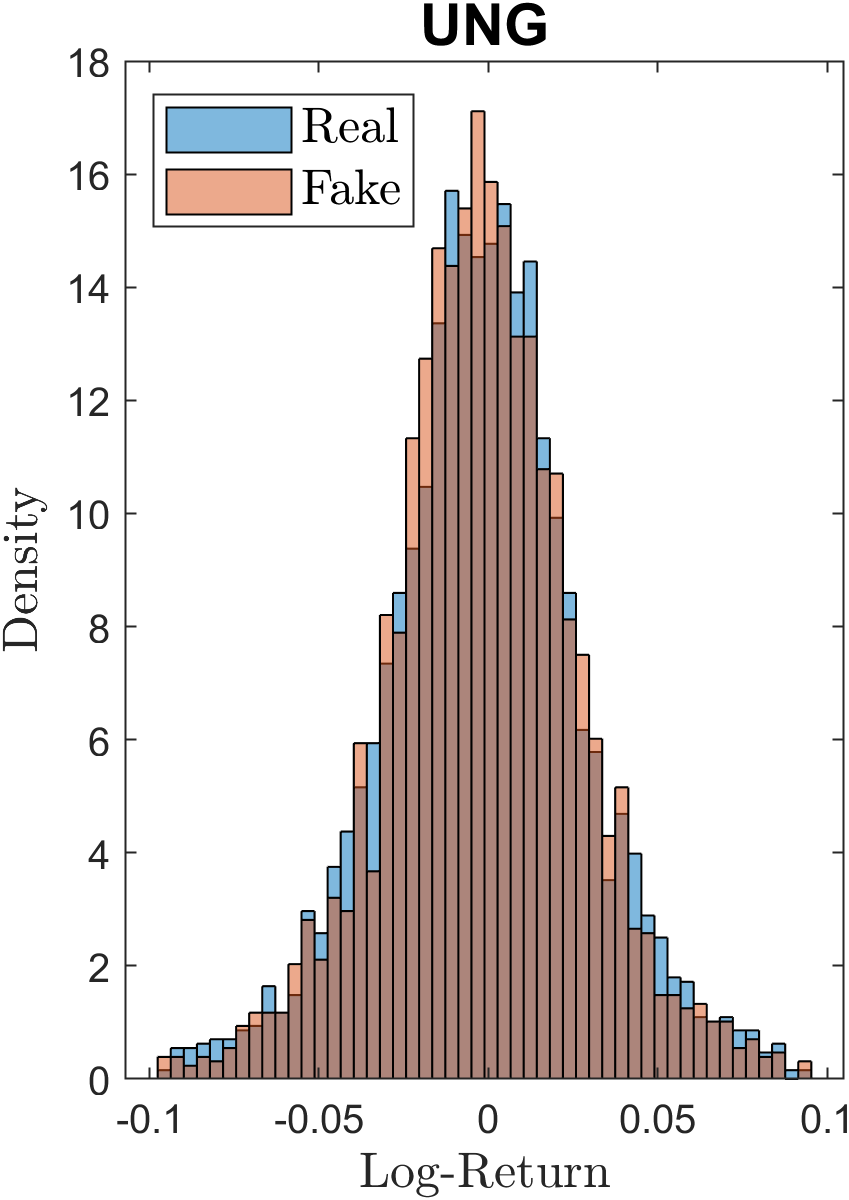}
\caption{UNG}
\end{subfigure}
\hfill
\begin{subfigure}[b]{0.28\textwidth}
\includegraphics[width=\textwidth]{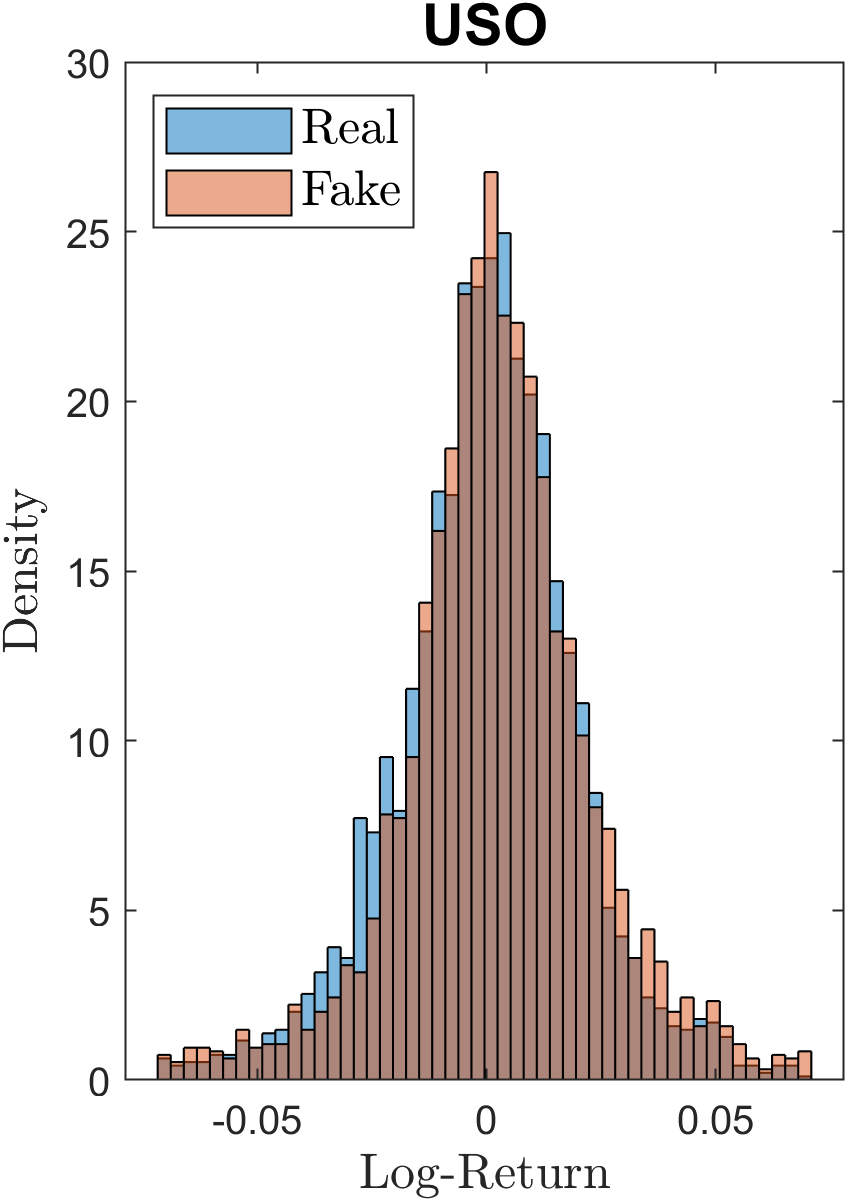}
\caption{USO}
\end{subfigure}
\hfill
\begin{subfigure}[b]{0.28\textwidth}
\includegraphics[width=\textwidth]{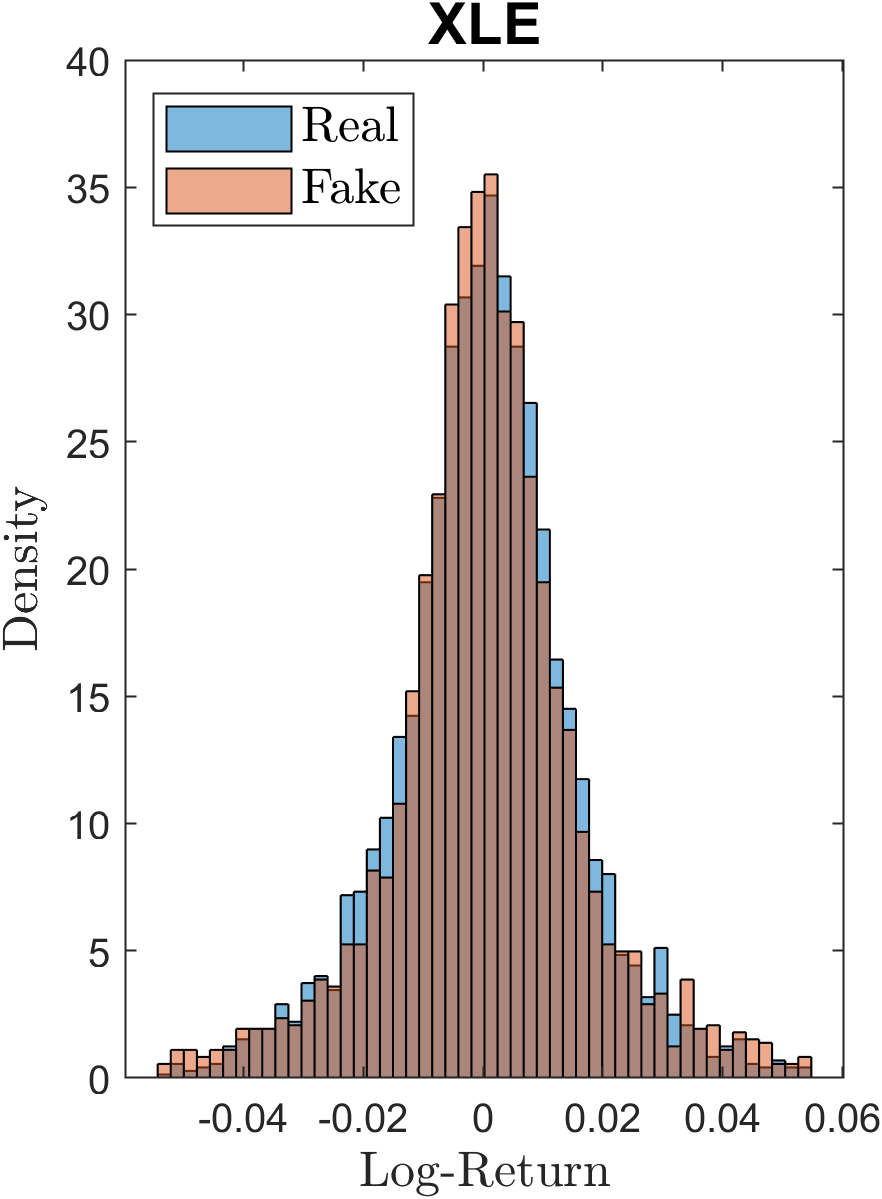}
\caption{XLE}
\end{subfigure}
\caption{Comparison of empirical (blue) and simulated (orange) daily return distributions for the nine assets}
\label{fig:hist_real_fake}
\end{figure}

To compare the real and simulated series at the return level, Figure~\ref{fig:hist_real_fake} displays the daily return distributions of the empirical data and the representative Quant GAN paths. The simulated returns do not perfectly reproduce the concentration of real returns around zero, but they capture parts of the left and right tail behavior. This is important for risk management, where extreme gains and losses play a central role and accurate tail modeling is essential~\citep{danielsson2000value, cont2001empirical, embrechts2013modelling}. These distributional comparisons suggest that the selected Quant GAN paths reproduce several important features of the real returns, but they do not by themselves establish whether long-memory structures are preserved. We therefore apply the R/S, DFA, and ARFIMA--FIGARCH diagnostics introduced in Section~\ref{sec:measure_LRD} to further investigate the LRD of the simulated data.

\subsubsection{Benchmark Accuracy Assessment}
Before reporting the benchmark results, it is useful to clarify the evaluation objective. Since the goal of this study is synthetic financial time-series generation, with emphasis on distributional and temporal-dependence fidelity rather than conventional point-forecasting performance, we assess comparative performance in terms of statistical fidelity, tail-risk preservation, volatility-dependence accuracy, and long-memory preservation. These criteria are directly relevant for stress testing, scenario generation, and long-horizon risk analysis. 

To provide a more rigorous evaluation against established benchmarks, we compare the representative Quant GAN path with four benchmark generators: an ARMA(1,1)-GARCH(1,1) model estimated by Gaussian quasi-maximum likelihood~\citep{bollerslev1987conditionally}, a Gaussian iid model matched to the empirical mean and variance, an iid historical bootstrap, and a moving-block bootstrap with block length 20~\citep{kunsch1989jackknife}. The comparison is conducted at the return level and averaged across the nine assets.

Let $r_{j,t}$ denote the empirical return of asset $j$ and let $\widetilde r_{j,t}^{(m)}$ denote the return generated by benchmark model $m$, where $j=1,\ldots,9$ and $t=1,\ldots,T$. For each asset and benchmark, we compute five diagnostics. First, distributional accuracy is measured by the first-order Wasserstein distance~\citep{villani2009optimal},
\begin{equation*}
W_{1,j}^{(m)}=\int_{0}^{1}\left|\widehat{Q}_{j}^{\rm real}(u)-\widehat{Q}_{j,m}^{\rm gen}(u)\right|\,du,
\end{equation*}
where $\widehat{Q}_{j}^{\rm real}$ and $\widehat{Q}_{j,m}^{\rm gen}$ are the empirical quantile functions of the real and generated returns. Second, tail accuracy is measured as
\begin{equation*}
T_{j}^{(m)}=\frac{1}{|\mathcal{P}|}\sum_{p\in\mathcal{P}}\left|\widehat{Q}_{j}^{\rm real}(p)-\widehat{Q}_{j,m}^{\rm gen}(p)\right|,\qquad
\mathcal{P}=\{0.01,0.05,0.95,0.99\}.
\end{equation*}
Third, volatility-dependence accuracy is evaluated through the absolute-return autocorrelation error,
\begin{equation*}
A_{j}^{(m)}=\frac{1}{20}\sum_{\ell=1}^{20}\left|\rho_{j}^{|r|}(\ell)-\widetilde{\rho}_{j,m}^{|r|}(\ell)\right|,
\end{equation*}
where $\rho_{j}^{|r|}(\ell)$ and $\widetilde{\rho}_{j,m}^{|r|}(\ell)$ denote the lag-$\ell$ autocorrelations of $|r_{j,t}|$ and $|\widetilde r_{j,t}^{(m)}|$, respectively. Finally, long-memory accuracy is measured by the absolute errors in the R/S and DFA Hurst estimates,
\begin{equation*}
E_{RS,j}^{(m)}=\left|\widehat{H}_{RS,j}-\widetilde{H}_{RS,j}^{(m)}\right|,\qquad
E_{DFA,j}^{(m)}=\left|\widehat{H}_{DFA,j}-\widetilde{H}_{DFA,j}^{(m)}\right|.
\end{equation*}
Table~\ref{tab:benchmark_accuracy} reports the cross-asset averages of these diagnostics, e.g., $\overline{W}_1^{(m)}=9^{-1}\sum_{j=1}^{9}W_{1,j}^{(m)}$, with analogous averages for the remaining metrics.

\begin{table}[htbp]
\centering
\begin{tabular}{l c c c c c}
\hline
Model & Wasserstein & Tail quantile & Vol. ACF & R/S $H$ & DFA $H$\\
 & distance & error & error & error & error\\
\hline
Quant GAN & $1.09\cdot 10^{-3}$ & $4.74\cdot 10^{-3}$ & 0.152 & 0.102 & 0.118\\
ARMA(1,1)-GARCH(1,1) & $1.84\cdot 10^{-3}$ & $3.65\cdot 10^{-3}$ & 0.0596 & 0.0302 & 0.0558\\
Gaussian iid & $2.51\cdot 10^{-3}$ & $3.59\cdot 10^{-3}$ & 0.163 & 0.0301 & 0.0347\\
Historical bootstrap & $4.20\cdot 10^{-4}$ & $9.60\cdot 10^{-4}$ & 0.162 & 0.0395 & 0.0582\\
Moving-block bootstrap & $6.10\cdot 10^{-4}$ & $1.62\cdot 10^{-3}$ & 0.0965 & 0.0321 & 0.0546\\
\hline
\end{tabular}
\caption{Average distributional, tail-risk, dependence, and long-memory accuracy relative to empirical returns}
\label{tab:benchmark_accuracy}
\end{table}

Table~\ref{tab:benchmark_accuracy} shows that Quant GAN improves over the Gaussian iid benchmark in distributional distance, but it does not dominate classical econometric or resampling benchmarks. The ARMA(1,1)-GARCH(1,1) benchmark yields the smallest volatility-autocorrelation error, reflecting the value of explicitly modeling conditional heteroskedasticity. The Gaussian iid and moving-block bootstrap benchmarks also perform competitively in the Hurst-error metrics, whereas Quant GAN exhibits larger R/S and DFA Hurst errors. This comparison reinforces the central conclusion of the paper: Quant GANs can generate realistic return distributions, but the representative paths examined here do not reproduce the long-memory and volatility-dependence structure as reliably as benchmark procedures that explicitly incorporate or inherit dependence from the observed data.

\subsubsection{Full-Ensemble Hurst Robustness Check}
\begin{figure}[htbp]
\centering
\includegraphics[width=\textwidth]{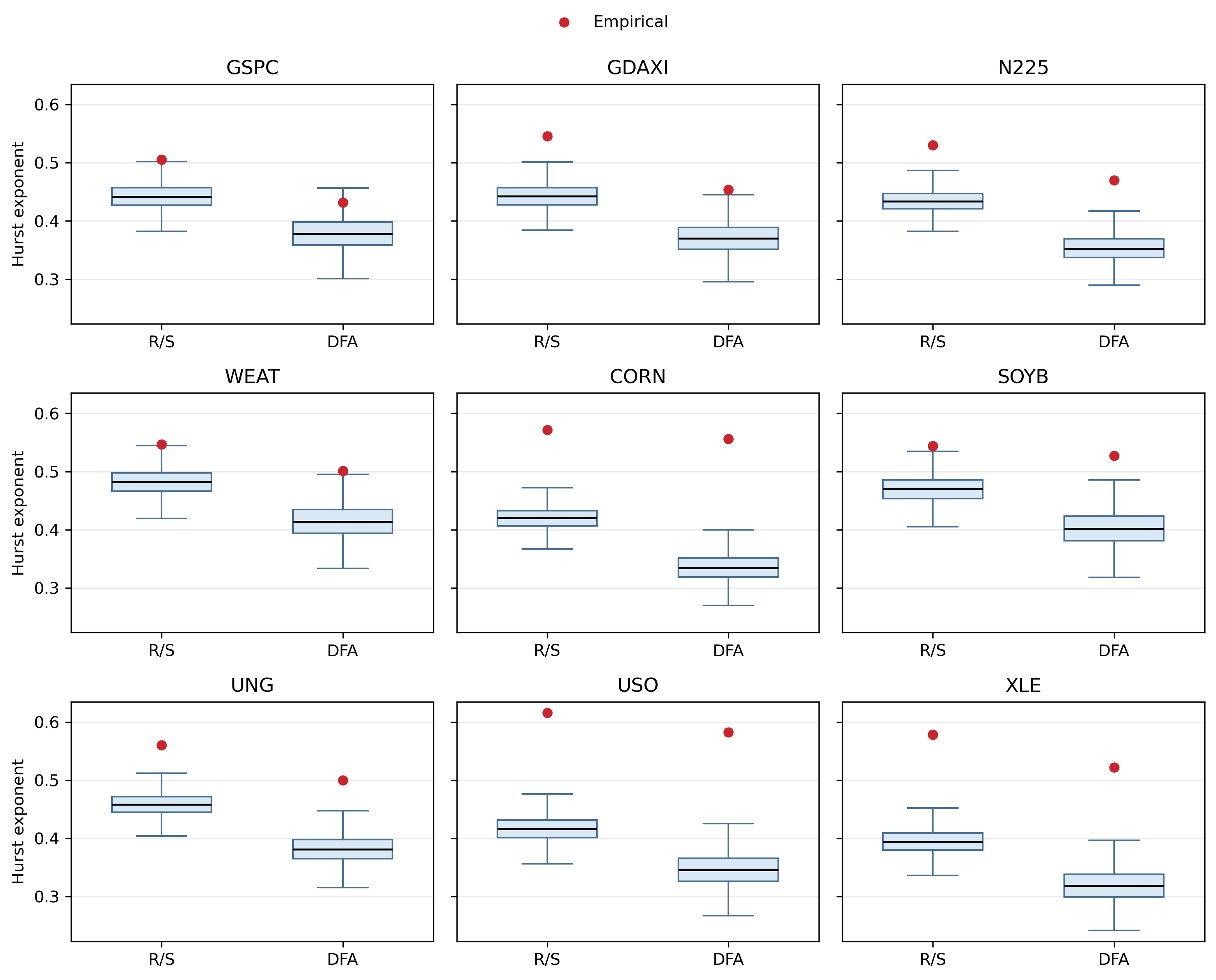}
\caption{Full-ensemble Hurst distributions for 10,000 Quant GAN-generated paths per asset. Red markers indicate the empirical Hurst estimates}
\label{fig:ensemble_hurst_boxplots}
\end{figure}

The preceding diagnostics use one representative Quant GAN path for each asset. To ensure that the long-memory conclusion is not driven by this path-selection rule, we further examine the full generated ensemble. For each asset, the 10,000 generated log-price paths are first converted into log-return paths by first differencing. We then estimate R/S and DFA Hurst exponents for every generated path and compare the resulting cross-path distribution with the corresponding empirical estimate. Table~\ref{tab:ensemble_hurst} reports the median generated Hurst exponent together with the 5\%--95\% interval across the 10,000 paths, while Figure~\ref{fig:ensemble_hurst_boxplots} visualizes the full distribution through box-whisker plots.

\begin{table}[htbp]
\centering
\begin{tabular}{c | l c c c c}
\hline
Category & Ticker & Real R/S $H$ & Generated R/S $H$ & Real DFA $H$ & Generated DFA $H$\\
& & & Median [5\%, 95\%] & & Median [5\%, 95\%]\\
\hline
& \textasciicircum GSPC & 0.506 & 0.442 [0.407, 0.479] & 0.432 & 0.378 [0.333, 0.426]\\
Stock & \textasciicircum GDAXI & 0.546 & 0.443 [0.408, 0.478] & 0.454 & 0.370 [0.327, 0.418]\\
& \textasciicircum N225 & 0.530 & 0.434 [0.402, 0.465] & 0.470 & 0.353 [0.317, 0.394]\\ \hline
& WEAT & 0.546 & 0.482 [0.443, 0.520] & 0.501 & 0.414 [0.367, 0.464]\\
Commodity & CORN & 0.572 & 0.420 [0.388, 0.452] & 0.556 & 0.334 [0.299, 0.376]\\
& SOYB & 0.544 & 0.470 [0.431, 0.509] & 0.527 & 0.402 [0.352, 0.453]\\ \hline
& UNG & 0.561 & 0.458 [0.426, 0.491] & 0.500 & 0.381 [0.344, 0.422]\\
Energy & USO & 0.616 & 0.417 [0.381, 0.453] & 0.582 & 0.346 [0.300, 0.395]\\
& XLE & 0.578 & 0.394 [0.360, 0.429] & 0.522 & 0.319 [0.274, 0.366]\\
\hline
\end{tabular}
\caption{Full-ensemble Hurst robustness check for 10,000 Quant GAN-generated paths per asset}
\label{tab:ensemble_hurst}
\end{table}

The ensemble-level results support the representative-path findings. For every asset, the median generated R/S and DFA Hurst exponents lie below the corresponding empirical estimates. This pattern is also visually apparent in Figure~\ref{fig:ensemble_hurst_boxplots}, where the empirical red markers are generally above the boxes of the generated-path distributions. Moreover, for most assets, even the upper 95\% quantile of the generated Hurst distribution remains below the empirical value. This confirms that the attenuation of long-memory behavior is a systematic feature of the generated ensemble rather than an artifact of the selected representative path.

\subsubsection{Long-Range Dependence and Memory Structure Analysis}
Table~\ref{tab:rs_quantgans} reports the R/S analysis applied to the Quant GAN-generated series. The estimated Hurst exponents $\widehat{H}$ are predominantly close to or below 0.5, indicating weak persistence or slight anti-persistence in the simulated data. This contrasts with the empirical series in Table~\ref{tab:rs}, where most assets display $\widehat{H} > 0.5$ and statistically significant evidence of LRD. The $R^2$ values for the simulated data remain high, generally above 0.95, so the difference is not driven by a poor scaling fit; rather, the inferred memory structure itself differs from that of the real markets.
\begin{table}[htbp]
\centering
\begin{tabular}{c | l r c c}
\hline
Category & Ticker & $\widehat{H}$ ($p$-value) & 95\% CI & $R^2$\\
\hline
 & \textasciicircum GSPC & 0.436 (0.991) & $[0.385, 0.486]$ & 0.980\\
Stock & \textasciicircum GDAXI & 0.470 (0.914) & $[0.424, 0.516]$ & 0.986\\
 & \textasciicircum N225 & 0.428 (0.971) & $[0.353, 0.503]$ & 0.956 \\ 
 \hline
 & WEAT & 0.537 (*)$^{a}$ & $[0.500, 0.574]$ & 0.993\\
Commodity & CORN & 0.482 (0.774) & $[0.430, 0.534]$ & 0.983\\
 & SOYB & 0.463 (0.935) & $[0.413, 0.513]$ & 0.983\\
\hline
 & UNG & 0.458 (0.894) & $[0.386, 0.523]$ & 0.964 \\
Energy & USO & 0.410 (0.999) & $[0.365, 0.455]$ & 0.982\\
 & XLE & 0.388 (0.999) & $[0.348, 0.429]$ & 0.984\\ 
\hline
\multicolumn{5}{l}{\footnotesize $^{a}$ Indicates a $p$-value less than 0.05.}\\
\hline
\end{tabular}
\caption{The results of rescaled range analysis on Quant GAN-generated data}
\label{tab:rs_quantgans}
\end{table}

For the equity market, all three indices generated by Quant GANs yield $\widehat{H}$ values below 0.5, namely 0.436 for the S\&P 500, 0.470 for the DAX, and 0.428 for the Nikkei 225, with large $p$-values and CIs that all include 0.5. This implies that the simulated equity series behave approximately like short-memory or near-random processes. In comparison, the empirical data show moderately persistent behavior, especially for the DAX ($\widehat{H}=0.546$, statistically significant) and the Nikkei 225 ($\widehat{H}=0.530$), while the S\&P 500 is close to the boundary at 0.5. Hence, Quant GANs tend to underestimate persistence in stock indices and smooth out the degree of LRD present in the real series. 

In the commodity market, the discrepancy is mixed but still evident. Among the three assets, only Wheat (WEAT) in the simulated data exhibits a statistically significant $\widehat{H}$ above 0.5 (0.537, $p<0.05$), which is broadly consistent with the empirical estimate (0.546). However, for Corn (CORN) and Soybeans (SOYB), the simulated Hurst exponents fall below 0.5 (0.482 and 0.463, respectively), whereas the real data show clear and highly significant persistence, particularly for CORN ($\widehat{H}=0.572$) and SOYB ($\widehat{H}=0.544$). This indicates that while Quant GANs can partially reproduce long-memory characteristics for certain commodities, they generally weaken the strength of persistence relative to the observed data. 

For the energy market, the divergence is most pronounced. The simulated series for UNG, USO, and XLE all yield $\widehat{H}$ values well below 0.5 (0.458, 0.410, and 0.388, respectively), accompanied by very high $p$-values, suggesting no evidence of LRD and even hints of anti-persistence. By contrast, the empirical energy assets exhibit some of the strongest persistence in the dataset, with all three real series showing $\widehat{H}$ substantially above 0.55 and highly significant $p$-values. This gap indicates that Quant GANs do not preserve the strong long-memory structure observed in energy prices and instead generate paths closer to short-memory or mean-reverting processes. 

Taken together, these results suggest that although Quant GAN-generated series exhibit good regression fit quality in the R/S framework, their LRD properties are attenuated relative to the empirical data. The model tends to push the Hurst exponent toward or below the random-walk benchmark of 0.5, especially in markets where the empirical data display strong persistence, highlighting a limitation in capturing deep temporal dependence structures.

Table~\ref{tab:dfa_quantgans} reports the Hurst exponents estimated by DFA for the Quant GAN-generated series. Similar to the R/S results, the $R^2$ values remain consistently high across assets, generally above 0.92, indicating satisfactory log--log fits. However, the estimated Hurst exponents for the synthetic data are systematically lower than those obtained from the empirical series in Table~\ref{tab:dfa}, suggesting that Quant GANs tend to underestimate persistence and LRD in the real markets. 
\begin{table}[htbp]
\centering
\begin{tabular}{c | l r c c}
\hline
Category & Ticker & $\widehat{H}$ ($p$-value) & 95\% CI & $R^2$\\
\hline
 & \textasciicircum GSPC & 0.368 (0.999) & $[0.308, 0.428]$ & 0.962\\
Stock & \textasciicircum GDAXI & 0.394 (0.999) & $[0.343, 0.446]$ & 0.975\\
 & \textasciicircum N225 & 0.342 (0.999) & $[0.262, 0.422]$ & 0.924 \\ 
 \hline
 & WEAT & 0.480 (0.743) & $[0.413, 0.547]$ & 0.972\\
Commodity & CORN & 0.433 (0.965) & $[0.360, 0.507]$ & 0.959\\
 & SOYB & 0.437 (0.973) & $[0.372, 0.501]$ & 0.968\\
\hline
 & UNG & 0.358 (0.998) & $[0.275, 0.440]$ & 0.926 \\
Energy & USO & 0.331 (0.999) & $[0.288, 0.375]$ & 0.975\\
 & XLE & 0.329 (0.999) & $[0.281, 0.378]$ & 0.969\\ 
\hline
\end{tabular}
\caption{The results of detrended fluctuation analysis on Quant GAN-generated data}
\label{tab:dfa_quantgans}
\end{table}

For the equity market, all three simulated indices exhibit Hurst exponents well below 0.5 with values of 0.368 for the S\&P 500, 0.394 for the DAX, and 0.342 for the Nikkei 225. The corresponding $p$-values are all close to 1, so the null hypothesis $H=0.5$ cannot be rejected, and the 95\% CIs are entirely or largely below 0.5, indicating an absence of statistically significant persistence. In contrast, the empirical DFA estimates are noticeably higher (ranging roughly from 0.43 to 0.47), with the Nikkei 225 even approaching the persistence threshold. This comparison shows that the synthetic equity series display a stronger tendency toward anti-persistence or near-random behavior than the real data. 

In the commodity market, the simulated Hurst exponents are closer to 0.5 but still generally lower than their empirical counterparts. WEAT yields $\widehat{H}=0.480$ with a non-significant $p$-value, and its CI straddles 0.5, implying behavior close to a random walk. CORN and SOYB produce values of 0.433 and 0.437, respectively, again with large $p$-values and intervals that include or lie slightly below 0.5. By comparison, the real data show stronger persistence, especially for CORN and SOYB, whose empirical Hurst exponents exceed 0.52 and are statistically significant. Hence, although Quant GANs roughly capture the neutral memory structure of WEAT, they substantially weaken the persistence observed in the other two commodities. 

For the energy market, the discrepancy is even more pronounced. The simulated series produce clearly sub-0.5 Hurst exponents--0.358 for UNG, 0.331 for USO, and 0.329 for XLE--with extremely large $p$-values and CIs predominantly below 0.5. These results suggest anti-persistent or near-white-noise dynamics in the generated data. By contrast, the empirical DFA estimates indicate moderate to strong persistence, particularly for USO and XLE, whose real Hurst exponents exceed 0.52 and are statistically significant. This divergence shows that, for energy assets, Quant GANs miss not only the magnitude of LRD but also its qualitative direction. 

Overall, the DFA analysis reinforces the conclusion drawn from the R/S results: while the regression fits remain statistically adequate, Quant GAN-generated series exhibit weaker memory and reduced persistence relative to the empirical financial data, with the largest underestimation occurring in the energy sector.

Table~\ref{tab:arfima_figarch_quantgans} summarizes the ARFIMA--FIGARCH estimation results for the Quant GAN-generated return series. In contrast to the empirical results in Table~\ref{tab:arfima_figarch}, the fractional differencing parameter in the mean equation, $\widehat{d}_m$, is almost uniformly indistinguishable from zero for the synthetic data, whereas the volatility fractional parameter, $\widehat{d}_v$, remains positive and statistically significant across all assets. This indicates that Quant GANs largely do not reproduce LRD in the conditional mean, but still preserve a degree of long memory in volatility dynamics.
\begin{table}[htbp]
\centering
\begin{tabular}{c | l r l r c}
\hline
Category & Ticker & $\widehat{d}_m$ ($p$-value) & 95\% CI for $\widehat{d}_m$ & $\widehat{d}_v$ $^{a}$ & 95\% CI for $\widehat{d}_v$ \\ \hline
 & \textasciicircum GSPC & $1.00\cdot 10^{-8}$ (0.500) & $[-2.77\cdot 10^{-3}, 2.77\cdot 10^{-3}]$ & 0.398 & $[0.396,0.399]$\\
Stock & \textasciicircum GDAXI & $1.00\cdot 10^{-8}$ (0.500) & $[-2.28\cdot 10^{-3}, 2.28\cdot 10^{-3}]$ & 0.390 & $[0.385, 0.396]$ \\ 
 & \textasciicircum N225 & $1.00\cdot 10^{-8}$ (0.500) & $[-7.04\cdot 10^{-4}, 7.04\cdot 10^{-4}]$ & 0.402 & $[0.401, 0.402]$ \\  \hline
 & WEAT & $1.00\cdot 10^{-8}$ (0.500) & $[-1.18\cdot 10^{-3}, 1.18\cdot 10^{-3}]$ & 0.401 & $[0.397, 0.405]$ \\ 
Commodity & CORN & $1.00\cdot 10^{-8}$ (0.500) & $[-2.61\cdot 10^{-4}, 2.61\cdot 10^{-4}]$ & 0.352 & $[0.351, 0.354]$ \\
 & SOYB & $1.00\cdot 10^{-8}$ (0.500) & $[-2.03\cdot 10^{-4}, 2.03\cdot 10^{-4}]$ & 0.399 & $[0.398, 0.400]$ \\ \hline
 & UNG & $5.56\cdot 10^{-4}$ (0.496) & $[-0.120, 0.120]$ & 0.901 & $[0.871, 0.930]$ \\
Energy & USO & 0.0111 (***) $^{b}$ & $[8.48\cdot 10^{-3}, 0.0137]$ & 0.390 & $[0.389, 0.391]$ \\ 
 & XLE & 0.0340 (0.0856) & $[-0.0147, 0.0828]$ & 0.325 & $[0.150, 0.501]$ \\ \hline
\multicolumn{5}{l}{\footnotesize $^{a}$ All $p$-values for hypothesis test $\textnormal{H}_0: d_v = 0$ vs. $\textnormal{H}_a: d_v > 0$ are less than 0.001.}\\
\multicolumn{5}{l}{\footnotesize $^{b}$ Indicates a $p$-value less than 0.001.}\\
\hline
\end{tabular}
\caption{The results of ARFIMA--FIGARCH fit on Quant GAN-generated data}
\label{tab:arfima_figarch_quantgans}
\end{table}

For the equity market, all three indices (S\&P 500, DAX, and Nikkei 225) yield $\widehat{d}_m$ values essentially equal to zero, with $p$-values of 0.500 and CIs tightly centered around zero. This implies that the null hypothesis $\textnormal{H}_0: d_m = 0$ cannot be rejected, and thus no evidence of long memory in returns is detected in the generated data. In comparison, the empirical results show a small but statistically significant $d_m$ for the S\&P 500, indicating mild persistence in the real series that disappears entirely in the synthetic counterpart. Regarding the volatility equation, however, $\widehat{d}_v$ remains around 0.39--0.40 for all three indices, closely matching the magnitude observed in the empirical data, especially for the S\&P 500 and Nikkei 225. This suggests that Quant GANs are more successful in capturing volatility clustering than mean persistence in equity returns. 

In the commodity market, the contrast with the real data becomes more pronounced. For WEAT and CORN, the empirical ARFIMA estimates indicate moderate positive $d_m$--particularly strong for WEAT--while the Quant GAN-generated series again produce $\widehat{d}_m$ values virtually equal to zero with narrow CIs. SOYB exhibits negligible $d_m$ in both real and synthetic data, making it the only commodity where the model aligns well with empirical behavior in the mean equation. For the variance dynamics, $\widehat{d}_v$ remains significantly positive in all three cases, but the magnitudes are generally slightly lower and more concentrated in the synthetic data. This reflects that the GAN tends to compress the dispersion of volatility memory across commodities, reducing the heterogeneity seen in the real market. 

The energy market displays the most heterogeneous behavior. For UNG, the synthetic $\widehat{d}_m$ remains statistically insignificant and centered near zero, which is broadly consistent with the empirical finding of weak or absent mean long memory. USO, however, represents an exception: the generated data produce a small but statistically significant $\widehat{d}_m = 0.0111$, whereas the empirical estimate is effectively zero. This indicates a slight overestimation of mean persistence by the GAN for this asset. XLE shows an intermediate case, with a positive but statistically insignificant $d_m$ and a relatively wide CI. In the volatility equation, $\widehat{d}_v$ remains strongly significant for all three energy assets, with UNG exhibiting particularly high persistence (around 0.90) in both real and synthetic series. Nevertheless, for USO and XLE the synthetic volatility memory is noticeably weaker and less dispersed than in the empirical data. 

Overall, the ARFIMA--FIGARCH comparison reveals a consistent structural tendency: Quant GANs suppress LRD in the mean process while retaining, to a considerable extent, long memory in volatility. The relationship $H = d + 0.5$ further clarifies this outcome--near-zero $\widehat{d}_m$ in the synthetic data corresponds to Hurst exponents close to 0.5, reinforcing the earlier DFA and R/S findings that the generated return series behave more like short-memory or near-random processes in their mean dynamics, even though their conditional variances continue to exhibit persistent clustering effects similar to those observed in real financial markets.

\subsection{Complexity Considerations}\label{sec:complexity}
The computational cost of the Quant GAN framework is mainly driven by adversarial training and path generation. Let $N$ denote the length of each return series, $E$ the number of training epochs, and $B$ the batch size. For a TCN layer with kernel width $k$, input channels $c_{\rm in}$, and output channels $c_{\rm out}$, one forward pass has order $O(BNk c_{\rm in}c_{\rm out})$. Therefore, the per-epoch cost of the generator and discriminator is approximately linear in the sequence length and in the number of convolutional filters, with the total training cost scaling as
\[
\mathcal{O}\left(EBN\sum_{\ell=1}^{L} k_\ell c_{\ell-1}c_\ell\right),
\]
up to constant factors associated with backpropagation and discriminator updates. This complexity is higher than that of classical diagnostic tools such as R/S and DFA, which require only repeated block-level calculations and log--log regressions, but it remains parallelizable because TCN operations can be evaluated across time points and paths.

From a numerical perspective, the empirical evaluation is nontrivial because it involves both large-scale path generation and repeated post-generation diagnostics. Table~\ref{tab:numerical_complexity} summarizes the main computational quantities. The study generates 10,000 synthetic paths for each asset. With $N=3318$ daily returns per path, this corresponds to $10{,}000\times 3{,}318=33{,}180{,}000$ generated return observations for each asset. Since the empirical application covers nine assets, the total generated sample size is $9\times 33{,}180{,}000=298{,}620{,}000$, i.e., approximately 298.62 million generated observations. Stored as dense double-precision arrays, these generated returns require about 265 MB per asset and 2.39 GB across all nine assets, excluding text-file overhead and intermediate objects.

\begin{table}[htbp]
\centering
\begin{tabular}{l c c}
\hline
Quantity & Per asset & Nine assets\\
\hline
Generated paths & 10,000 & 90,000\\
Return observations per path & 3,318 & 3,318\\
Generated return observations & 33.18 million & 298.62 million\\
Approximate GAN training time & $\approx$1 h & $\approx$9 h if run serially\\
Dense double-array storage & 265 MB & 2.39 GB\\
R/S and DFA ensemble fits & 20,000 & 180,000\\
Full-ensemble R/S--DFA validation time & $\approx$37 s & $\approx$5.6 min\\
\hline
\end{tabular}
\caption{Numerical computational workload of the Quant GAN empirical evaluation}
\label{tab:numerical_complexity}
\end{table}

The reported training time is approximate and reflects our use of the default Quant GAN implementation and hyperparameter configuration without modification; exact runtime depends on the hardware and software environment. The last two rows emphasize that the computational burden does not end after path generation. The full-ensemble robustness check estimates both R/S and DFA Hurst exponents for every generated path, resulting in 20,000 Hurst estimations per asset and 180,000 estimations across the full empirical design. In the CPU-based implementation used for the revision, this full-ensemble R/S--DFA validation required approximately 5.6 minutes for all nine assets. Additional representative-path evaluation further requires distributional diagnostics, volatility-dependence measures, Wasserstein and tail-error calculations, and ARFIMA--FIGARCH estimation. These requirements imply that practical use of Quant GANs for financial simulation should report not only distributional accuracy but also memory-preservation diagnostics and computational cost, especially when the target application involves stress testing or long-horizon risk evaluation.

\section{Discussion and Conclusion}\label{sec:conclusion}
This study provides both methodological and economic insights into the nature of LRD in financial markets and the ability of deep generative models to replicate such structures. Across equity, commodity, and energy assets, the empirical analyses indicate that long memory is more pronounced in conditional volatility than in mean returns. This pattern is consistent with the economic intuition that while price changes themselves may approximate weak-form efficiency, the intensity of market fluctuations can remain persistent because of gradual information diffusion, heterogeneous investor behavior, and prolonged macroeconomic uncertainty. In this sense, LRD in volatility can be interpreted as a channel through which shocks propagate across time rather than disappearing instantaneously.

From a practical standpoint, the presence of volatility-based long memory has direct implications for risk management. Models that neglect slow-decaying dependence structures may systematically underestimate long-horizon risk, leading to biased Value-at-Risk and Expected Shortfall estimates. Portfolio allocation strategies that rely on short-memory assumptions may also fail to capture the persistence of turbulent periods, thereby exposing investors to prolonged drawdowns. The findings suggest that incorporating fractional or long-memory components into volatility models can materially improve forecasting accuracy and stress-testing reliability, especially in multi-asset or cross-sector portfolios.

The evaluation of Quant GANs reveals a nuanced picture. The model reproduces heavy-tailed return distributions and certain aspects of volatility clustering, but it does not consistently capture the magnitude of LRD observed in empirical volatility. This limitation is economically meaningful: synthetic data that underestimate persistence may lead to overly optimistic assessments of diversification benefits or tail risk. Consequently, practitioners employing GAN-generated financial data for scenario analysis or regulatory reporting should exercise caution, particularly when the objective involves long-term forecasting or systemic risk evaluation.

From a policy perspective, persistent volatility dynamics imply that financial shocks may have enduring effects on market stability. Regulatory frameworks that assume rapid mean reversion in risk measures may underestimate the duration of stressed conditions. The limited ability of current generative models to replicate such persistence also highlights the importance of transparent model validation standards when synthetic data are used in supervisory or compliance contexts.

In conclusion, this study underscores both the robustness of volatility-driven long-range dependence in diverse financial markets and the current limitations of deep generative models in learning slow temporal structures. Future research may benefit from integrating explicit fractional dynamics, multi-scale attention mechanisms, or hybrid econometric-deep learning architectures to better capture long-memory behavior. Additional extensions could examine formal structural-break tests beyond the COVID-19 segmentation used here, compare alternative generative architectures and hyperparameter configurations, and evaluate forecasting-oriented objectives when the research question is predictive performance rather than synthetic time-series fidelity. For the econometrics community, particularly promising directions include combining generative models with causal inference, structural identification, and interpretable nonlinear dynamics. Such extensions would help distinguish persistent statistical dependence from economically meaningful transmission channels, identify whether long-memory features arise from regime shifts or structural shocks, and make synthetic financial data more transparent for policy and risk-management use. Progress in this direction would not only enhance the realism of synthetic financial data but also strengthen their applicability in risk management, asset allocation, and financial regulation.

\bibliography{paper_r2}
\bibliographystyle{apalike}
\end{document}